\documentclass[11pt,a4paper]{article}
\usepackage{jheppub}
\pdfoutput=1

\usepackage[utf8]{inputenc}
\usepackage{amsfonts,amsmath,amssymb,amstext,amsbsy}
\usepackage{amsfonts}
\usepackage{graphicx,graphics,subfigure,epsfig}
\usepackage{xcolor,cancel,youngtab}
\usepackage{xspace}
\usepackage[normalem]{ulem} %\usepackage[normalem]{ulem} if want italics
\usepackage{lscape}
\usepackage{wasysym}
\usepackage{float}
\usepackage{mathrsfs}
\usepackage{slashed}
\usepackage{multirow,rotating}
\usepackage{braket,slashed,bm}
\usepackage{float}
\usepackage{textcomp}               
\usepackage{braket,slashed,bm}
\usepackage{array}
\usepackage{scalerel}
\allowdisplaybreaks

\usepackage{hyperref}
\hypersetup{
  colorlinks=true,
  allcolors=darkpurple,
  pdfborder={0 0 0},
  linktocpage=true%false
}
\definecolor{darkpurple}{rgb}{0.5,0,0.5}

\usepackage{xcolor,cancel,youngtab}

\usepackage[T1]{fontenc} % if needed

\usepackage{mathrsfs}
\usepackage{booktabs}
\usepackage{adjustbox}
\usepackage{mathtools}
\usepackage{comment}
\usepackage{soul}
\usepackage{tikz}
\usetikzlibrary{arrows,decorations.pathmorphing,backgrounds,positioning,fit,petri,automata,shadows,calendar,mindmap, decorations.markings,calc}

\usepackage{dsfont}
\usepackage{color}
\usepackage{pdflscape,pgfplots,pgfplotstable}
\pgfplotsset{compat=1.18}
\usepackage{pifont}
\usepackage{bbold}
\usepackage[italic]{hepnames}
\usepackage[output-decimal-marker={.},uncertainty-separator = {\pm}, separate-uncertainty = true]{siunitx}
\usepackage{preprint_defs} % To declare new commands

\DeclareUnicodeCharacter{2212}{\ensuremath{-}}

\makeatletter
\g@addto@macro\bfseries{\boldmath}
\newcommand\Label[1]{&\refstepcounter{equation}(\mathrm{\theequation})\ltx@label{#1}&}
\makeatother

\graphicspath{{pictures/},{figs/},{figs/numerical_results/},{figs/decay_modes/}}

\newcommand*{\LLNF}{\ensuremath{ {N}}}
\newcommand*{\NUSMEFT}{\ensuremath{\nu}SMEFT}
\newcommand*{\LHC}{\texttt{LHC}}
\newcommand*{\Pythia}{\texttt{PYTHIA8}}
\newcommand*{\Dune}{\texttt{DUNE}}
\newcommand*{\DUNE}{\texttt{DUNE}}
\newcommand*{\ATLAS}{\texttt{ATLAS}}
\newcommand*{\CMS}{\texttt{CMS}}
\newcommand*{\ANUBIS}{\texttt{ANUBIS}}
\newcommand*{\CODEXB}{\texttt{CODEX-b}}
\newcommand*{\FACET}{\texttt{FACET}}
\newcommand*{\FASEROne}{\texttt{FASER}}
\newcommand*{\FASERTwo}{\texttt{FASER2}}
\newcommand*{\MAPPOne}{\texttt{MAPP1}}
\newcommand*{\MAPPTwo}{\texttt{MAPP2}}
\newcommand*{\MATHUSLA}{\texttt{MATHUSLA}}

\newcommand{\lam}[1]{{\lambda}}

\allowdisplaybreaks
	
\title{Long-lived neutral fermions at the DUNE near detector}

\author[a]{Julian~Y.~G\"unther,}
\emailAdd{guenther@physik.uni-bonn.de}
\affiliation[a]{Bethe Center for Theoretical Physics \& Physikalisches Institut der 
	Universit\"at Bonn,\\ Nu{\ss}allee 12,	53115 Bonn, Germany}

\author[b,c]{Jordy~de~Vries,}
\emailAdd{j.devries4@uva.nl}
\affiliation[b]{Institute for Theoretical Physics Amsterdam and Delta Institute for Theoretical Physics, University of Amsterdam, Science Park 904, 1098 XH Amsterdam, The Netherlands}
\affiliation[c]{Nikhef, Theory Group, Science Park 105, 1098 XG, Amsterdam, The Netherlands}

\author[a]{Herbi~K.~Dreiner,}
\emailAdd{dreiner@uni-bonn.de}

\author[d,e]{Zeren Simon Wang,}
\emailAdd{wzs@mx.nthu.edu.tw}
\affiliation[d]{Department of Physics, National Tsing Hua University, Hsinchu 300, Taiwan}
\affiliation[e]{Center for Theory and Computation, National Tsing Hua University, Hsinchu 300, Taiwan}

\author[b,c,f]{Guanghui~Zhou}
\emailAdd{zhough.uva@gmail.com}
\affiliation[f]{CAS Key Laboratory of Theoretical Physics, Institute of Theoretical Physics,
Chinese Academy of Sciences, Beijing 100190, P.~R.~China}

\abstract{At the Deep Underground Neutrino Experiment (DUNE), a proton beam hits a fixed target leading to large production rates of mesons.
These mesons can decay and potentially provide a source of long-lived neutral fermions.
Examples of such long-lived fermions are heavy neutral leptons which can mix with the standard-model active neutrinos, and the bino-like lightest neutralino in R-parity-violating supersymmetry.
We show that the Standard Model Effective Field Theory extended with right-handed singlet neutrinos can simultaneously describe heavy neutral leptons and bino-like neutralinos in a unified manner.
We use the effective-field-theory framework to determine the sensitivity reach of the DUNE near detector in probing various scenarios of long-lived neutral fermions.
}

\begin{document}
\maketitle

%\tableofcontents

%!TEX root= dune_eft.tex
\section{Introduction}\label{sec:intro}
\setcounter{footnote}{0}

At the Deep Underground Neutrino Experiment 
(\texttt{DUNE})~\cite{DUNE:2020lwj,DUNE:2020ypp,DUNE:2020jqi,DUNE:2021cuw,DUNE:2021mtg}, currently under construction, a wide range of physics goals will be pursued~\cite{DUNE:2015lol}.
Key examples are the study of neutrino oscillations to determine the neutrino-mass ordering and the $CP$-violating Dirac mass mixing phase, the search for proton decay, as well as the detection of supernova neutrino bursts. 
In addition, \texttt{DUNE} will serve as a beam-dump-type experiment where a proton beam impinges on a fixed 
target, 
generating a large number of mesons. These mesons can in rare cases potentially decay into light particles 
that appear in broad classes of Beyond-the-Standard-Model (BSM) theories. To have avoided detection so 
far, such particles must interact feebly with Standard Model (SM) fields causing them to be long-lived. 
See Refs.~\cite{Alimena:2019zri,Lee:2018pag,Curtin:2018mvb,Knapen:2022afb} for some recent reviews on 
long-lived particle (LLP) searches. 

At \texttt{DUNE}, LLPs can be produced from rare meson decays, travel a macroscopic distance, and then decay in a detector far from the original fixed-target collision.
While the \texttt{DUNE} far detector, with a distance of 1300\,km from the beam target, is expected to have very limited sensitivities for LLPs, the \texttt{DUNE} near detector (\texttt{DUNE-ND}) is much closer to the beam target (574\,m) and will be equipped with state-of-the-art event-reconstruction capabilities~\cite{DUNE:2021tad}. The \texttt{DUNE-ND} will 
therefore be an ideal device for detecting LLPs. Phenomenological studies on LLP 
detection at the \texttt{DUNE-ND} have already been performed in \textit{e.g.}~Refs.~\cite{Batell:2022xau, Costa:2022pxv, 
Dev:2021qjj, Co:2022bqq, Kelly:2020dda, Krasnov:2019kdc, Ballett:2019bgd, Berryman:2019dme, Coloma:2020lgy, Ovchynnikov:2022rqj, Moghaddam:2022tac, Plows:2022gxc,Coloma:2023oxx}.

In this work, as an example of LLPs, we focus on long-lived neutral fermions (LLNFs). A famous example is provided by heavy neutral leptons (HNLs), which are among the most sought-after candidates of LLNFs.
They are SM singlets that mix with active neutrinos and can be the origin of neutrino 
masses~\cite{Minkowski:1977sc,  Yanagida:1979as, Gell-Mann:1979vob, Mohapatra:1979ia, Schechter:1980gr}. 
In minimal scenarios, the HNLs only interact with SM fields through mixing with the active neutrinos.
In more general scenarios, the phenomenology of the HNLs can be very different.
Examples include SM extensions with an extra $U(1)$ gauge symmetry and an associated gauge boson ($Z'$)~\cite{Davidson:1978pm, Mohapatra:1980qe, Deppisch:2019kvs, Chiang:2019ajm, Davidson:1987mh,Appelquist:2002mw} or left-right 
symmetric models~\cite{Mohapatra:1974gc, Pati:1974yy, Mohapatra:1980yp, Keung:1983uu}, where the HNLs interact with right-handed gauge bosons.

A very useful framework to describe these possible HNLs interactions in a unified way is the neutrino-extended standard model effective field theory ($\nu$SMEFT)~\cite{delAguila:2008ir, Aaboud:2017qph, Liao:2016qyd, Chala:2020vqp,Bischer:2019ttk}.
The $\nu$SMEFT Lagrangian contains all operators involving only HNLs (assumed to be singlets under the SM gauge group) and SM fields that are invariant under Lorentz and SM gauge symmetries.
The main assumption is that the only relatively light BSM particles are the HNLs and all other BSM particles are heavy degrees of freedom and can be integrated out.
The $\nu$SMEFT framework can be used to efficiently describe a wide range of possible HNL scenarios and has already been used to describe HNL phenomenology at the LHC~\cite{DeVries:2020jbs, Cottin:2021lzz, Beltran:2021hpq}, neutrinoless double beta decay and related searches~\cite{Dekens:2020ttz, Zhou:2021lnl}, meson decays~\cite{Li:2020lba,Beltran:2022ast,Beltran:2023nli,Beltran:2023ksw}, $B$-factories such as Belle II~\cite{Zhou:2021ylt,Han:2022uho}, as well as fixed-target experiments~\cite{DeVries:2020jbs}.

Here, we point out that $\nu$SMEFT can describe not only the HNLs but also other SM singlets predicted in various BSM theories.
For instance, the lightest neutralino in the R-parity-violating minimal supersymmetric model (MSSM) (RPV-SUSY)~\cite{Dreiner:1997uz, Allanach:2003eb, Barbier:2004ez, Mohapatra:2015fua}, if bino-like, can be very light, \textit{i.e.}~at the GeV-scale or even massless~\cite{Choudhury:1999tn, Dedes:2001zia, Gogoladze:2002xp, Dreiner:2009ic, Domingo:2022emr}.
Such a light neutralino is allowed by all laboratory, as well as cosmological and observational constraints~\cite{Grifols:1988fw, Ellis:1988aa, Lau:1993vf, Dreiner:2003wh, Dreiner:2013tja, Profumo:2008yg, Dreiner:2011fp, Dreiner:2023gir}.
In these scenarios, dark matter is not comprised of the lightest neutralino~\cite{Chun:1999cq, Belanger:2002nr, Hooper:2002nq, Bottino:2002ry, Belanger:2003wb, AlbornozVasquez:2010nkq, Calibbi:2013poa, Colucci:2018yaq}, and as long as these light binos decay (which could be triggered by non-vanishing RPV operators), overclosing of the Universe is avoided~\cite{Bechtle:2015nua}.
Such light binos are SM singlets by definition and can be produced from meson decays through RPV operators~\cite{Dedes:2001zia, Dreiner:2001kc, 
Dreiner:2002xg, Dreiner:2009er}.
These decays are mediated by scalar fermions which are currently bounded from below at the TeV scale~\cite{Dercks:2017lfq, ATLAS:2018nud, CMS:2017brl, CMS:2019vzo, CMS:2019zmd, ATLAS:2020xgt, Dreiner:2023bvs}.
One can integrate out the scalar fermions and derive effective operators involving quarks, a lepton, and a light bino only~\cite{deVries:2015mfw, Domingo:2022emr}.
These operators are already included in the $\nu$SMEFT Lagrangian, if we interpret 
the right-handed SM singlets in the model as the light bino in the RPV-SUSY. 
Furthermore, if the RPV couplings are small enough, the light neutralino can easily be long-lived, leading to displaced-vertex signatures at colliders~\cite{deVries:2015mfw,Cottin:2022gmk,Wang:2019orr} including proposed 
far detectors~\cite{Dercks:2018eua,Dercks:2018wum,Dreiner:2020qbi,Dreiner:2022swd}, beam-dump experiments~\cite{Dedes:2001zia, Dreiner:2002xg, deVries:2015mfw, SHiP:2015vad}, $B$-factories~\cite{Dey:2020juy, Dib:2022ppx}, and even neutrino experiments~\cite{Candia:2021bsl}.

In this paper, we investigate the sensitivity reach of the \texttt{DUNE-ND} to both HNLs and light neutralinos within the framework of the $\nu$SMEFT.
The paper is organized as follows.
We first introduce the $\nu$SMEFT framework in Sec.~\ref{sec:nuSMEFT_framework}, followed by Sec.~\ref{sec:theoretical_scenarios} listing and discussing the theoretical scenarios we study.
The production and decay of the LLNFs are then discussed in Secs.~\ref{sec:LLNF_production} and \ref{sec:LLNF_decay}, respectively.
In Sec.~\ref{sec:DUNE} we describe the \DUNE{} experiment including the \texttt{DUNE-ND}, and elaborate on our simulation procedure.
We then present our numerical results in Sec.~\ref{sec:results} and conclude the paper in Sec.~\ref{sec:conclusions}.
In addition to the main text, in Appendix~\ref{sec:appendix_decayratecalc}, we derive the relevant meson decay constants and form factors, and calculate all employed decay rates.
In Appendix~\ref{sec:appendix_phenomenological_values}, we list the numerical values of the decay constants we use and the parameterization of the form factors.
Finally, we provide Appendix~\ref{sec:appendix_rpv_derivation} where we present the derivation of 4-fermion effective interactions in the RPV-SUSY with the heavy sfermion fields integrated out.

%!TEX root= dune_eft.tex
\section{The \texorpdfstring{$\nu$}{}SMEFT framework}\label{sec:nuSMEFT_framework}
Sterile neutrinos $\nu_R$ are spin-1/2, right-handed gauge-singlet fields under the SM gauge group.
Without introducing additional fields, the only renormalizable interaction with SM fields is through the neutrino portal.
For $N_\nu$ sterile neutrinos the renormalizable interactions are given by
\bea
\mathcal{L} = \mathcal{L}_{SM}-\bigg[\ep_{ab}\big(L_a\big)^T_i \tilde H_b \big(Y_\nu\big)_{ij}\nu_{Rj} + \frac{1}{2}\bar{\nu}^c_{Ri} \bar{M}_{Rij}\nu_{Rj} + \hc\bigg]\,,\label{eq:sn_extended_lagrangian}
\eea
where $\mathcal{L}_{SM}$ is the SM Lagrangian.
$H$ is the SM complex scalar Higgs $SU(2)_L$ doublet, which in the unitary gauge becomes
\bea
	H = \frac{v}{\sqrt{2}}\rm  \left(\begin{array}{c}
		0 \\
		1 + \frac{h}{v}
	\end{array} \right)\,,\label{eq:higgs_unitary_gauge}
\eea
in terms of the Higgs boson $h$ and the Higgs vacuum expectation value $v\simeq\SI{246}{\giga\eV}$.
In the Yukawa coupling we use $ \tilde H =i\tau_2H^*$.
$(L_a)_i$ denote the SM $SU(2)_L$ lepton doublet fields, with $SU(2)_L$ gauge indices $a,b\in\{1,2\}$ for the fundamental representation, and generation indices $i\in\{1,2,3\}$.
$\ep_{kl}$ denotes the anti-symmetric tensor with $\ep_{12}=+1$.
$\nu_{Rj}$ are the sterile neutrino fields with generation indices $j\in\{1,2,\dots,N_\nu\}$.
$Y_\nu$ is then a $3\times N_\nu$ matrix consisting of Yukawa couplings and $\bar{M}_R$ is a complex symmetric Majorana mass matrix.
$\psi^c$ is the charge conjugate field of $\psi$, where we define $\psi^c\equiv C\bar{\psi}^T$, with the charge conjugation matrix $C=-i\gamma^2\gamma^0$.
For chiral spinors $\psi_{L,R}=P_{L,R}\psi$ with the projection operators $P_{L,R}=(1\mp\gamma_5)/2$, we define $\psi^c_{L,R}\equiv (\psi_{L,R})^c\equiv C\overline{\psi_{L,R}}^T=P_{R,L}\psi^c$.

The Lagrangian in Eq.~\eqref{eq:sn_extended_lagrangian} corresponds to a minimal model where sterile neutrinos only interact through mixing with left-handed active neutrinos.
It is not unlikely that sterile neutrinos only appear sterile, but actually have additional interactions that are decoupled at low energies.
A possible scenario would be left-right symmetric models where sterile neutrinos are charged under non-SM gauge symmetries and interact through the exchange of right-handed gauge bosons.
Assuming that such interactions arise from the exchange of particles with masses well above the electroweak (EW) scale, at lower energies they can be described with the $\nu$SMEFT Lagrangian
\bea
\mathcal{L}_{\nu\mathrm{SMEFT}}=&\mathcal{L} + \sum\limits_{d>4}\sum\limits_{i} C_i^{(d)}
\mathcal{O}_i^{(d)}\,.
\label{eq:lagrangian_nsmeft}
\eea
Here, $\mathcal{O}_i^{(d)}$ is a dimension $d$ operator consisting just of SM fields and $\nu_{Ri}$, and $C_i^{(d)}$ are Wilson coefficients that scale as $(\Lambda)
^{4-d}$.
The experiments we consider involve energy scales well below the electroweak scale, and the $\nu$SMEFT Lagrangian can be matched to an effective field theory (EFT) where heavy degrees of freedom (\textit{e.g.}~Higgs, top, $W^\pm$, $Z$, and potentially some sterile neutrinos) are integrated out.
The resulting EFT satisfies $SU(3)_c\times U(1)_{\text{em}}$ gauge symmetry.
We give more details below.

For simplicity, we consider only one sterile neutrino ($N_\nu=1$), as including more sterile neutrinos does not significantly impact the observables under consideration.
With only one sterile neutrino, the dimension-5 $\nu$SMEFT operators contribute to the Majorana masses of $\nu_L$ and $\nu_R$ fields after electroweak symmetry breaking.
They can be absorbed into the active and sterile neutrino masses and we do not consider them here.

{\renewcommand{\arraystretch}{1.3}\begin{table}[t!]\small
		\center
		\begin{tabular}{||c|c||c|c||}
	\hline		
   & $\psi^2 H^3$  &  &  $\psi^4 $\\
			\hline
			$\mathcal{O}^{}_{L\nu H}$ & $(\bar{L}\nu_R)\tilde{H}(H^\dagger H)$ & $\mathcal{O}^{}_{du\nu e}$ & $ (\bar{d}_R
			\gamma^\mu u_R)(\bar{\nu}_R \gamma_\mu e)$  
			\\ \cline{1-2} &  $\psi^2 H^2 D$ &  $\mathcal{O}^{}_{Qu\nu L}$ & $(\bar{Q}u_R)(\bar{\nu}_RL)$  \\ \cline{1-2}  
			$\mathcal{O}^{}_{H\nu e}$ & $(\bar{\nu }_R\gamma^\mu e_R)({\tilde{H}}^\dagger i D_\mu H)$ & $\mathcal{O}^{}_{L\nu Qd}$ 
			& $(\bar{L}\nu_R )\epsilon(\bar{Q}d_R)$
			\\ \cline{1-2}  &  $\psi^2 H F$  & $\mathcal{O}^{}_{LdQ\nu }$ & $(\bar{L}d_R)\epsilon(\bar{Q}\nu_R )$ \\ \cline{1-2}
			$\mathcal{O}^{}_{\nu W}$ &$(\bar{L}\sigma_{\mu\nu}\nu_R )\tau^I\tilde{H}W^{I\mu\nu}$  & $\mathcal{O}^{}_{L\nu Le}$ 
			&$(\bar{L}\nu_R)\epsilon (\bar{L}e_R)$ \\ 
			$\mathcal{O}^{}_{\nu B}$ &$(\bar{L}\sigma_{\mu\nu}\nu_R )\tilde{H}B^{\mu\nu}$& & \\
			\hline
		\end{tabular}
		\caption{Dimension-6 operators that involve one $\nu_R$ field.
		Each fermion field has a generation index, except $\nu_R$, which, if necessary, are explicitly indicated. For example, $\mathcal{O}^{21 1}_{du\nu e}$ refers to the operator $(\bar{s_R}\gamma^\mu u_R)(\bar{\nu}_R \gamma_\mu e_R)$. We work in the basis where all the SM fields are mass
eigenstates, except the left-handed down-type quarks $d^i_L$, for which 
we have $d^i_L = V^{ij}d^{j,\text{mass}}_L$ with $V$ being 
the CKM matrix and $d^{j, \text{mass}}_L$ the down-like quark 
mass eigenstates.
		} \label{tab:O6R}
\end{table}}

At the level of the dimension-6 operators, we restrict ourselves to the operators involving only one sterile neutrino field; they are shown in Table~\ref{tab:O6R}.
Not all operators are relevant for our purposes.
The $\mathcal{O}_{L\nu H}$ operator can be effectively absorbed in the dimension-4 $L\nu H$ interaction in Eq.~\eqref{eq:sn_extended_lagrangian}, up to terms with 
explicit Higgs bosons, which play no role in our analysis.
The dipole operators $\mathcal{O}^{(6)}_{\nu W}$ and $\mathcal{O}^{(6)}_{\nu B}$ are strongly constrained by searches for neutrino dipole moments~\cite{Butterworth:2019iff,Canas:2015yoa} and mainly lead to displaced-vertex signatures of a single photon which are relevant only for a limited set of the proposed experiments~\cite{Barducci:2022gdv,Liu:2023nxi,Barducci:2023hzo}.
We therefore only keep the $O_{H\nu e}$ operator and the five four-fermion operators.

We apply one-loop QCD anomalous dimensions to  evolve these operators from $\Lambda$ to the electroweak scale $v$, except $\mathcal {O}^{(6)}_{H\nu e}$, which does not evolve under the one-loop QCD.
For the remaining operators, we find~\cite{Dekens:2020ttz} 
\bea\label{eq:STrge}
\frac{d C^{(6)}_{Qu\nu L}}{d\ln\mu} = 
\left(\frac{\al_s}{4\pi}\right)\gamma_{S}\,C^{(6)}_{Qu\nu L}\,, 
\qquad \frac{dC^{(6)}_{S}}{d\ln\mu} = 
\left(\frac{\al_s}{4\pi}\right)\gamma_{S}\,C^{(6)}_{S }\,,
\qquad \frac{dC^{(6)}_{T}}{d\ln\mu} = 
\left(\frac{\al_s}{4\pi}\right)\gamma_{T}\,C^{(6)}_{T}
\,,\;\;\;
\eea
where $\gamma_S = -6 C_F$ and $\g_T = 2C_F$, with $C_F = (N_c^2-1)/(2N_c)$ and $N_c=3$, the number of colors. $C^{(6)}_S$ and $C^{(6)}_T$ are defined as 
\bea
C_{S}^{(6)} =-\frac{1}{2} C_{LdQ\nu}^{(6)}+C_{L\nu 
Qd}^{(6)}\,,\qquad C_{T}^{(6)} =-\frac{1}{8} C_{LdQ\nu}^{(6)}\,.
\eea

At the electroweak scale we integrate out the heavy SM fields and we 
match to the neutrino-extended low-energy EFT ($\nu$LEFT).
For our purpose, the most important operators are Fermi-like charged-current operators
\bea\label{d6CC}
\mathcal L^{(6)}_{CC}& =& \frac{2 G_F}{\sqrt{2}} \sum_{i,j,k} \Bigg\{ 
\bar u_L^{i} \gamma^\mu d_L^{j} \left[  \bar e_{L}^{k}   \gamma_\mu (-2V_{ij}) \,  \nu_{L}^{k} + \bar e_{R}^{k}   
\gamma_\mu  c^{CC1}_{\textrm{VLR},ijk} \,  \nu_{R} \right]+
\bar u_R^{i}  \gamma^\mu d_R^{j} \,\bar e_{R}^{k} \,  \gamma_\mu   c^{CC}_{\textrm{VRR}, ijk} \,\nu_{R}\nn \\
& & +
\bar u_L^{i}   d_R^{j}  \,\bar e_{L}^{k} \,  c^{CC}_{ \textrm{SRR}, ijk}  \nu_{R} + 
\bar u_R^{i}   d_L^{j}  \, \bar e_{L}^{k}  \,  c^{CC}_{ \textrm{SLR}, ijk}    \nu_{R} +  \bar u_L^{i}  \sigma^{\mu\nu} d_R^{j} \,  
\bar e_{L}^{k}   \sigma_{\mu\nu}  c^{CC}_{ \textrm{T}, ijk} \, \nu_{R}\nn\\
& &   +\bar{\nu}^i_L\gamma^\mu e^j_L \bar{e}_R^{k} \gamma_\mu c^{CC2}_{\textrm{VLR}, ijk}\nu_R  +{\rm h.c.} \Bigg\} 
- \frac{4G_F}{\sqrt{2}}\sum_{i,j}\bar{\nu}^i_L\gamma^\mu e^i_L \bar{e}^j_L \gamma_\mu \nu^j_L \label{lowenergy6_l0},
\eea
in terms of chiral charged up-type quarks, $u^i_{L,R}$, down-type quarks $d^i_{L,R}$, charged leptons $e^i_{L,R}$, and left-handed neutrinos $\nu_L^i$, where $i=\{1,2,3\}$ denote generation indices.

For the neutral-current interactions, we obtain\footnote{Except for the SM weak 
interactions, no neutral (axial-)vector-like currents are present in the $\mathcal{L}_{\nu\mathrm{LEFT}}$. 
None of the dimension-6 operators introduced in Table~\ref{tab:O6R} include 
such currents, neither directly nor after rearranging the currents with 
Fierz-identities.}
\begin{eqnarray}\label{d6NC}
		\mathcal L^{(6)}_{\rm NC}&=&\frac{-4 G_F}{\sqrt{2}}\sum_{i,j}
		\bar \nu^i_L \gamma^\mu \nu^i_L\Bigg\{ 
		\bar e^j_{L}  \gamma_\mu (-\frac{1}{2}+\sin^2\theta_W)  e^j_L+ 
		\bar e^j_{R}  \gamma_\mu (\sin^2\theta_W) e^j_R\nonumber \\ %\allowdisplaybreaks
		&&+   \bar u^j_L \gamma^\mu (\frac{1}{2}-\frac{2}{3}\sin^2\theta_W)u^j_L\,+\bar u^j_{R}\,  \gamma_\mu   (-\frac{2}{3}\sin^2\theta_W)u^j_R \nonumber\\ 
		&&+  \bar d^j_L \gamma^\mu (-\frac{1}{2}+\frac{1}{3}\sin^2\theta_W)d^j_L\,+\bar d^j_{R}\,  \gamma_\mu   (\frac{1}{3}\sin^2\theta_W)d^j_R 
		+ \frac{1}{4}(2-\delta_{ij})\bar \nu^j_L \gamma^\mu \nu^j_L \Bigg\}\,\\
		&&+\frac{2G_F}{\sqrt{2}}\sum_{i,j,k}\Bigg\{ \bar{u}_R^{i}  u_L^{j} \bar{\nu}_L^{k}  c^{NC}_{\textrm{SLR}, ijk}\nu_R+ 
		\bar{d}_L^{i} d_R^{j} \bar{\nu}_L^{k}  c^{NC}_{\textrm{SRR},ijk}\nu_R   
		+\bar{d}_L^{i} \sigma^{\mu\nu}d_R^{j}  \bar{\nu}_L^{k} \sigma_{\mu\nu} c^{NC}_{\textrm{T}, ijk}\nu_R +\rm h.c. \Bigg\}, \nonumber
\end{eqnarray}
with $\theta_W$ being the electroweak mixing angle.

The dimensionless Wilson coefficients, $c$, are then obtained from a tree-level matching calculation to the $\nu$SMEFT Lagrangian at the electroweak scale~\cite{Dekens:2020ttz,Zhou:2021ylt}.
For charged currents we use
\bea\label{match6CC}
c_{\textrm{VLR}, ijk}^{CC1} &=& \left[-v^2C_{H\nu e, k}^{}\right]^\dagger V_{ij}\,,\qquad\qquad \qquad \;\;\,c_{\textrm{VLR}, ijk}^{CC2} = \left[-v^2C_{H\nu e, k}\right]^\dagger \delta_{ij}\,,\nn\\
c_{\textrm{ VRR}, ijk}^{CC} &=& v^2\left(C_{du\nu e, jik}^{}\right)^\dagger\,,\qquad\qquad \qquad \quad \quad\;\;\,c_{\textrm{ T}, ijk}^{CC} = \frac{v^2}{8} C_{LdQ\nu,kji }^{}\,,\nn\\
c_{\textrm{SRR},ijk}^{CC}&=& -v^2C_{L\nu Qd, kij}^{}+\frac{v^2}{2} C_{LdQ\nu,kji }^{}\,,\qquad
c_{\textrm{ SLR}, ijk}^{CC}= v^2\left(C_{Qu\nu L, lik}^{}\right)^\dagger V_{lj}\,,
\eea
where repeated flavor indices are summed over.
For neutral currents, we find
\bea\label{match6NC}
c_{\textrm{ SLR}, ijk}^{NC} &=& v^2\left(C_{Qu\nu L,jik}^{}\right)^\dagger \,, \qquad 
c_{\textrm{ SRR},ijk}^{NC} =  v^2C_{L\nu Qd,k lj}^{}V_{li}^*-\frac{v^2}{2} C_{LdQ\nu,kjl }V_{li}^*\,,\nn\\
 c_{\textrm{ T},ijk}^{NC}&=& -\frac{v^2}{8} C_{LdQ\nu,kjl }^{} V_{li}^*\,.
\eea

Finally, we evolve both charged- and neutral-current operators from the EW scale down to the QCD scale $\Lambda_{\text{QCD}}\simeq $ 1 GeV,  using the one-loop QCD running.
The vector-type interactions do not evolve.
Only the scalar- and tensor-type operators in Eqs.~\eqref{match6CC} and \eqref{match6NC} evolve analogously to Eq.~\eqref{eq:STrge}. 
This evolution leads to an enhancement of the scalar-type interactions of roughly 60\%, while the tensor-type ones are suppressed by 15\%~\cite{Liao:2020roy}.

%!TEX root= dune_eft.tex
\section{Theoretical scenarios}\label{sec:theoretical_scenarios}

Instead of an exhaustive analysis of the very large parameter space of the $\nu$SMEFT, we consider four representative scenarios in which a single long-lived neutral fermion \LLNF{} is kinematically relevant:
\begin{enumerate}
\item  We first study the minimal case, where we turn off all higher-dimensional operators and the sterile neutrino only interacts with the SM fields through the SM weak interaction after mixing.
Active neutrinos $\nu^k_L$ are related to neutrino mass eigenstates $\nu_i$ through
\bea
\nu^k_L= U_{k l} \nu_l, \label{eq:neutrino_mass_mixing}
\eea
where $l=1,2,3,4$ and $k= e,\mu,\tau$.
We focus on the mixing angle $U_{e4} \neq 0$ for the sterile neutrino.
We keep $U_{e4}$ as a free parameter (considering the stringent bounds on $U_{e4}$ we can freely interchange $\nu_R$ and $\nu_4$).
The first interaction in Eq.~(\ref{d6CC}) then includes the term
\begin{equation}
   \mathcal L^{(6)}_{CC} =\frac{2 G_F}{\sqrt{2}} \sum_{i,j}
c_{\textrm{VLL},ij}^{CC}\big(\bar u_L^{i} \gamma^\mu d_L^{j}\big)\big(   \bar e_{L}   \gamma_\mu \,  \nu_{R}\big)
\end{equation}
where we introduced the coefficient $c_{\textrm{VLL},ij}^{CC}=-2 V_{ij} U_{e4}$.

\item In the second scenario, we introduce leptoquarks, which are hypothetical scalar 
or vector particles that allow for the transition of leptons into quarks and vice versa~\cite{Buchmuller:1986zs}.
A multitude of leptoquark representations are possible (for an extensive review, see
Ref.~\cite{Dorsner:2016wpm}).
As an example, we focus on matter interacting with the leptoquark representation $\tilde R_2$ with SM gauge group $\big(SU(3)_C,\,SU(2)_L,$ $U(1)_{Y}\big)$ transformation properties $\big({\mathbf{3}},~{\mathbf{2}},~1/6 \big)$.
This leads to the Lagrangian
\bea
{\cal L}_{\mathrm{LQ}}=-{y}^{RL}_{jk}\bar{d}_{Rj}\tilde R^a_2\epsilon^{ab}L_{Lk}^{b}+y^{\overline {LR}}_{i}\bar{Q}^{a}_{Li}\tilde R^a_2\nu_{R} 
+{\mathrm{h.c.}}\,, \label{eq:LQ_lagrangian}
\eea
where $a,b$ are $SU_L(2)$ gauge indices, $i,j,k$ are flavor indices, and ${y}^{RL}_{jk}$ and $y^{\overline {LR}}_{i}$ are complex Yukawa coupling matrices.
The mass of the leptoquark is restricted by direct and indirect searches~\cite{Workman:2022ynf} to be $m_{\text{LQ}}>\orderof(\SI{1}{\tera\eV})$.
Integrating out the leptoquark leads to the matching relation
\begin{equation}
C_{LdQ\nu,kji} = \frac{1}{m^2_{LQ}}\,y^{\overline {LR}}_ 
{i}\,{y}^{RL*}_{jk}\,.
\end{equation}
Below the EW scale, we fix the lepton generational index to electrons $k=e$ and match the interactions to $\nu$LEFT with
\bea\label{eq:LQmatching}
c_{\textrm{VLL},ij}^{CC}&=&-2 V_{ij} U_{e4}\,, \nn\\
c_{\textrm{SRR},ije}^{CC}&=& 4c_{\textrm{T},ije}^{CC} = \frac{v^2}{2}\frac{1}{m_{LQ}^2}y^{\overline {LR}}_{i}({y}^{RL}_{je})^*	 \,,\nn\\
c_{\textrm{SRR},ije}^{NC}&=&4c_{\textrm{T},ije}^{NC}=-V_{li}^*c_{\textrm{SRR},lje}^{CC}\,.\quad \qquad
\eea

\item In left-right symmetric models the SM gauge group is extended by an $SU(2)_R$ and sterile neutrinos arise naturally.
This introduces new gauge fields which interact with right-handed neutrinos and which can mix with SM $SU(2)_L$ gauge bosons after spontaneous symmetry breaking.
LHC and low-energy precision experiments constrain the extra gauge bosons to be heavier than several TeV~\cite{ATLAS:2019fgd, Dekens:2021bro}.
After integrating them out, we again induce contributions to higher-dimensional $\nu$SMEFT operators.
Instead of performing a detailed analysis, we consider $c_{\textrm{VLR}}^{CC1}$ to be the only non-zero Wilson coefficient in combination with minimal mixing and fix $k=e$.
We summarize the non-vanishing couplings here:
\bea\label{eq:VLRmatching}
c_{\textrm{VLL}, ij}^{CC}&=&-2 V_{ij} U_{e4}\,,\nn\\
c_{\textrm{VLR}, ije}^{CC1} &\neq& 0\,,
\eea
and set $c_{\textrm{VLR}, ije}^{CC1} >0$ for concreteness.
A more complete analysis of displaced-vertex searches for left-right symmetric models at \Dune{} and elsewhere is work in progress (see also Ref.~\cite{Alves:2023znq}).

\item The $\nu$SMEFT framework can also be used to  match other frameworks involving a SM-singlet fermion to the EFT, which is not related to neutrino physics.
That is, our framework is \textit{not} restricted to neutrino physics.
One such framework is RPV-SUSY with the lightest neutralino being very light and hence dominantly bino-like.
The bino couples via $U(1)_Y$ gauge interactions to pairs of either squarks and quarks or sleptons and leptons.
The virtual sfermions couple here via a non-zero RPV-operator $\lambda_{ijk}^\prime L_iQ_j \bar{D}_k$ to SM quarks and leptons. 
Integrating out the heavy sfermion fields (see App.~\ref{sec:appendix_rpv_derivation} for details) and applying $k=e$, we obtain the effective interactions\footnote{We neglect here the potential mixing between the neutralinos and the neutrinos
obtained for example via bilinear R-parity breaking as this dominantly involves the higgsino component of the neutralino~\cite{Dreiner:2023yus}.} 
\bea\label{eq:RPVmatching}
c_{\textrm{VLL}, ij}^ {CC}&=& 0\,,\nn\\
c_{\textrm{SRR},ije}^{CC}= -36c_{\textrm{T},ije}^{CC}&=& \frac{3}{4}\frac{g^\prime}{G_F}\frac{\big(\lambda^{\prime}_{eij}\big)^*}{ m_{\mathrm{SUSY}}^2}\,,\nn\\
c_{\textrm{SRR},ije}^{NC}= -36c_{\textrm{T},ije}^{NC}&=&-c_{\textrm{SRR},lje}^{CC}V_{li}^*\,,
\eea
where $g'$ is the $U(1)_{Y}$ coupling constant and $m_{\mathrm{SUSY}}$ is the sfermion mass-scale.
\end{enumerate}

%!TEX root= dune_eft.tex
%%%%%%%%%%%%%%%%%%%%%%%%%%%%%%%%%%%%%%%%%%%%%%%%
\section{Production of the neutral fermion \LLNF{}}\label{sec:LLNF_production}
%%%%%%%%%%%%%%%%%%%%%%%%%%%%%%%%%%%%%%%%%%%%%%%%

In this section, we discuss the production of Majorana fermions, \LLNF{}, through meson decays.\footnote{The production of \LLNF{} in the \SI{}{\giga\eV} mass range from the decays of Higgs-, $W$-, and $Z$-bosons or via the direct production at the target is sub-dominant and hence not considered~\cite{Curtin:2018mvb,Helo:2018qej,Hirsch:2020klk}.}
The branching ratios into $N$ are small for mesons with a relatively short lifetime (such as vector mesons, and $\eta$ and $\eta'$ mesons) and we neglect their contributions.
Further, the production rate of the $B_c$ mesons at \texttt{DUNE} is rather small, so we also ignore these contributions.
Finally, we do not include the production of $N$ in pion decays because our simulation tool of choice, \Pythia{}, does not predict forward pion-production very well (see also Ref.~\cite{Fieg:2023kld} for a recent work on tuning \Pythia{} for the forward physics spectra).
This essentially implies that our analysis is only reliable for Majorana fermion masses above the pion mass threshold.
To summarize, we consider production of a single \LLNF{} in the decays of pseudoscalar mesons: kaons ($K^\pm$, $K_{\subls}$), $D$-mesons ($D^\pm$, $D^0$, $D_s^\pm$), and $B$-mesons ($B^\pm$, $B^0$, $B_s^0$).

%%%%%%%%%%%%%%%%%%%%%%%%%%%%%%%%%%%%%%%%%%%%%%%%
\subsection{The minimal scenario}\label{sec:LLNF_production_minimal}
%%%%%%%%%%%%%%%%%%%%%%%%%%%%%%%%%%%%%%%%%%%%%%%%

In the minimal scenario, we do not consider $\nu$SMEFT operators.
The \LLNF{} can only interact with the SM fields through mixing with the active neutrinos. 
The only non-zero coupling is 
\begin{equation}
    c_{\textrm{VLL}, ij}^{CC} = -2V_{ij} U_{e4},
\end{equation}
where $i$ and $j$ are the generational indices of up- and down-type generation, respectively.
Such a coupling induces leptonic and semi-leptonic decays of a meson $M^{(\pm)}$ into an \LLNF{}, an electron, and possibly an accompanying lighter meson $M^\prime$:
\bea
M^\pm &\rightarrow& \LLNF{}+e^\pm\,,\label{eq:charged_prod_leptonic}\\
M&\rightarrow& \LLNF{}+e^\pm +M^{\prime}\,.\label{eq:charged_prod_semileptonic}
\eea
Depending on the mass of \LLNF{} and the contributing mesons $M$ and $M^\prime$, both 
leptonic and semi-leptonic modes should be taken into account when evaluating the number of LLNF.
The \LLNF{}-production rate depends not only on the meson decay branching ratio, but also the number of initial mesons $N_M$ at a given experimental facility.
Hence, a branching ratio for a decay mode $M_1\to N+X_1$ smaller than that for another mode $M_2\to N+X_2$ could still contribute significantly to the total number of $N$'s produced.

\begin{figure}[t]
	\centering
	\includegraphics[width=0.95\textwidth]{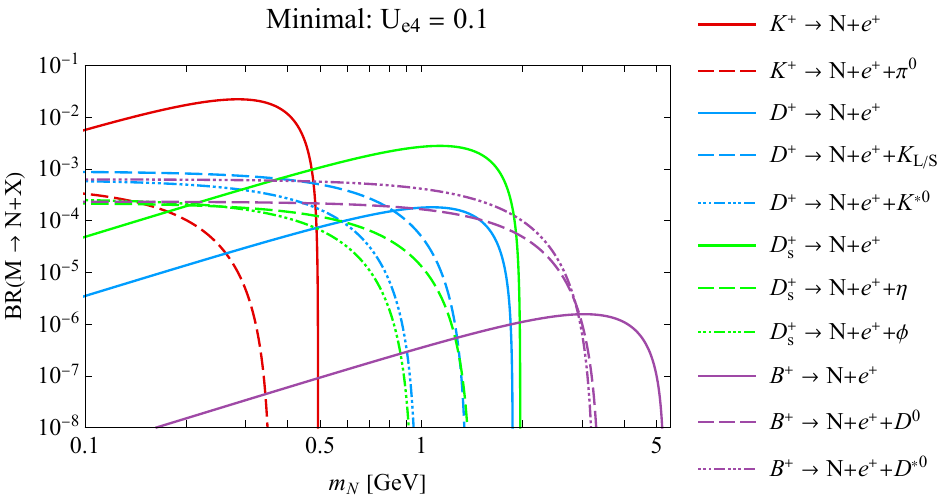}
	\caption{Branching ratios for a subset of dominant production modes (see Table~\ref{tab:minimal_production_modes}) of a \orderof(\SI{}{\giga\eV}) LLNF \LLNF{} \textit{only} mixing with the electron neutrino $\nu_e$.
	No higher-dimensional effective operator is switched on.
        The mixing parameter $U_{e4}$ is set to 0.1.
	}\label{fig:minimal_production}
\end{figure}
For the leptonic modes as given in Eq.~\eqref{eq:charged_prod_leptonic}, we include the following $M^\pm$'s via CKM element $V_{ij}$:$\,K^\pm\,\big(V_{us}\big)$, $D^\pm\,\big(V_{cd}\big)$, $D^\pm_s\,\big(V_{cs}\big)$, and $B^\pm\,\big(V_{ub}\big)$.
The included semi-leptonic modes are summarized in Table~\ref{tab:minimal_production_modes}.
We have considered only the lowest excited states of e.g.~kaons in the decay channels of the bottom mesons. In principle multiple higher resonances exist and could give non-negligible contributions (see, for instance, Ref.~\cite{Boiarska:2019jym} where it is shown that these channels are relevant for the production of a light scalar from $B$-meson decays that mixes with the SM Higgs boson at the LHC).
These contributions only affect our results marginally because our production modes are dominated by the two-body decays of the $B$-mesons. We have therefore not included these excited states in our simulations. 
The same conclusion holds for the effective-operator scenarios as well.
{\renewcommand{\arraystretch}{1.3}\begin{table}[H]\small
	\center
	\begin{tabular}{|c||c|cc||c||c|cc||c||c|cc||}
	\cline{5-12}
	\multicolumn{4}{c||}{} &$V_{ij}$ & $M$ & $M^\prime$ &$M^{*\prime}$ &$V_{ij}$ & $M$ & $M^\prime$ &$M^{*\prime}$\\
	\cline{5-12}
	\multicolumn{4}{c||}{} & \multirow{3}{*}{$V_{us}$}	& $K^\pm$ 	& $\pi^0$ &  	&\multirow{3}{*}{$V_{ub}$}	& $B^\pm$ 	& $\pi^0$& $\rho^0$\\
	\multicolumn{4}{c||}{} & & $K_{\subls}$ & $\pi^\mp$ & 	& &$B^0$ 	& $\pi^\mp$& $\rho^\mp$ \\
	\multicolumn{4}{c||}{} & & & & & & $B^0_s$   & $K_{\subls}$& $K^{*0}$\\
	\hline
	$V_{ij}$ & $M$ & $M^\prime$ &$M^{*\prime}$&$V_{ij}$ & $M$ & $M^\prime$ &$M^{*\prime}$&$V_{ij}$ & $M$ & $M^\prime$ &$M^{*\prime}$\\
	\hline
	\multirow{3}{*}{$V_{cd}$}	& $D^\pm$ 	& $\pi^0$& $\rho^0$& \multirow{3}{*}{$V_{cs}$}	& $D^\pm$	& $K_{\subls}$& $K^{*0}$ &\multirow{3}{*}{$V_{cb}$}	& $B^\pm$	& $D^0$&$D^{*0}$\\
	& $D^0$ 	& $\pi^\mp$& $\rho^\mp$ & 	& $D^0$  	& $K^\mp$& $K^{*\mp}$&  & $B^0$  	& $D^\mp$& $D^{*\mp}$ \\
	& $D^\pm_s$ & $K_{\subls}$& $K^{*0}$& 	& $D^\pm_s$ & $\eta^{(\prime)}$& $\phi$&  & $B^0_s$ 	& $D^\mp_s$& $D^{*\mp}_s$ \\
	\hline
	\end{tabular}
	\caption{All included \LLNF{} production modes of the form $M\rightarrow \LLNF{}+e^\pm + M^{(*)\prime}$ in the minimal scenario.
                We give also the responsible CKM matrix element $V_{ij}$ for the meson decay. 
	}\label{tab:minimal_production_modes}
\end{table}}

The branching ratios for a subset of these decay modes are depicted in Fig.~\ref{fig:minimal_production} for a minimal mixing $U_{e4}=0.1$ (see App.~\ref{sec:appendix_decayratecalc} for explicit calculations). 
Our results agree well with previous computations~\cite{Bondarenko:2018ptm,DeVries:2020jbs,Coloma:2020lgy}.

%%%%%%%%%%%%%%%%%%%%%%%%%%%%%%%%%%%%%%%%%%%%%%%%
\subsection{Effective operators}\label{sec:LLNF_production_EFT}
%%%%%%%%%%%%%%%%%%%%%%%%%%%%%%%%%%%%%%%%%%%%%%%%

In Scenarios 2-4, \textit{cf.}~Sec.~\ref{sec:theoretical_scenarios}, we turn on specific EFT-operators.
Depending on the quark-flavor indices of the operators, different leptonic and semi-leptonic \LLNF{} production modes are allowed.
The relevant branching ratios have already been calculated for $D$- and 
$B$-meson decays via the charged currents of Eq.~\eqref{d6CC}~\cite{DeVries:2020jbs}. However, in Ref.~\cite{DeVries:2020jbs}, the branching ratios for kaon decays were not included and neutral currents, \textit{cf.}~Eq.~\eqref{d6NC}, present in the leptoquark and RPV scenarios, were not taken into account\footnote{See also Ref.~\cite{Beltran:2023ksw} for a recent work on kaon decays to the HNLs in the $\nu$SMEFT and $\nu$LEFT.}.
In addition to the decay modes given in Eqs.~\eqref{eq:charged_prod_leptonic}-\eqref{eq:charged_prod_semileptonic}, we here consider
\bea
M^0 &\rightarrow& \LLNF{}+\nu_{e}\,,\label{eq:neutral_prod_leptonic}\\
M&\rightarrow& \LLNF{}+\nu_{e}+M^{\prime}\,.\label{eq:neutral_prod_semileptonic}
\eea
For several $\nu$SMEFT operators, neutral decays involving different quark flavors are possible because of the insertion of the CKM matrix in the matching expressions in Eq.~\eqref{match6NC}.
However, some of these are suppressed by small CKM entries.

We select several flavor benchmarks for each theoretical scenario and stress that \LLNF{} can be either the HNL or another particle such as the light bino in RPV-SUSY. 
We have fixed the lepton index $k=e$ so that we only consider electrons and electron 
neutrinos, as the phenomenology for muons is very similar, while $\tau$ leptons are too heavy for \Dune{}.
The production of $B$-mesons at \Dune{} is comparatively low and we therefore do not consider flavor couplings in which the production of \LLNF{} is mainly mediated by $B$-mesons.
The possible up- and down-type generational quark indices $ij$ for the charged currents are set to $11$, $12$ and $21$.
Including the related neutral couplings with quark indices $lj$ suppressed by $|V_{il}|^2$, the possible meson decay modes for kaons, charm, and bottom mesons are listed in Table~\ref{tab:eft_production_modes}.

\begin{figure}[t]
\centering
	\includegraphics[width=0.49\textwidth]{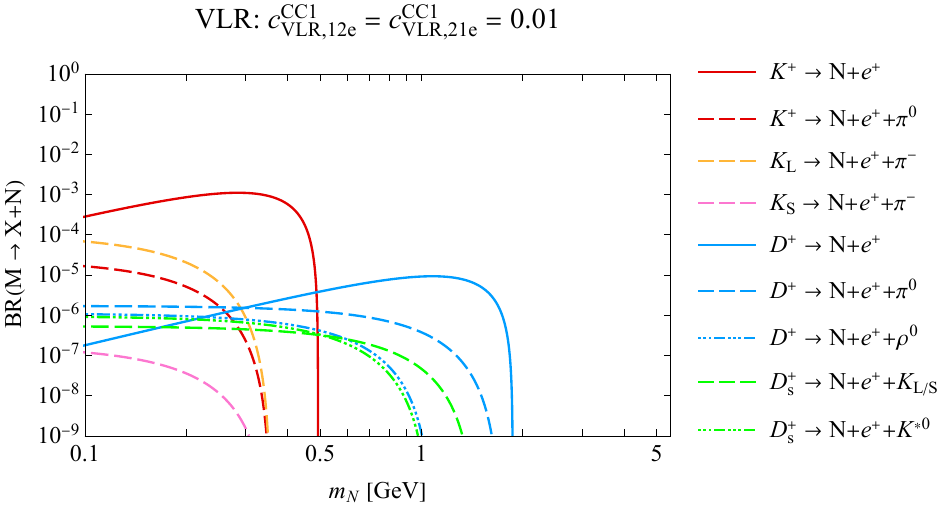}
	\includegraphics[width=0.49\textwidth]{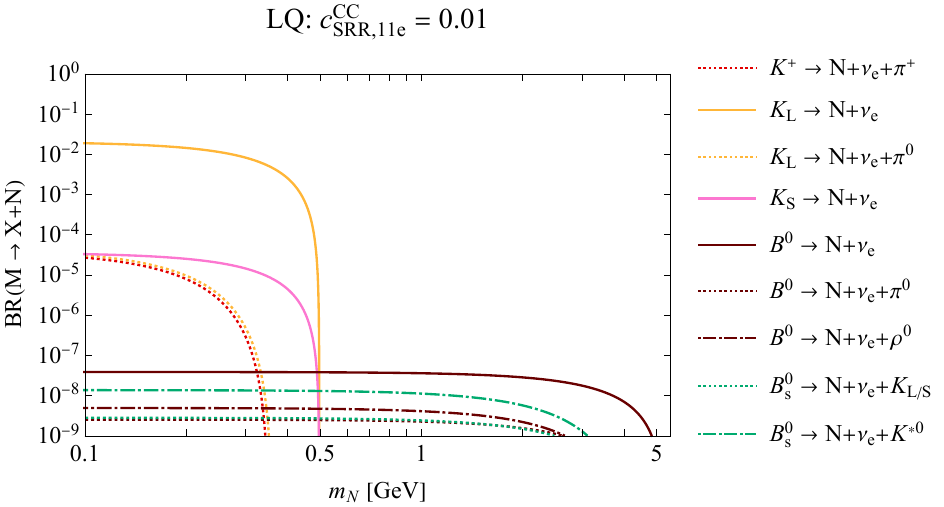}
	\includegraphics[width=0.49\textwidth]{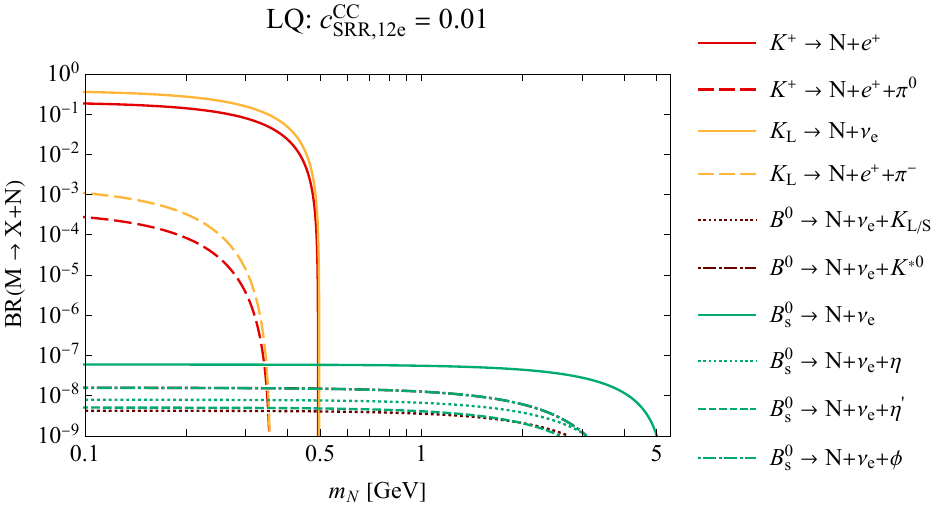}
	\includegraphics[width=0.49\textwidth]{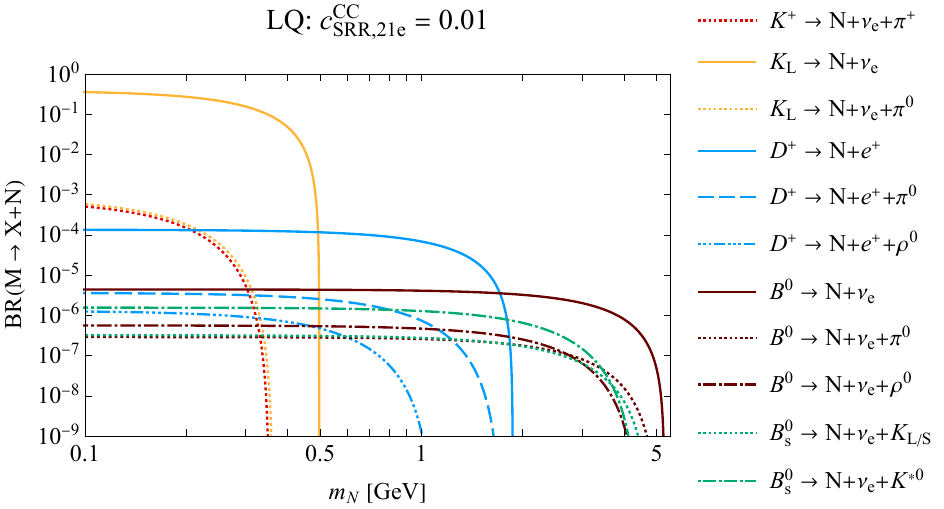}
	\caption{Branching ratios for the production of a GeV-scale $N$ via the non-zero EFT-couplings of each scenario (specified in Sec.~\ref{sec:theoretical_scenarios}) for various flavor benchmarks, with the related Wilson coefficients all set to 0.01.
            We neglect the contribution from a non-zero minimal mixing $U_{e4}$.
	The upper left depicts the meson decay modes of the left-right 
	symmetric model-like scenario.
	The remaining three plots are for the leptoquark scenario, split into the flavor couplings $ijk$=111 (upper right), $ijk$=121 (lower left), and $ijk$=211 (lower right).
	}\label{fig:eft_production}
\end{figure}

{\renewcommand{\arraystretch}{1.3}\begin{table}[H]\small
	\center
	\begin{tabular}{|c||c|ccc||c|c|ccc||}
	\hline
	\multirow{2}{*}{$ij$}& $M\to N+e^\pm$ &\multicolumn{3}{c||}{$M\to N+e^\pm+M^{(*)\prime}$}& \multirow{2}{*}{$V_{il}$} & $M\to N+\nu_e$ &\multicolumn{3}{c||}{$M\to N+\nu_e+M^{(*)\prime}$}\\
	\cline{2-5}
	\cline{7-10}
		& $M$ & $M$ & $M^\prime$ &$M^{*\prime}$ & & $M$ & $M$ & $M^\prime$ &$M^{*\prime}$\\
	\hline
	\hline
	\multirow{5}{*}{11} & & & & &\multirow{2}{*}{$V_{us}$} & \multirow{2}{*}{$K_{\subls}$} & $K^\pm$ & $\pi^\pm$ & \\
	& & & & & & & $K_{\subls}$ & $\pi^0$ & \\
	\cline{6-10}
	& & & & & \multirow{3}{*}{$V_{ub}$} & \multirow{3}{*}{$B^0$} & $B^\pm$ & $\pi^\pm$ & $\rho^\pm$\\
	& & & & & & & $B^0$ & $\pi^0$ & $\rho^0$\\
	& & & & & & & $B^0_s$ & $K_{\subls}$ & $K^{*0}$\\
	\hline
	\multirow{5}{*}{12} & \multirow{5}{*}{$K^\pm$} & & & &\multirow{2}{*}{$V_{ud}$} & \multirow{2}{*}{$K_{\subls}$} & $K^\pm$ & $\pi^\pm$ & \\
	& & \multirow{2}{*}{$K^\pm$} & \multirow{2}{*}{$\pi^0$} & & & & $K_{\subls}$ & $\pi^0$ & \\
	\cline{6-10}
	& & \multirow{2}{*}{$K_{\subls}$} & \multirow{2}{*}{$\pi^\mp$} & & \multirow{3}{*}{$V_{ub}$} & \multirow{3}{*}{$B^0_s$} & $B^\pm$ & $K^\pm$ & $K^{*\pm}$\\
	& & & & & & & $B^0$ & $K_{\subls}$ & $K^{*0}$\\
	& & & & & & & $B^0_s$ & $\eta^{(\prime)}$ & $\phi$\\
	\hline
	\multirow{5}{*}{21} & \multirow{5}{*}{$D^\pm$} & & & &\multirow{2}{*}{$V_{cs}$} & \multirow{2}{*}{$K_{\subls}$} & $K^\pm$ & $\pi^\pm$ & \\
	& & $D^\pm$ & $\pi^0$ & $\rho^0$ & & & $K_{\subls}$ & $\pi^0$ & \\
	\cline{6-10}
	& & $D^0$ & $\pi^\mp$ & $\rho^\mp$ & \multirow{3}{*}{$V_{cb}$} & \multirow{3}{*}{$B^0$} & $B^\pm$ & $\pi^\pm$ & $\rho^\pm$\\
	& & $D^\pm_s$ & $K_{\subls}$ & $K^{*0}$ & & & $B^0$ & $\pi^0$ & $\rho^0$\\
	& & & & & & & $B^0_s$ & $K_{\subls}$ & $K^{*0}$\\
	\hline
	\end{tabular}
	\caption{All included \LLNF{} production modes of the form $M\rightarrow \LLNF{}+\big(e^\pm/\nu_e\big) + \,\big( M^{(*)\prime}\big)$.
                We state the modes for the generational indices $ij$ of the charged modes.
                For the neutral modes present in the leptoquark and RPV scenarios suppressed by the CKM matrix element $V_{il}$, the quark family indices are $lj$. 
                We do not list the possible pion decay modes such as $\pi^\pm \to N+e^\pm$ for $ij=11$, as pion-production in the very forward region is not well validated in \Pythia{}.
	}\label{tab:eft_production_modes}
\end{table}}
For details of the associated decay rates we refer to App.~\ref{sec:appendix_decayratecalc}. 
Neglecting possible interference arising from a non-zero $U_{e4}$, a subset of the branching ratios of these production modes are depicted in Fig.~\ref{fig:eft_production}, as a function of the mass of $N$, where we set the values of the relevant dimensionless Wilson coefficients to be 0.01.
The upper left plot is for the left-right symmetric model-like scenario with the Wilson coefficients $c_{VLR,12e}^{CC1}$ and $c_{VLR,21e}^{CC1}$ leading to kaon and charm-meson decays, respectively.
The other three plots are for the leptoquark scenario for three different flavor generation indices.
The number of decay modes in the leptoquark scenario is greater than in the left-right model-like scenario, because of the numerous neutral modes available (suppressed by the CKM matrix elements though).
The RPV-SUSY scenario plots are not presented here for the reason given below.
From the relations between the Wilson coefficients and the UV-complete model parameters, \textit{cf.~}Eq.~\eqref{eq:LQmatching} and Eq.~\eqref{eq:RPVmatching}, we observe that the leptoquark and RPV scenarios with $U_{e4}=0$ \textit{only} differ in the magnitude of the tensor currents, and the contributions from these tensor currents to the depicted decays are minor.
As a result, the branching ratios of the RPV scenario are almost identical to those of the leptoquark scenario.

%!TEX root= dune_eft.tex
\section{Decay of the neutral fermion \LLNF{}}\label{sec:LLNF_decay}

\subsection{The minimal scenario}\label{sec:LLNF_decay_minimal}

The decay rates of an \LLNF{} interacting with the SM fields through minimal mixing have been calculated extensively; see \textit{e.g.}~Refs.~\cite{Johnson:1997cj, Gribanov:2001vv, Gorbunov:2007ak, Atre:2009rg, Bondarenko:2018ptm, Coloma:2020lgy, DeVries:2020jbs}.
Here, we mainly follow Ref.~\cite{DeVries:2020jbs}. 
The Lagrangians in Eqs.~\eqref{d6CC} and \eqref{d6NC} allow for invisible decays (only final-state neutrinos) and visible leptonic or semi-leptonic decays.
For a relatively light \LLNF{}, the latter are dominated by decays into a single final-state meson plus a lepton.
We consider the pseudoscalar mesons: $\pi$, $K$, $\eta$, $\eta^\prime$, $D$, $D_s$, $\eta_c$, and separately the vector mesons: $\rho$, $K^*$, $\omega$, $\phi$, $D^*$, $D^*$, $J/\psi$.
For a heavier \LLNF{}, decays into multi-meson states become relevant and to estimate the total hadronic decay width we follow Ref.~\cite{Bondarenko:2018ptm} to calculate the decay rate into spectator quarks times appropriate loop corrections to account for the hadronization effects
\bea\label{eq:N_decay_multi_meson}
\Gamma\big(\LLNF{}\rightarrow 
(e/\nu_e)+\text{hadrons}\big)= \big[1+\Delta_{\text{QCD}}(m_\LLNF{})\big]
\times\Gamma_{\textrm{tree}}\big(\LLNF{}\rightarrow (e/\nu_e)+\bar{q}q\big)\,,
\eea
which includes both charged and neutral weak currents.
The loop corrections are obtained from a comparison to hadronic $\tau$ decays and reproduced below~\cite{Gorishnii:1991hw,Bondarenko:2018ptm}
\bea
\Delta_{\text{QCD}}(m_\LLNF{}) = \frac{\alpha_s}{\pi}+5.2\frac{\alpha_s^2}{\pi^2}+26.4\frac{\alpha_s^3}{\pi^3}\,,
\eea
where $\alpha_s=\alpha_s(m_N)$.
The hadronic decay width into multi-meson final states becomes relevant for \LLNF{} masses above \SI{1}{\giga\eV} roughly and we write as an approximation
\bea\label{eq:N_total_decay_width}
\Gamma_{\LLNF{}}=\Gamma_{\LLNF{}\rightarrow\textrm{leptons}}&&+\Theta\big(\SI{1}{\giga\eV}-m_{\LLNF{}}\big)\,\Gamma_{\LLNF{}\rightarrow\textrm{single meson}}\nonumber\\
&&+\Theta\big(m_{\LLNF{}}-\SI{1}{\giga\eV}\big)\,\big[1+\Delta_{\text{QCD}}(m_\LLNF{})\big]\,\Gamma_{\LLNF{}\rightarrow\bar{q}q}\,.
\eea
App.~\ref{sec:appendix_decayratecalc} and Refs.~\cite{Helo:2010cw,Gribanov:2001vv} provide more detail on the explicit decay-rate calculations.

\subsection{Effective operators}\label{sec:LLNF_decay_EFT}

The $\nu$SMEFT operators contribute to a greater number of decay processes of an LLNF.
In the theoretical scenarios under consideration, the most important additional decay modes involve a single final-state meson and a lepton\footnote{Multi-meson final states can become important but are hard to control theoretically for scalar and tensor operators and therefore not included.
The estimates of Ref.~\cite{DeVries:2020jbs} indicate that such modes are subdominant for \LLNF{} with a mass below 2 GeV and thus safe to neglect for \Dune{}}.

\begin{figure}
    \centering
    \includegraphics[width=0.9\textwidth]{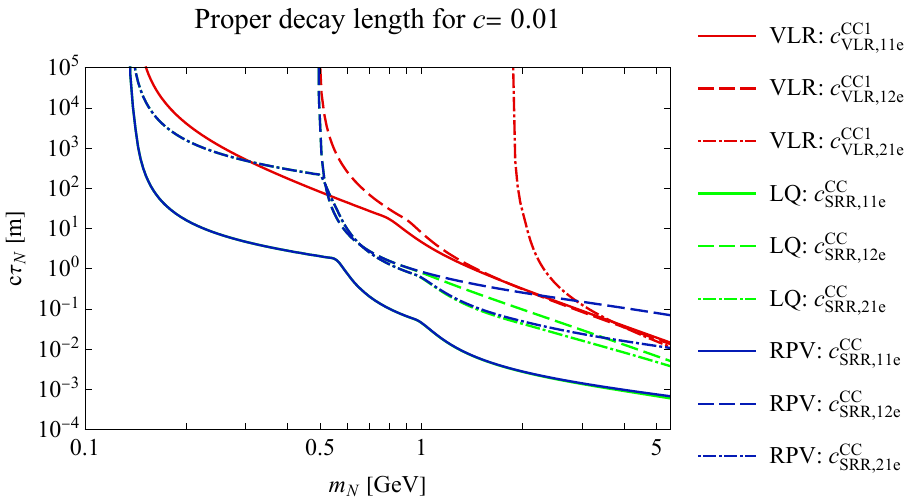}
    \caption{Proper decay length of \LLNF{} for all theoretical scenarios and different flavor couplings.
    We set the EFT couplings to $c=0.01$ and turn off the minimal mixing ($U_{e4}=0$).
    }
    \label{fig:eft_decay}
\end{figure}

We list the allowed decays modes for each scenario and flavor benchmark in Table~\ref{tab:eft_decay_modes}, assuming the $N$ is massive enough in each case. 
The proper decay length for each scenario and each flavor benchmark with a charged-current Wilson coefficient $c=0.01$ and no minimal mixing, $U_{e4}=0$, is plotted in 
Fig.~\ref{fig:eft_decay}.
The remaining Wilson coefficients are then set according to Eqs.~\eqref{eq:LQmatching}-\eqref{eq:RPVmatching}.
For the left-right symmetric model, the resulting decay lengths agree with those 
computed in Ref.~\cite{DeVries:2020jbs}.
However for the leptoquark scenario, the decay lengths differ, as the neutral decay modes via the EFT couplings were not taken into account in that work.
Because of the larger number of the available final states, the total decay rate of $\LLNF{}$ increases and hence the decay length decreases accordingly.
The numerical difference depends on the available modes based on the LLNF mass $m_N$.
The explicit expressions for decays with two-body final states are provided in App.~\ref{sec:appendix_decayratecalc}.
 
We stress that without minimal mixing $U_{e4}$, the decay lengths for the leptoquark 
and RPV scenario are identical for $m_n<\SI{1}{\giga\eV}$.
For masses larger than \SI{1}{\giga\eV}, the decay lengths for flavor coupling $ijk=11e$ remain the same between the two scenarios, while for $ijk=12e$ and $ijk=21e$, the decay lengths start to differ as the contributions from the tensor couplings in these scenarios become more important.

{\renewcommand{\arraystretch}{1.3}\begin{table}[H]\small
		\center
		\begin{tabular}{|c||c|lc||c||}
			\hline  
			&\multicolumn{3}{c||}{Leptoquark / RPV}	& Left-Right \\
			\hline
			$ij$  & $M^\pm$ & $M^0$ & $\big[V_{il}\big]$  &$M^\pm$  \\
			  \hline
			\multirow{3}{*}{11} & \multirow{3}{*}{$\pi^\pm,\,\rho^\pm$} & $\pi^0,\,\rho^0,\,\eta,\,\eta^\prime,\,\omega$ & $\big[V_{ud}\big]$  & \multirow{3}{*}{$\pi^\pm,\,\rho^\pm$}   \\
			 	 &  & $K_{\subls},\,K^{*0}$  & $\big[V_{us}\big]$ &  \\	
			 	 &  & $B^{0},\,B^{*0}$  & $\big[V_{ub}\big]$ &  \\	
			 	 \hline
			\multirow{3}{*}{12} & \multirow{3}{*}{$K^\pm,\,K^{*\pm}$} & $K_{\subls},\,K^{*0}$  & $\big[V_{ud}\big]$& \multirow{3}{*}{$K^\pm,\,K^{*\pm}$}   \\
			 	 &  & $\eta,\,\eta^\prime,\,\phi$  & $\big[V_{us}\big]$&  \\	
			 	 &  & $B_s^{0},\,B_s^{*0}$  & $\big[V_{ub}\big]$&  \\	
			 	 \hline
			\multirow{3}{*}{21} & \multirow{3}{*}{$D^\pm,\,D^{*\pm}$} & $\pi^0,\,\rho^0,\,\eta,\,\eta^\prime,\,\omega$ & $\big[V_{cd}\big]$ & \multirow{3}{*}{$D^\pm,\,D^{*\pm}$}   \\
			 	 &  & $K_{\subls},\,K^{*0}$  & $\big[V_{cs}\big]$ &  \\	
			 	 &  & $B^{0},\,B^{*0}$  & $\big[V_{cb}\big]$&  \\
			\hline
		\end{tabular}
		\caption{Possible decay modes for an LLNF $\LLNF{}$ for each EFT scenario.
                We sort the modes by the generation indices $ij$ of the charged quark current.
                Possible neutral quark currents have the indices $lj$ and are suppressed by the stated CKM matrix element $V_{il}$. 
        We fix the lepton index to $k=e$, so that the decays have the form $\LLNF{}\to M^\pm + e^\mp$ or $\LLNF{}\to M^0 + \nu_e$.}
		\label{tab:eft_decay_modes}
\end{table}}

%!TEX root= dune_eft.tex
\section{The \DUNE{} experiment}\label{sec:DUNE}

\begin{figure}
\centering
\includegraphics[width=0.85\textwidth]{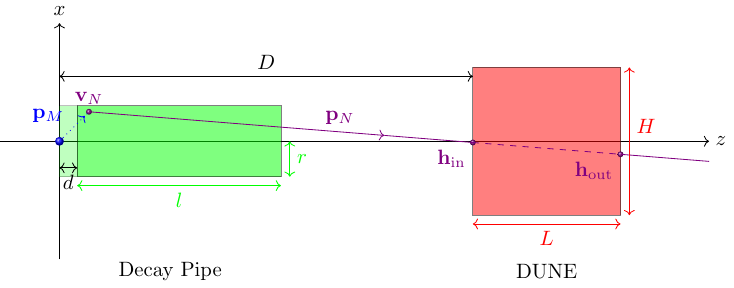}
	\caption{
	Schematic drawing of the \Dune{} near detector (red) and the 
	decay pipe (green).
 An example event of a LLNF produced at a displaced 
	vertex is included.
 A meson (dashed blue) traveling in direction 
	$\vec{p}_M$ can decay at a displaced vertex $\vec{v}_{\LLNF{}}$ into an LLNF \LLNF{} (purple).
    The LLNF traveling in direction $\vec{p}_{\LLNF{}}$ can then be observed in the near neutrino detector. 
	The geometric dimensions are given in Table~\ref{tab:detector_beam_parameters}.
	}\label{fig:Dune_schematic}
\end{figure}

In this section, we briefly introduce the \Dune{} experimental setup.
The experiment is comprised of a beamline facility at Fermilab in Illinois, USA, and an underground cryogenic detector infrastructure at the Sanford Underground Research Facility (SURF) in South Dakota.
The objective of the beamline facility is to provide a neutrino beam ideal for studying neutrino oscillation physics~\cite{DUNE:2016hlj,DUNE:2015lol,DUNE:2016evb,DUNE:2016rla}. 
To that end, a proton beam (\numrange[range-phrase = --]{60}{120} \SI{ }{\giga\eV}) is extracted from the Fermilab Main Injector.
The beam collides with a graphite target generating a secondary 
beam of charged particles.
For a beam-energy of \SI{120}{\giga\eV} the design allows for \SI{1.1e21} protons on target per year.
The charged particles subsequently are sign-selected and focused with horns.\footnote{In this work, to simplify the simulation, we do not consider the effect of the magnetic horns.
The focusing effect would improve the number of mesons and hence the number of \LLNF{}'s that could be detected by \Dune{}, so that our results are conservative.} 
A \SI{194}{\meter} long and \SI{4}{\meter} wide decay pipe displaced \SI{27}{\meter} downstream from the target provides space for the secondary beam to decay into neutrinos, or other LLNFs.
Additionally, the beamline facility includes a conventional setup for a near-detector system (NDS) \SI{574}{\meter} downstream from the target.
The NDS consists of a beamline measurement system, a data acquisition system, and a 
fine-grained tracker (FGT) near neutrino detector (NND) with a tracking volume of \SI{78.4}{\cubic\meter}.
Important parameters such as geometrical sizes and reference beam parameters are listed in Table~\ref{tab:detector_beam_parameters}.
A schematic drawing of the beamline facility is shown in Fig.~\ref{fig:Dune_schematic}.
For our numerical simulation, we fix the proton beam energy to be 120 GeV, and consider ten years' operation time corresponding to $1.1\times 10^{22}$ protons on target.

The focused neutrino beam travels approximately \SI{1300}{\kilo\meter} towards the SURF. 
Because of the large distance between the target and the \Dune{} far detector, the latter is not suited for our topic of interest.
We will instead focus on the \Dune{} near detector as a possibility to detect long-lived \LLNF{}'s.

\begin{table}
\centering
\renewcommand{\arraystretch}{1.3}
\begin{tabular}{|lr||lr|}
\hline
Reference beam parameters 	&		&	geometrical variables &  \\
\hline\hline
proton beam energy	&	\SI{120}{\giga\eV}	&	decay pipe length $l$ & \SI{194}{\meter}\\
cycle time 	&	\SI{1.2}{\second}	&	decay pipe radius $r$ & \SI{2}{\meter}\\
protons per cycle 	&	\SI{7.5e13}{}	&	distance: target - decay pipe $d$ & \SI{27}{\meter}\\
spill duration 	&	\SI{1.0e-5}{\second}	&	near detector length $L$ & \SI{6.4}{\meter}\\
Protons on target per year 	&	\SI{1.1e21}{}	&	near detector width $H$ & \SI{3.5}{\meter}\\
 	&		&	distance: target - near detector $D$ & \SI{574}{\meter}\\
\hline
\end{tabular}
\caption{		Reference beam parameters and geometrical details of the beamline 
				facility at Fermilab~\cite{DUNE:2016evb,DUNE:2016rla}.
				}\label{tab:detector_beam_parameters}
\end{table}

\subsection{Monte-Carlo simulation}\label{subsec:MC_simulation}

Considering the scenarios described in Sec.~\ref{sec:theoretical_scenarios}, we are interested in the secondary beam of charged particles.
The mesons produced in the proton-proton collisions may decay into LLNFs, \LLNF{}'s. 
We perform Monte-Carlo (MC) simulations in order to evaluate the search sensitivity  of the \Dune{} near detector to an \LLNF{} produced from a meson and subsequently decaying at a displaced vertex into a lighter meson.

First, we determine the number of \LLNF{}'s produced in meson decays over the runtime of the \Dune{} experiment, which can be calculated in \NUSMEFT{} via the number $N_M$ of any given meson $M$ and its decay branching ratio into \LLNF{}:
\begin{equation}
N_{M,\LLNF}^{\text{prod}} = N_M \cdot \text{Br}\big(M\rightarrow \LLNF+X\big),
\end{equation}
where $X$ denotes any possible combination of leptons and lighter mesons.
The branching ratios have been calculated in Ref.~\cite{DeVries:2020jbs} and in Sec.~\ref{sec:LLNF_production}.
For the numbers of mesons produced at the experiment, we follow Ref.~\cite{Krasnov:2019kdc}, where the numbers of secondary hadrons are calculated by
\begin{equation}\label{eq:number_of_mesons}
N_M = N_{\text{POT}}\cdot M_{pp}\cdot \chi_q \cdot f_{q\to M}.
\end{equation}
Here, $N_{\text{POT}}$ is the number of protons on target over the runtime of the experiment.
$M_{pp}$ is the multiplicity of the proton-target reaction and indicates on average how many mesons $M$ are produced in each primary proton-proton collision.
$\chi_q$ is the fraction of quarks $q$ generated at the primary vertex at a given beam energy with
\begin{equation}
\chi_q\equiv\frac{\sigma_{pp\rightarrow q}}{\sigma_{pp_{\text{total}}}},
\end{equation}
where $\sigma_{pp\to q}$ and $\sigma_{pp_{\text{total}}}$ denote the $q$-generation and inelastic scattering cross section, respectively.
$f_{q\rightarrow M}$ is the fragmentation factor of the quark $q$ into a meson $M$.
The implemented values for $\chi_q$ and $f_{q\to M}$ are listed in Tables~\ref{tab:quark_generation} and \ref{tab:quark_production_fractions}, respectively.

\begin{table}
\centering
\renewcommand{\arraystretch}{1.2}
\begin{tabular}{|c|ccc|}
\hline
Quark $q$ & $s$ & $c$ & $b$\\
\hline
$\chi_q$ & 1/7  & \SI{1e-4}{}	& \SI{1e-10}{}	\\
\hline
\end{tabular}
	\caption{
	Fraction of quarks generated in the proton-target 
	collisions at \Dune{} \cite{Lourenco:2006vw,Andersson:1983ia}.
	}\label{tab:quark_generation}
\end{table}

\begin{table}
\centering
\renewcommand{\arraystretch}{1.2}
\begin{tabular}{|c|ccc|ccc|ccc|}
\hline
Quark $q$ & \multicolumn{3}{c|}{$s$} & \multicolumn{3}{c|}{$c$} & \multicolumn{3}{c|}{$b$}\\
\hline
Meson $M$ & $K^-$ & $K_L$ & $K_S$ & $D^+$ & $D^0$ & $D_s$ & $B^+$ & $B^0$ & $B^0_s$\\
$f_{q\to M}$ & 1/3 & 1/3 & 1/3 & 0.207 & 0.632 & 0.088 & 0.405 & 0.405 & 0.101\\
\hline
\end{tabular}
	\caption{
	Different quark fragmentation factors for 
	kaons~\cite{Gorbunov:2007ak,Lourenco:2006vw}, 
	charm mesons~\cite{SHiP:2018xqw}, and bottom 
	mesons~\cite{ParticleDataGroup:2018ovx}.
	}\label{tab:quark_production_fractions}
\end{table}
For every meson except kaons, the multiplicity factor is already included 
in $\chi_q$, so we set $M_{pp}=1$ in these cases.
For kaons and a beam energy of \SI{120}{\giga\eV}, we take $M_{pp}=11$~\cite{Gorbunov:2007ak}.
The resulting meson numbers for a runtime of ten years are given in Table~\ref{tab:meson_numbers}.

\begin{table}
\centering
\renewcommand{\arraystretch}{1.2}
\begin{tabular}{|l|ccc|}
\hline
Meson $M$  & $K^-$ & $K_L$ & $K_S$ \\
Number of Mesons & \SI{5.76e21}{} & \SI{5.76e21}{} & \SI{5.76e21}{}\\
\hline\hline
Meson $M$  & $D^+$ & $D^0$ & $D_s$ \\
Number of Mesons & \SI{2.28e17}{} & \SI{6.95e17}{} & \SI{9.68e16}{}\\
\hline\hline
Meson $M$  & $B^+$ & $B^0$ & $B^0_s$\\
Number of Mesons  & \SI{4.46e11}{} & \SI{4.46e11}{} & \SI{1.11e11}{}\\
\hline
\end{tabular}
	\caption{Total numbers of mesons produced at \Dune{} over a 
	runtime of ten years with \SI{1.1e22}{} POT, over a $4\pi$ solid angle coverage.
	}\label{tab:meson_numbers}
\end{table}

In order for the LLNF \LLNF{} to be observable, we require a `visible' final state. 
`Invisible' decay modes are those for which we expect \Dune{} not to be able to reconstruct the decay vertex.
For the case of an \LLNF{} as a sterile neutrino in the minimal scenario, it mixes with electron neutrinos and the `invisible' modes are $\LLNF{} \to \nu_e+\nu_i+\bar{\nu}_i$ with $i\in\{e, \mu,\tau\}$.
Every other mode is considered as `visible'. 
Additionally, the LLNF must decay inside the near detector.
For an individual \LLNF{} produced by meson $M$, the probability for it to decay inside the fiducial volume can be calculated with the exponential decay law:
\begin{equation}\label{eq:individual_decay_probability}
P_{M,i}\big[\LLNF\text{ in f.v.}\big] = \exp\bigg[-\frac{L_{T,i}}{\lambda_i}\bigg] \cdot \bigg(1-\exp\bigg[-\frac{L_{I,i}}{\lambda_i}\bigg]\bigg),
\end{equation}
where ``f.v.'' stands for ``fiducial volume'', $L_{T,i}$ is the length that \LLNF{} travels to reach the detector, $L_{I,i}$ is the length it would take for the $\LLNF$ to travel through the detector if it did not decay inside the fiducial volume, and $\lambda_i$ is the boosted decay length of the $i$-th \LLNF\ simulated.
The decay length is determined with the speed $\beta_i$, the boost factor $\gamma_i$, the speed of light $c$, as well as the LLNF proper lifetime $\tau_{\LLNF{}}$, with $\lambda_i=\beta_i\gamma_i c\tau_{\LLNF{}}$.
We employ MC techniques and simulate $N_{\text{MC}} = 10^6$ proton-on-target events with the event generator \Pythia{}~\cite{Bierlich:2022pfr,Sjostrand:2014zea} in order to obtain the average decay probability inside the detector for the \LLNF{}'s produced from the meson $M$:
\begin{equation}
\Braket{P_M\big[\LLNF\text{ in f.v.}\big]}= \frac{1}{N_{\text{MC}}}\sum_{i}^{N_{\text{MC}}} P_{M,i}\big[\LLNF\text{ in f.v.}\big].
\end{equation}
The number of observable $\LLNF$ in the fiducial volume is then estimated with
\begin{equation}
N_{\LLNF}^{\text{obs}} = \text{Br}\big(\LLNF\rightarrow \text{visible}\big) \cdot \sum_M N_{M,\LLNF}^{\text{prod}} \cdot \Braket{P_M\big[\LLNF\text{ in f.v.}\big]}\,,
\end{equation}
where $\text{Br}\big(\LLNF\rightarrow \text{visible}\big)$ denotes the visible decay branching ratio of $N$.
We note, that we assume $100\%$ detection efficiencies when calculating $N_{\LLNF}^{\text{obs}}$.

\subsection{Three-dimensional detector approximation}\label{subsec:three_dimensional_approximation}

\begin{figure}
\centering
\includegraphics[width=0.5\textwidth]{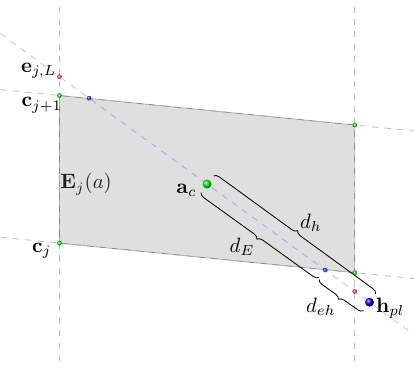}
	\caption{Sketch for determining, whether a particle trajectory hits a 	bounding side of the fiducial volume.
                The plane $\vec{P}$ is hit at $\vec{h}_{pl}$.
                Together with the center of the designated area $\vec{a}_c$ and the corner points $\vec{c}_j$, we can determine the intersection $\vec{e}_{j,L}$	(small red and blue dots) between the edges $\vec{E}_j(a)$ and a line through $\vec{a}_{c}$ and $\vec{h}_{pl}$. 
	Subsequently, for the $\vec{e}_{j,L}$ closest to $\vec{h}_{pl}$ we can determine the distances $d_{E,j}$, $d_{eh,j}$, and $d_h$ between the three vertices $\vec{a}_{c}$, $\vec{e}_{j,L}$, and $\vec{h}_{pl}$.
        If $d_h \leq d_{E,j}$ the designated area is hit.
	}\label{fig:edge_hit_schematic}
\end{figure}

To calculate the individual decay probability within the detector of an LLNF, $\LLNF{}_i$, produced by a meson $M$, we require the lengths $L_{T,i}$ and $L_{I,i}$, which can be determined with the help of three spacial vertices: the production vertex of the $\LLNF{}_i$, $\vec{v}_{\LLNF,i}$\footnote{3-vectors are written in boldface.
For LLNFs produced from kaons, because of the kaons' long lifetime, the production vertex is in most cases not the primary vertex.
On the other hand, for charm and bottom meson decays, the production vertex can be approximated as coinciding with the primary vertex.}, the vertex $\vec{h}_{\mathrm{in}}$ where the trajectory of $\LLNF{}_i$ enters the fiducial volume of the detector, and the vertex $\vec{h}_{\mathrm{out}}$ where the trajectory of $\LLNF{}_i$ would exit the fiducial volume if $\LLNF{}_i$ were not to decay inside the detector volume. 
For every generated particle $P$ of the $i$-th event \Pythia{} provides a production vertex $\vec{v}_{P,i}$ and the orientation in terms of the momentum $\vec{p}_{P,i}$ of every generated particle $P$ of the $i$-th event.
We use the position $\vec{v}_{\LLNF,i}$ and $\vec{p}_{\LLNF,i}$, to describe the trajectory of any neutral particle $\LLNF{}_i$ (unaffected by any electromagnetic field) with a (real scalar) parameter $s \geq 0$ as (see Fig.~\ref{fig:Dune_schematic})
\begin{equation}
\vec{T}(s)=\vec{v}_{\LLNF,i}+s\,\vec{p}_{\LLNF,i}\,.
\end{equation}
\Dune{} can be approximated as a cuboid.
We define the fiducial volume of the \Dune{} near detector as enclosed by six areas defined via corner vertices $\vec{c}_j$ ($j=1,\dots,8$).
For every area, we take three of the four corners of that area ($\vec{c}_j,$ 
$\vec{c}_{j^\prime},$ and $\vec{c}_{j^{\prime\prime}}$) and calculate 
two directional vectors
\begin{equation}
\vec{o}_{1} = \vec{c}_{j^\prime} - \vec{c}_j\,,\hspace{5mm}\vec{o}_{2} = \vec{c}_{j^{\prime\prime}} - \vec{c}_j\,,
\end{equation}
where $\vec{o}_1\nparallel \vec{o}_2$.
The plane in which the area lies can be described with two (real scalar) parameters $t$ and $u$ as
\begin{equation}
\vec{P}(t,u)=\vec{c}_j + t\, \vec{o}_{1}+u \,\vec{o}_{2}.
\end{equation}
Solving $\vec{T}(s) = \vec{P}(t,u)$ for $s$, $t$ and $u$, we can determine the point $\vec{h}_{pl}$ where the plane $\vec{P}$ is hit.
To determine if $\vec{h}_{pl}$ lies within the area of a bounding side of the fiducial volume or if the area is missed by the trajectory $\vec{T}(s)$, we perform the following steps of calculation (see Fig.~\ref{fig:edge_hit_schematic}):\footnote{In principle, for a cuboid it would be sufficient to check if $t,u\in(0,1)$, for $\vec{T}(s)$ to hit the fiducial volume. 
Our procedure can be extended to other geometric shapes and also approximate cylindrical detectors such as \texttt{ANUBIS} (see Refs.~\cite{Dreiner:2020qbi,DeVries:2020jbs}). 
}

\begin{enumerate}
\item We determine the center $\vec{a}_c$ of the area as an average of all four corners $\vec{c}_j$, and the edge line equations $\vec{E}_j(a)=\vec{c}_j + a\cdot \big(\vec{c}_{j+1}-\vec{c}_{j}\big)$.
\item An imagined line between $\vec{a}_c$ and $\vec{h}_{pl}$ crosses every $\vec{E}_j(a)$. 
If $a\in(0,1)$ for the cross points $\vec{e}_{j,L}$, this cross point truly hits the edge of the area.
Such a calculation results in two $\vec{e}_{j,L}$.
\item To find the $\vec{e}_{j,L}$ closer to $\vec{h}_{pl}$, we calculate the distance between center and both edge cross points $d_{E,j} = |\vec{e}_{j,L}-\vec{a}_c|$, as well as the distance between the hit point and edge cross point $d_{eh,j} = |\vec{e}_{j,L}-\vec{h}_{pl}|$.
The $\vec{e}_{j,L}$, for which the sum of both distances is smaller, is the cross point closer to $\vec{h}_{pl}$.
\item Lastly, if $d_{h}= |\vec{h}_{pl}-\vec{a}_c|<d_{E,j}$, the hit point $\vec{h}_{pl}$ lies within the area.
In other words, $\vec{T}(s)$ hits the fiducial volume in that area.
\end{enumerate}
These steps are repeated for every area until either all areas are checked, or two areas are found to be hit at $\vec{h}_{1/2}$, where $\vec{h}_1$ is the vertex closer to $\vec{p}_{\LLNF,i}$ and $\vec{h}_2$ is the vertex further.
The required distances are then calculated as $L_{I,i} = |\vec{h}_{2}-\vec{h}_{1}|$ and $L_{T,i} = |\vec{h}_{1}-\vec{v}_{\LLNF,i}|$.

\subsection{Kaons at the \LHC{}}\label{subsec:kaon_LHC}

We are interested in comparing the sensitivity reach of \Dune{} to that of other experiments, especially (proposed) \LHC{} far detectors such as \texttt{FASER}~\cite{Feng:2017uoz,FASER:2018eoc,Feng:2022inv} and \texttt{MATHUSLA}~\cite{Chou:2016lxi,Curtin:2018mvb,MATHUSLA:2020uve}.
For $D$- and $B$-mesons and only involving charged-current processes, these numerical results have already been obtained in Ref.~\cite{DeVries:2020jbs}.
Implementing the geometry of these detectors in the simulation procedure explained in Secs.~\ref{subsec:MC_simulation} and \ref{subsec:three_dimensional_approximation}, we can extend the numerical results by including neutral couplings and production through kaon decays.
To evaluate the sensitivity for such decays we need the total production cross sections of kaons at the \LHC{}.
We perform an estimate with the MC simulation tool \texttt{EPOS-LHC}~\cite{Pierog:2013ria}, which is available in the \texttt{CRMC} simulation package~\cite{CRMC}, and obtain the following total kaon numbers at the \LHC{} over the full $4\pi$ solid angle, for a center-of-mass energy 14 TeV, and an integrated luminosity of 3 ab$^{-1}$:
\begin{eqnarray}
    K_L = 1.30\times 10^{18},\qquad K_S = 1.31\times 10^{18},\qquad K^\pm = 2.38\times 10^{18}\,.
\end{eqnarray}

We should treat the decay of kaons at the \LHC{} with care. 
Heavier mesons, such as $D$- and $B$-mesons decay promptly and we can legitimately approximate the \LLNF{} as being produced at the interaction point (IP). 
Kaons, especially $K_L$ and $K^\pm$, however, can travel a macroscopic distance before decaying.
The algorithm specified in Sec.~\ref{subsec:three_dimensional_approximation} can account for the displaced \LLNF{}-production vertex.
However, contrary to \DUNE{}, the \LHC{} does not include a decay pipe in which the long-lived mesons could decay.
Depending on the experiment we investigate, the kaons could either enter a calorimeter or be absorbed by, for example, the surrounding rock wall.
We do not simulate these possibilities.
Instead, we restrict the allowed meson decay vertices $\vec{v}_{P,i}$ to be within 
a specific `allowed decay region'.
Such decay regions can be divided into different types:
\begin{itemize}
\item 
The first type of the decay regions is based on the hadron calorimetry (HCAL) of \CMS{}~\cite{CMS:2008xjf} or \ATLAS{}~\cite{2137105,Gingrich:2007ia}.
\ANUBIS{}~\cite{Bauer:2019vqk} is supposed to be installed in a service shaft above \ATLAS{}.
A kaon that travels in the direction of \ANUBIS{} necessarily enters the HCAL of \ATLAS{}.
We choose not to account for the effects of the HCAL on the kaons and therefore only consider kaons that decay before reaching the HCAL in our Monte-Carlo simulation.
The inner radius of the barrel calorimeter HCAL at \ATLAS{} is \SI{2.28}{\meter}~\cite{2137105}.
The end-cap calorimeter is placed $\pm$\SI{4.277}{\meter} from the IP in the direction of the beam axis.
Hence, the allowed decay region for \ANUBIS{} is restricted to a cylinder around the beam axis with the stated dimensions. 
Similarly, \MATHUSLA{} would be located on the surface above the \CMS{} experiment, which involves an HCAL.
Here, the inner radius of the barrel calorimeter HCAL is \SI{1.77}{\meter}, whereas the end-cap calorimeter is placed approximately $\pm$\SI{3}{\meter} from the IP.
Similar to \ANUBIS{}, the allowed decay region for kaons at \MATHUSLA{} is a cylinder with the specified dimensions.

\item 
Next, we discuss LLP detectors in the far forward region, namely, \FASEROne{}, \FASERTwo{}, and \FACET{}~\cite{Cerci:2021nlb}.
These detectors are positioned downstream of the \ATLAS{} or \CMS{} IP.
The beamlines close to the \ATLAS{} and \CMS{} IP's are almost identical~\cite{Bruning:2004ej}.
Important for our purposes are the so-called `Target Absorber Secondaries' (TAS) and `Target Absorber Neutral' (TAN).
These are absorbers for charged and neutral secondary particles in order to protect magnets and other machine elements of the beamline.
Additionally, they can be used to reduce the background at LLP detectors~\cite{FASER:2019aik}.
Applied to our studies, the TAS and TAN limit the allowed decay region of charged and neutral kaons, respectively.
The TAS is placed approximately \SI{19}{\meter} after the IP, and the TAN's position is \SI{141}{\meter} downstream from the IP.
As \FACET{} would be located closer to the IP than the respective TAN, namely, at \SI{101}{\meter}, we only consider kaons which decay within \SI{101}{\meter}, instead of the TAN's position.

\item
For \CODEXB{}~\cite{Gligorov:2017nwh}, a specific lead shield is installed between the IP and the detector, covering the complete fiducial volume.
The distance between the IP and the lead shield is approximately \SI{5}{\meter}. Kaons are required to decay before reaching the shield in our simulation.

\item 
The fourth decay region is relevant for \MAPPOne{} and \MAPPTwo{}~\cite{Pinfold:2019nqj,Pinfold:2019zwp}.
Because of the location and shape of these detectors, we use the natural rock between the beam cavern and the detector cavern as the absorber for charged and neutral particles.
Hence, kaons should decay inside the beam cavern, which is approximately \SI{3.8}{\meter} wide. 

\end{itemize} 

%!TEX root= dune_eft.tex
\section{Numerical results}\label{sec:results}

We now present the results of the numerical simulation for the theoretical scenarios discussed in Sec.~\ref{sec:theoretical_scenarios}.
We begin with the minimal scenario where we only consider the active-sterile neutrino mixing and then subsequently discuss various scenarios where the long-lived neutral fermions have additional interactions.
For each scenario we compare the reach of \DUNE{} to that of other (proposed) experiments.

\subsection{Minimal scenario}

\begin{figure}[t]
\centering
	\includegraphics[width=0.8\textwidth]{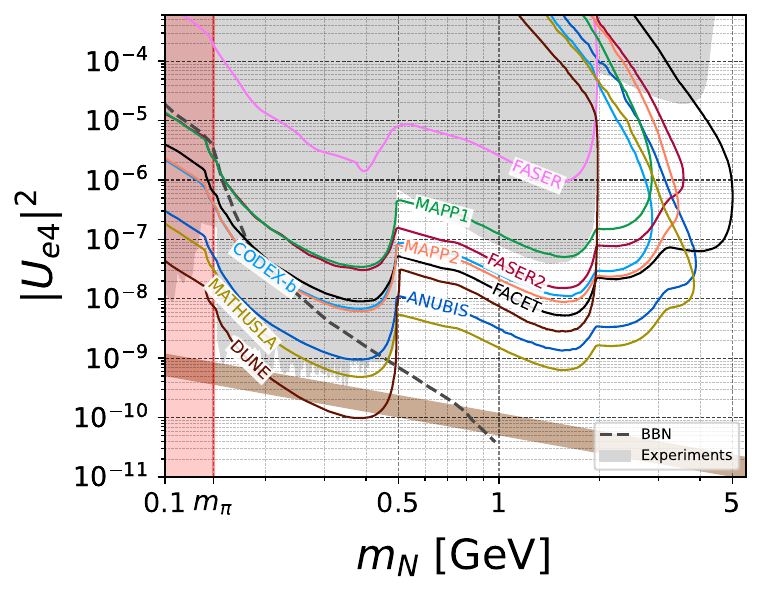}
	\caption{Sensitivity isocurves of \Dune{} and (proposed) \LHC{} far detectors for the minimal scenario, shown in the plane $|U_{e4}|^4$ vs.~$m_{\LLNF{}}$.
                 The brown band indicates a target region for the type-I seesaw relation $|U_{e4}|^2=m_{\nu}/m_\LLNF{}$, for $\SI{0.05}{\eV} < m_\nu < \SI{0.12}{\eV}$.
                 The light gray area and dark gray curve represent current limits on $U_{e4}$ from experiments and the lower bounds from Big Bang Nucleosynthesis~\cite{Sabti:2020yrt}, respectively.
                 The red area on the left corresponds to $m_{\LLNF{}}$ below the pion threshold, where, in principle, contributions from pion decays could enhance the sensitivities but are not considered in the present work.
    }\label{fig:minimal_scenario}
\end{figure}

In the minimal scenario, $\LLNF{}$ denotes specifically the sterile neutrino. The production 
and decay of light sterile neutrinos are both governed by the minimal mixing between sterile 
and active neutrinos. We do not employ the canonical see-saw relation $|U_{e4}|^2 = m_{\nu_e} 
/m_{\LLNF{}}$ and instead treat $U_{e4}$ as a free parameter, together with the mass of the 
LLNF, $m_{\LLNF{}}$. We present the results of the numerical simulation (see Sec.~\ref{sec:DUNE} 
for detail) in the $|U_{e4}|^2$ vs.~$m_{\LLNF{}}$ plane in Fig.~\ref{fig:minimal_scenario}. 
The \DUNE{} sensitivity reach is compared to a range of other (proposed) experiments. The 
results are displayed as three-event isocurves, which correspond to the 95$\%$ confidence 
level (C.L.) sensitivity reach, assuming zero background events (see Ref.~\cite{Coloma:2023oxx} for a more 
nuanced discussion regarding backgrounds). The figure also shows a 
brown band representing the target region for the type-I seesaw relation $|U_{e4}|^2=m_{\nu}/m_
\LLNF{}$, where $\SI{0.05}{\eV} < m_\nu < \SI{0.12}{\eV}$~\cite{Canetti:2010aw,Aghanim:2018eyx}.
Further, we include current limits on $|U_{e4}|^2$~\cite{Fernandez-Martinez:2023phj} from 
various experiments, shown in the light gray area, combining the results from \texttt{PIENU}~
\cite{PIENU:2017wbj}, \texttt{KENU}~\cite{Bryman:2019bjg}, \texttt{NA62}~\cite{NA62:2020mcv}, 
\texttt{T2K}~\cite{T2K:2019jwa}, \texttt{BEBC}~\cite{Barouki:2022bkt}, and \texttt{DELPHI}~
\cite{DELPHI:1996qcc}. We also display the lower bounds expected from Big Bang
Nucleosynthesis~\cite{Sabti:2020yrt}, as a dashed dark grey curve; note that these limits depend on the 
cosmological model considered. One shortcoming of our simulations is that we do not include sterile 
neutrino production through pion decays. Therefore, our results underestimate the sensitivity reach 
for $m_\LLNF{}< m_\pi \simeq \SI{140}{\mega\eV}$, and this region is marked in red.
 
The \Dune{} isocurve can be split into two segments, where different production modes govern the 
sensitivity reach. For $m_\pi<m_{\LLNF{}}<m_K$ it is dominated by sterile neutrinos produced in 
kaon decays, because the number of kaons produced at \Dune{} is roughly four orders of magnitude 
larger than that of charmed mesons (see Table~\ref{tab:meson_numbers}). For $m_K<m_{\LLNF{}}<
m_{D_s}$, kaon decays into the LLNF are no longer kinematically allowed, hence the kink in the 
sensitivity curves. The production is then dominated by the decays of charmed mesons. The peak
sensitivity here is a bit above $\SI{1}{\giga\eV}$, because there the branching ratio for $D^\pm$ and $D_s$ 
into a $N+e$ final state is the largest (\textit{cf.} Fig.~\ref{fig:minimal_production}). For 
sterile neutrinos with masses above $m_{D_s}$, only \LLNF{} production modes via bottom meson 
decays are allowed. However, the number of bottom mesons produced at \Dune{} is so small, that 
the three-event isocurve does not extend into the region of this plot. Our sensitivity estimates 
are, apart from the region below the pion mass, in good agreement with, for example, 
Refs.~\cite{Krasnov:2019kdc, Coloma:2020lgy} confirming the validity of our simulation.

As shown in Fig.~\ref{fig:minimal_scenario}, compared to the (proposed) detector concepts at the \LHC{}, \Dune{} is expected to perform better than many of them for the minimal scenario. 
In the `charmed'-region of the isocurve, only \ANUBIS{} and \MATHUSLA{} extend beyond the parameter 
region covered by \Dune{}, by less than one order of magnitude in $|U_{e4}|^2$. In the 
`kaon'-region, \Dune{} is expected to probe $|U_{e4}|^2$ values smaller than what \MATHUSLA{} and 
\ANUBIS{} can do by almost one order of magnitude. It can even cover the see-saw band range for 
$m_{\LLNF{}}$ between \SI{0.2}{\giga\eV} and \SI{0.5}{\giga\eV}. On the other hand, because of the
much larger center-of-mass energy, the isocurve of every \LHC{} detector concept except \FASEROne{} 
extends beyond \Dune{} for masses above $m_{D_s}$, complementing \Dune{}'s  strong sensitivities 
at lower mass regions.

\subsection{Flavor benchmarks}\label{subsec:flavor_benchmarks}

In each theoretical scenario we select different flavor benchmarks as discussed in 
Sec.~\ref{sec:LLNF_production_EFT} and Sec.~\ref{sec:LLNF_decay_EFT}. We select one 
EFT operator $(c_P)_{ij}$ \textit{mainly} responsible for the production and another 
operator $(c_D)_{i^\prime j^\prime}$ for the decay of the \LLNF{}, where $i^{(\prime)}$ and $j^{(\prime)}$ are generation 
indices of the up- and down-type quarks of the charged $\nu$LEFT current, respectively. In the 
left-right symmetric model scenario, the operator is given by $c_{VLR}^{CC1}$ with various 
flavor indices. For the leptoquark and the RPV-SUSY scenarios, we choose a specific flavor of 
the charged scalar operator $c^{CC}_{SRR}$, and further include the three related operators as
described in Eq.~(\ref{eq:LQmatching}) and Eq.~(\ref{eq:RPVmatching}). Contributions from 
active-sterile neutrino mixing $U_{e4}$ are taken into consideration for the leptoquark and 
left-right symmetric model-like scenario using the canonical see-saw relation 
\begin{equation}\label{eq:canonical_see-saw}
U_{e4}= \sqrt{\frac{m_{\nu_e}}{m_\LLNF}},\,
\end{equation}
with a representative active neutrino mass $m_{\nu_e}=\SI{0.05}{\eV}$ for both \Dune{} and 
the \LHC{} far detectors. For \Dune{}, we additionally perform simulations where we turn off 
minimal mixing. We observe essentially no difference between the leptoquark scenario without 
minimal mixing and the RPV-SUSY scenario for every benchmark. The different relations between
the scalar and tensor couplings in such scenarios [Eq.~\eqref{eq:LQmatching} and 
Eq.~\eqref{eq:RPVmatching}], both in signs and in coupling strength, do not significantly 
affect the estimated number of observed \LLNF{} events. This aligns with the discussions in 
Sec.~\ref{sec:LLNF_production_EFT} and Sec.~\ref{sec:LLNF_decay_EFT}. Both the production and 
decay modes for an LLNF in the leptoquark scenario without minimal mixing and the RPV-SUSY 
scenario are comparable (see also Fig.~\ref{fig:eft_decay}), so that differences in the 
numerical results are minor. Thus, we only show the leptoquark results.

Furthermore, we reinterpret existing limits on the active-sterile mixing $U_{e4}$ from 
meson decays in terms of the benchmark scenarios \cite{Beltran:2023nli}.
For sterile neutrinos with a mass $m_\pi<m_N<m_D$, we use
the $U_{e4}$-limits from \texttt{NA62}~\cite{NA62:2020mcv}, \texttt{T2K}~\cite{T2K:2019jwa}, 
and \texttt{BEBC}~\cite{Barouki:2022bkt}, where \LLNF{} is produced in the decay of kaons or 
$D$-mesons. We assume $U_{e4}$ is fixed according to Eq.~\eqref{eq:canonical_see-saw},
with $m_{\nu_e}=\SI{0.05}{\eV}$ in the reinterpretation. The limits on the EFT operators 
obtained in this procedure are included in the following.

%%%%%%%%%%%%%%%%%%%%%%%%%%%%%%%%%%%%%%%%%%%%%%%%%%%%%%%%%%%%%%%%%%%%%%%%%%%%%%%%
%%%%%%%%%%%%%%%%%%%%%%%%%%%%%%%%%%%%%%%%%%%%%%%%%%%%%%%%%%%%%%%%%%%%%%%%%%%%%%%%
\subsubsection{Flavor benchmark 1}
%%%%%%%%%%%%%%%%%%%%%%%%%%%%%%%%%%%%%%%%%%%%%%%%%%%%%%%%%%%%%%%%%%%%%%%%%%%%%%%%
%%%%%%%%%%%%%%%%%%%%%%%%%%%%%%%%%%%%%%%%%%%%%%%%%%%%%%%%%%%%%%%%%%%%%%%%%%%%%%%%

\begin{table}[t]
\centering
\renewcommand{\arraystretch}{1.3}
\begin{tabular}{|l||lr|l|}
\hline 
 & \multicolumn{2}{c|}{Leptoquark / RPV} & \multicolumn{1}{c|}{Left-Right} \\
 \hline
\hline
$(c_P)_{ij}$ & \multicolumn{2}{c|}{$c^{CC}_{SRR,12}$} & \multicolumn{1}{c|}{$c^{CC1}_{VLR,12}$} \\ 
$(c_D)_{i^\prime j^\prime}$ & \multicolumn{2}{c|}{$c^{CC}_{SRR,11}$} & \multicolumn{1}{c|}{$c^{CC1}_{VLR,11}$} \\ 
\hline\hline
\LLNF{} Production Modes via $\mathbf{M}$ & & & \\
\hline
$\mathbf{M}\rightarrow \LLNF{}+e^\pm(+X)$& \multicolumn{2}{l|}{$K^\pm,\,K_{\subls}$} & $K^\pm,\,K_{\subls}$\\
\hline
$\mathbf{M}\rightarrow \LLNF{}+\nu_e(+X)$ & $K^\pm,\,K_{\subls}$ & $\big[V_{us},\,V_{ud}\big]$ & \\
\hspace{10mm} via CKM $\big[V_{il}\big]$ & $B^\pm,\,B^0,\,B_s^0$ & $\big[V_{ub}\big]$ &\\
\hline\hline
\LLNF{} Decay Modes into $\mathbf{M}^\prime$ & & & \\
\hline
$\LLNF{}\rightarrow \mathbf{M}^\prime+e^\pm$ & \multicolumn{2}{l|}{$\pi^\mp,\,\rho^\mp,\,K^\mp,\,K^{*\mp}$} & $\pi^\mp,\,\rho^\mp,\,K^\mp,\,K^{*\mp}$\\
\hline
& $\pi^0,\,\rho^0,\,\omega$ & $\big[V_{ud}\big]$ & \\
$\LLNF{}\rightarrow \mathbf{M}^\prime+\nu_e$& $\eta,\,\eta^\prime$ & $\big[V_{us},\,V_{ud}\big]$ & \\
\hspace{10mm}via CKM  $\big[V_{il}\big]$ & $K_{\subls},\,K^{*0},\,\phi$ & $\big[V_{us}\big]$ &\\
&$B^0,\,B^{*0},\,B_s^0,\,B_s^{*0}$ & $\big[V_{ub}\big]$ &\\
\hline
\end{tabular}
	\caption{Summary of flavor benchmark 1 for theoretical scenarios 2, 3 
	and 4 as described in Sec.~\ref{sec:LLNF_production} and Sec.~\ref{sec:LLNF_decay}.
	Listed are the non-zero couplings $(c)_{P/D}$ as well as the induced $\LLNF{}$ production and decay modes.
        $X$ represents any possible additional final-state particle in the meson decays.
        Modes induced by related neutral quark currents are	suppressed by the corresponding CKM matrix element.
        We do not list the contributions from a non-zero $U_{e4}$ (see Sec.~\ref{sec:LLNF_production_minimal} and Sec.~\ref{sec:LLNF_decay_minimal} for details).
	}\label{tab:flavor_benchmark_summary_1}
\end{table}

\begin{figure}[t]
\centering
	\includegraphics[width=0.49\textwidth]{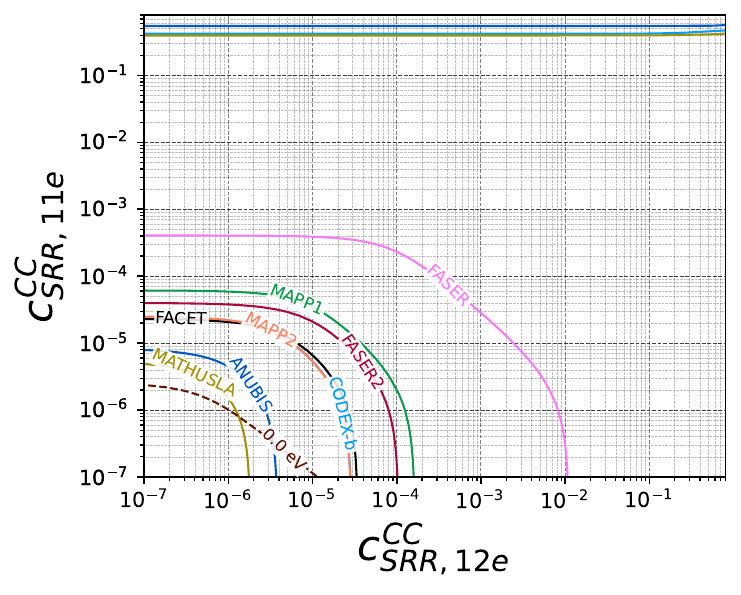}
	\includegraphics[width=0.49\textwidth]{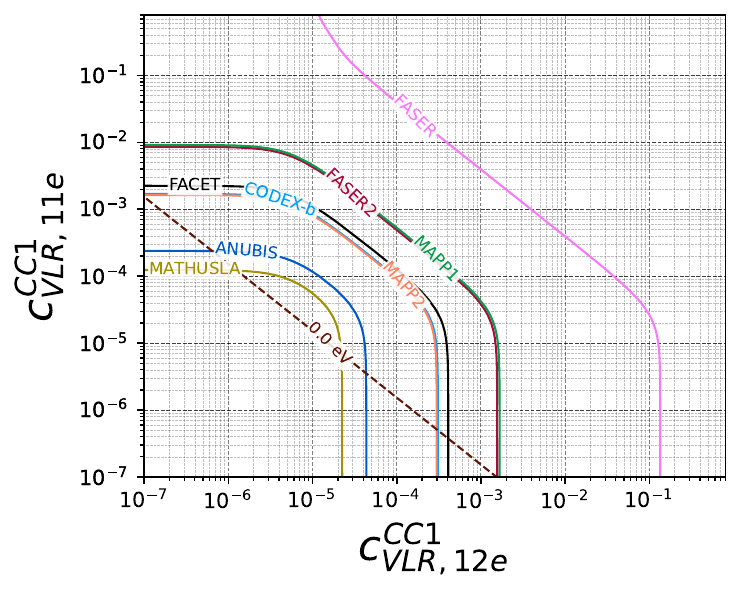}
	\includegraphics[width=0.49\textwidth]{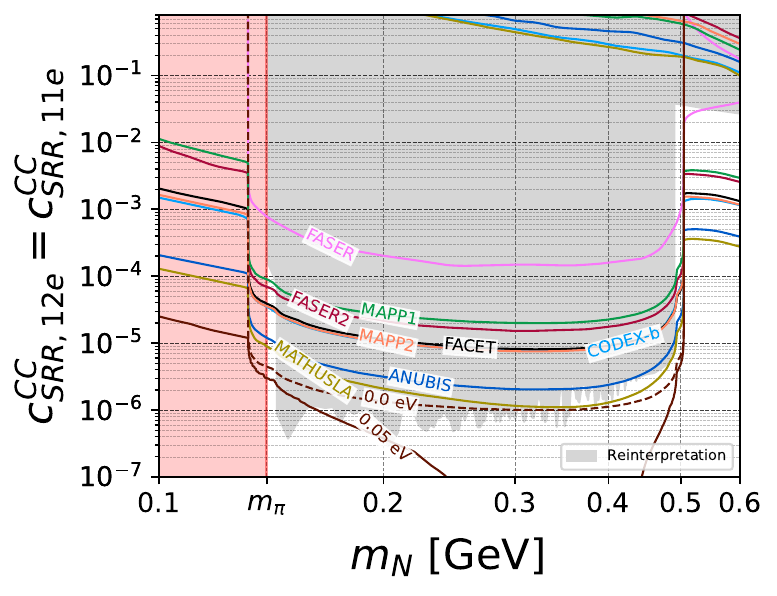}
	\includegraphics[width=0.49\textwidth]{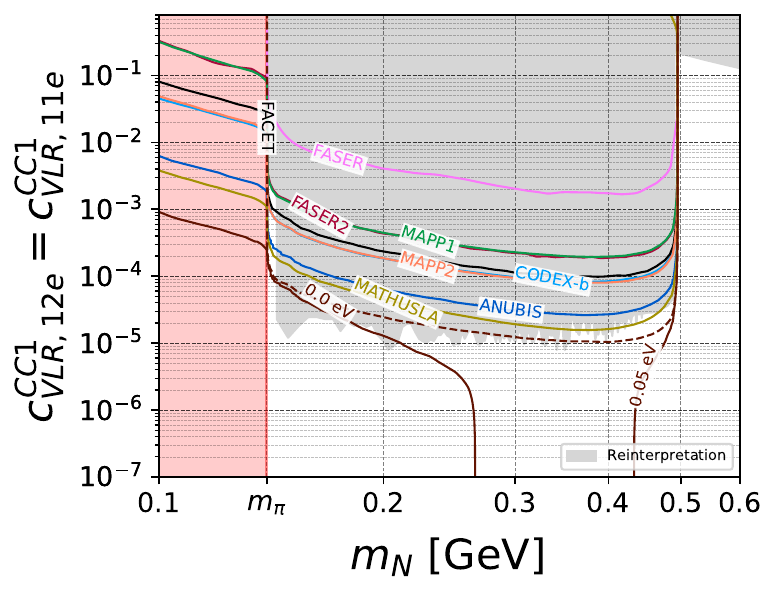}
	\caption{Sensitivity isocurves for flavor benchmark 1 as summarized in Table~\ref{tab:flavor_benchmark_summary_1}.
        \textit{Upper}: $c_D$ vs.~$c_P$ for a fixed mass of $m_{\LLNF{}}=\SI{0.3}{\giga\eV}$. \textit{Lower}: $c_P=c_D$ vs.~$m_{\LLNF{}}$.
        \textit{Left}: the leptoquark and RPV scenarios.
        \textit{Right}: the left-right symmetric model-like scenario.
        For all detectors, we show HNL sensitivity curves with the type-I seesaw relation 
        for an active neutrino mass $m_\nu=\SI{0.05}{\eV}$. Additionally, for \Dune{}, we 
        display the results for $m_\nu=\SI{0.0}{\eV}$, which also represent the sensitivity 
        of the RPV-SUSY scenario, to a good approximation. We  reinterpret existing limits 
        on $U_{e4}$ following Ref.~\cite{Beltran:2023nli}.
        The light gray area represents the restricted parameter space obtained from the reinterpretation.
        }
        \label{fig:flavor_benchmark_1}
\end{figure}

In the first flavor benchmark, we consider for the production $(c_P)_{12}\in\{c^{CC}
_{SRR,12}\,; c^{CC1}_{VLR,12}\}$ and for the decay $(c_D)_{11}\in \{c^{CC}_{SRR,11}\,; 
c^{CC1}_{VLR,11}\}$, for each respective scenario. The resulting meson and \LLNF{} 
decay modes are summarized in Table~\ref{tab:flavor_benchmark_summary_1}. Meson decay 
modes, as given in Eqs.~\eqref{eq:charged_prod_leptonic}-\eqref{eq:charged_prod_semileptonic}, 
are induced by kaons, while we include kaons and $B$-meson decays via 
Eqs.~\eqref{eq:neutral_prod_leptonic}-\eqref{eq:neutral_prod_semileptonic}, if neutral 
couplings are non-zero in the specific scenario. Contributions from a non-zero $U_{e4}$ 
are not listed specifically, but included as described in 
Sec.~\ref{sec:LLNF_production_minimal}. We present 
results for the first benchmark in Fig.~\ref{fig:flavor_benchmark_1}, where  we show 
3-signal-event isocurves as the sensitivity reach at 95\% C.L..

The upper half of the figure contains two plots in the $(c_D)_{11}$ vs.~$(c_P)_{12}$ plane, 
for a fixed mass $m_{\LLNF{}}=\SI{0.3}{\giga\eV}$. The lower half contains two plots in the 
plane $(c_D)_{11}=(c_P)_{12}$ vs.~$m_{\LLNF{}}$. In these plots, as discussed in 
Sec.~\ref{subsec:flavor_benchmarks}, we have followed Ref.~\cite{Beltran:2023nli} and
reinterpreted existing HNL searches in 
order to obtain recast bounds on the considered $\nu$LEFT scenario; these results are shown 
in light gray. The left side in each half shows results for the leptoquark (and RPV) 
scenarios, while the right side is for the left-right symmetric model-like scenario. 

In the 
upper row, the \Dune{} isocurves for a neutrino mass $m_\nu = \SI{0.05}{\eV}$ are missing. 
The reason for this is that in the minimal scenario, for such an active neutrino mass and $m_N= \SI{300}
{\mega\eV}$, the number of observed neutrino events is always greater than three independent of the EFT couplings. 
This also explains why the \Dune{}-isocurve here behave differently 
than the \LHC{}-detector curves, as the latter do include minimal mixing. 

For this flavor benchmark, \Dune{} essentially outperforms all \LHC{} detectors across the 
parameter space. This is clearly seen in the bottom panels, where the \Dune{} curves probe 
the smallest couplings. \MATHUSLA{} and \ANUBIS{} are not much weaker in sensitivity, 
though.\footnote{The sensitivity reaching down to zero coupling is here related solely to 
the minimal mixing contribution.}
Because of neutral-current processes and the resulting $B$-meson decays into \LLNF{}, the \LHC{} detectors isocurves extend beyond $m_K$ to the right and complement \Dune{}'s sensitivity reach.
This can be observed in the lower-left plot of Fig.~\ref{fig:flavor_benchmark_1}.
We find for LLNF mass below the kaon threshold, essentially only \Dune{} and \MATHUSLA{} can 
probe parameter space beyond the existing bounds.

Comparing the leptoquark scenario to the left-right symmetric model-like scenario, we find the sensitivity results of the left-right symmetric model-like scenario to be weaker than those of the leptoquark or RPV-SUSY scenario; the main reason is that the meson decay branching ratios for the LLNF production are enhanced in case of scalar-current interactions.
This difference is even more pronounced for smaller neutrino masses, because of the chiral suppression of the vector operators in the left-right symmetric model-like scenario.

Additionally, in the lower plots of Fig.~\ref{fig:flavor_benchmark_1}, the \Dune{} curve with 
$m_{\nu}=\SI{0.0}{\eV}$ in the left-right symmetric model-like scenario (the right panel), shows a 
similar $m_\LLNF{}$-dependence to the \LHC{}-detector sensitivity curves (with $m_{\nu}=\SI{0.05}{\eV}$). However, in the leptoquark scenario, the 
$m_\nu=\SI{0.0}{\eV}$ \Dune{} curve exhibits a slightly different $m_N$-dependence 
than the \LHC{}-detector sensitivities with non-zero minimal mixing. This is explained by interference between $\LLNF{}$ production modes arising from $(c_P)_{12}$ and $U_{e4}$ which are included for the LHC detectors but not for the \Dune{} $m_\nu=\SI{0.0}{\eV}$ line\footnote{We consider $(c_P)_{12}>0$, so that the interference enhances the sensitivity for a non-zero $U_{e4}$. In principle, $(c_P)_{12}<0$ or $U_{e4}<0$ is allowed, such that the interference term could instead weaken the sensitivity reach for these detectors.}. This interference is smaller in the left-right scenario and therefore the \Dune{} $m_\nu=\SI{0.0}{\eV}$ line has the same $m_N$ dependence as that of the LHC detectors in the bottom-right panel.

The \Dune{} RPV-SUSY results for this benchmark coincide with the $m_\nu=\SI{0.0}{\eV}$ 
results shown in the lower-left plot, and have been worked out recently in
Ref.~\cite{Dreiner:2023gir}, where a quick and simple recast was performed; we find the 
results to be comparable to the results from the full simulations of this work.

Very roughly, \DUNE{} will have a sensitivity range to $\nu$LEFT scalar couplings at the $C^{CC}_{SRR}\sim 10^{-6}$-level and to vector couplings at the $C^{CC1}_{VLR}\sim10^{-5}$-level.
This approximately corresponds to a BSM physics scales of $\Lambda \sim \sqrt{v^2/C^{CC}_{SRR}} = \mathcal O(250)$\SI{}{\tera\eV} and $\Lambda \sim \sqrt{v^2/C^{CC1}_{VLR}} = \mathcal O(80)$\SI{}{\tera\eV}, respectively, for \LLNF{} masses around \SI{0.3}{\giga\eV}.
These limits are stronger than many existing and proposed probes of $\nu$SMEFT couplings; see \textit{e.g.}~Refs.~\cite{DeVries:2020jbs, Fernandez-Martinez:2023phj, Alves:2023znq}, and, for this benchmark scenario, comparable to limits from \texttt{NA62}.

%%%%%%%%%%%%%%%%%%%%%%%%%%%%%%%%%%%%%%%%%%%%%%%%%%%%%%%%%%%%%%%%%%%%%%%%%%%%%%%%
%%%%%%%%%%%%%%%%%%%%%%%%%%%%%%%%%%%%%%%%%%%%%%%%%%%%%%%%%%%%%%%%%%%%%%%%%%%%%%%%
\subsubsection{Flavor benchmark 2}
%%%%%%%%%%%%%%%%%%%%%%%%%%%%%%%%%%%%%%%%%%%%%%%%%%%%%%%%%%%%%%%%%%%%%%%%%%%%%%%%
%%%%%%%%%%%%%%%%%%%%%%%%%%%%%%%%%%%%%%%%%%%%%%%%%%%%%%%%%%%%%%%%%%%%%%%%%%%%%%%%

\begin{table}[t]
\centering
\renewcommand{\arraystretch}{1.3}
\begin{tabular}{|l||lr|l|}
\hline 
 & \multicolumn{2}{c|}{Leptoquark / RPV} & \multicolumn{1}{c|}{Left-Right} \\
 \hline
\hline
$(c_P)_{ij}$ & \multicolumn{2}{c|}{$c^{CC}_{SRR,21}$} & \multicolumn{1}{c|}{$c^{CC1}_{VLR,21}$} \\ 
$(c_D)_{i^\prime j^\prime}$ & \multicolumn{2}{c|}{$c^{CC}_{SRR,11}$} & \multicolumn{1}{c|}{$c^{CC1}_{VLR,11}$} \\ 
\hline\hline
\LLNF{} Production Modes via $\mathbf{M}$ & & & \\
\hline
$\mathbf{M}\rightarrow \LLNF{}+e^\pm(+X)$& \multicolumn{2}{l|}{$D^\pm,\,D^0,\,D^\pm_s$} & $D^\pm,\,D^0,\,D^\pm_s$\\
\hline
$\mathbf{M}\rightarrow \LLNF{}+\nu_e(+X)$ & $K^\pm,\,K_{\subls}$ & $\big[V_{us},\,V_{cs}\big]$ & \\
\hspace{10mm} via CKM $\big[V_{il}\big]$ & $B^\pm,\,B^0,\,B_s^0$ & $\big[V_{ub},\,V_{cb}\big]$ &\\
\hline\hline
\LLNF{} Decay Modes into $\mathbf{M}^\prime$ & & & \\
\hline
$\LLNF{}\rightarrow \mathbf{M}^\prime+e^\pm$ & \multicolumn{2}{l|}{$\pi^\mp,\,\rho^\mp,\,D^\mp,\,D^{*\mp}$} & $\pi^\mp,\,\rho^\mp,\,D^\mp,\,D^{*\mp}$\\
\hline
\multirow{2}{*}{$\LLNF{}\rightarrow \mathbf{M}^\prime+\nu_e$}& $\pi^0,\,\rho^0,\,\eta,\,\eta^\prime,\,\omega$ & $\big[V_{ud},\,V_{cd}\big]$ & \\
\multirow{2}{*}{\hspace{10mm}via CKM  $\big[V_{il}\big]$} & $K_{\subls},\,K^{*0}$ & $\big[V_{us},\,V_{cs}\big]$ &\\
&$B^0,\,B^{*0},\,B_s^0,\,B_s^{*0}$ & $\big[V_{ub},\,V_{cb}\big]$ &\\
\hline
\end{tabular}
	\caption{Summary of flavor benchmark 2 for theoretical scenarios 2, 3, and 4 as described in Sec.~\ref{sec:LLNF_production} and Sec.~\ref{sec:LLNF_decay}.
	}\label{tab:flavor_benchmark_summary_2}
\end{table}

We now consider production through $(c_P)_{21}\in \{c^{CC}_{SRR,21}\,; c^{CC1}_{VLR,21}\}$, 
while the decay coupling is unchanged [$(c_D)_{11}$]. The LLNFs are produced from charged-current 
$D$-meson decays, while production through kaon and $B$-meson decays can occur through neutral 
currents in the leptoquark and RPV scenarios. A summary is given in 
Table~\ref{tab:flavor_benchmark_summary_2}.

\begin{figure}[t]
\centering
	\includegraphics[width=0.49\textwidth]{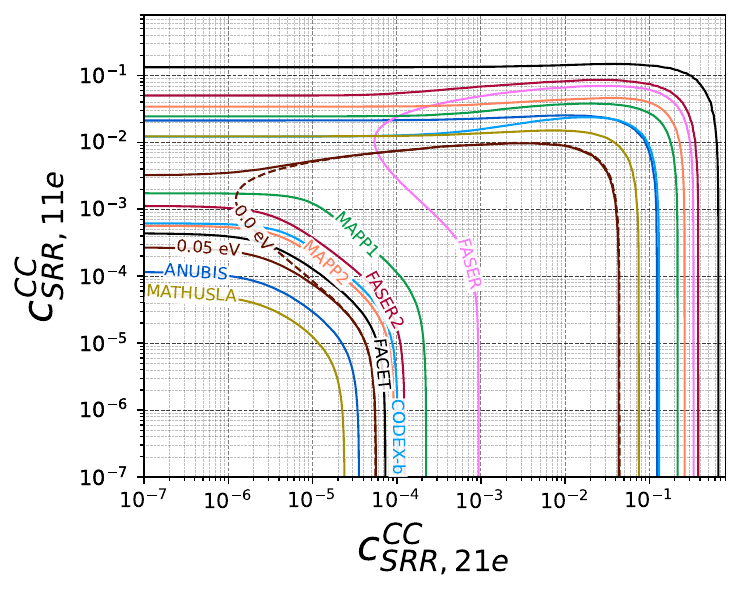}
	\includegraphics[width=0.49\textwidth]{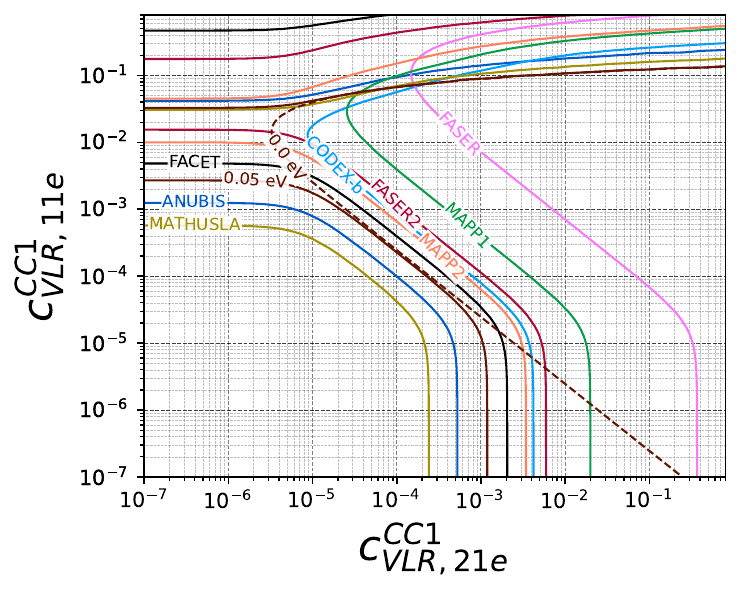}
	\includegraphics[width=0.49\textwidth]{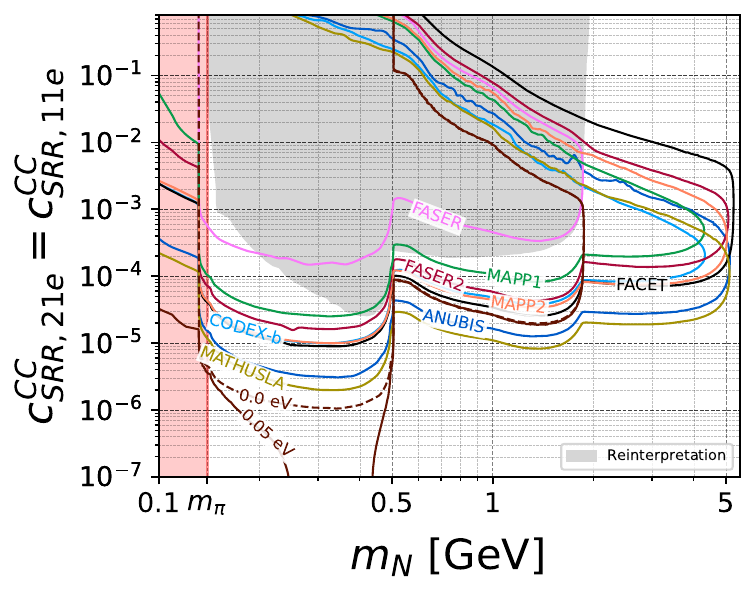}
	\includegraphics[width=0.49\textwidth]{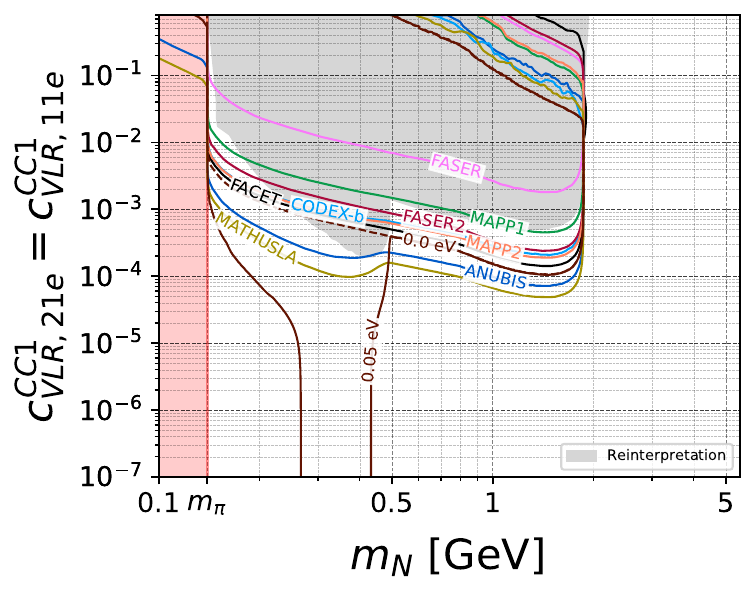}
	\caption{\DUNE{} sensitivity isocurves for flavor benchmark 2.
            The three-event isocurves (at 95$\%$ C.L.) depend on the individual couplings strengths $c_P$ and $c_D$, as well as the LLNF mass $m_{\LLNF{}}$.
                 In the upper row, we fix $m_{\LLNF{}}=\SI{1.0}{\giga\eV}$ and display the results as $c_P$ vs.~$c_D$.
                 In the lower row, we restrict $c_P=c_D$ and show the isocurves in the $c_P$ vs.~$m_{\LLNF{}}$ plane.
                 On the left side, the leptoquark and RPV scenarios are depicted, while the left-right symmetric model-like scenario is shown on the right. 
	          We include together sensitivity isocurves for several previously investigated \LHC{} detector proposals~\cite{DeVries:2020jbs} that we expand on here.
                The light gray area represents excluded parameter space obtained by reinterpreting existing $U_{e4}$ limits.
         }\label{fig:flavor_benchmark_2}
\end{figure}

Results for benchmark scenario 2 are given in Fig.~\ref{fig:flavor_benchmark_2} using the same panel formats as in Fig.~\ref{fig:flavor_benchmark_1}.
The upper plots spanned by $(c_D)_{11}$ and $(c_P)_{21}$ correspond to a fixed LLNF mass of \SI{1}{\giga\eV}.
For this mass, we can include the \Dune{} sensitivity for a non-zero minimal mixing corresponding to the $\SI{0.05}{\eV}$ line, which has a similar shape as the curves for the \LHC{} detectors.

For a $\SI{1}{\giga\eV}$ LLNF, the \MATHUSLA{} and \ANUBIS{} experiments outperform \Dune{} in sensitivity by roughly an order of magnitude in both production and decay EFT couplings.
The reason is that the production rate of $D$-mesons at \Dune{} is only an order-of-magnitude greater than at the \LHC{} experiments (whereas \Dune{} has a three-to-four order-of-magnitude gain for kaons). The larger angular coverage of \MATHUSLA{} and \ANUBIS{} in comparison to \DUNE{} then results in the increased sensitivity for the LHC experiments.
This is further reflected in the bottom panels where for \LLNF{}'s heavier than kaons, the \DUNE{} isocurves are above the \MATHUSLA{} and \ANUBIS{} lines (but still below many other experiments).

The production of LLNFs through kaon decays (even when setting $U_{e4} =0$) still leads to \DUNE{} being the most sensitive experiment for \LLNF{}'s lighter than a kaon. 
The mass-dependence for the leptoquark-model (and RPV model) with $m_\nu=\SI{0.0}{\eV}$ is now very similar to the \LHC{}-detectors where $U_{e4}\neq 0$ because interference effects play no significant role in this benchmark. The production through kaon decays from the EFT operators however does not work in the left-right scenario. As such, without minimal mixing, \MATHUSLA{} and \ANUBIS{} outperform \DUNE{} in this scenario even for light LLNFs as shown in the bottom-right panel.

Finally, the recast bounds shown in the light gray area are much weaker than several of the experiments considered here. This is mainly caused by the fact that the EFT operators in this benchmark do not contribute to the two-body decay $K^\pm\to N+e^\pm$, and therefore the \texttt{NA62} limits do not apply.

\subsubsection{Flavor benchmark 3}

\begin{table}[t]
\centering
\renewcommand{\arraystretch}{1.3}
\begin{tabular}{|l||lr|l|}
\hline 
 & \multicolumn{2}{c|}{Leptoquark / RPV} & \multicolumn{1}{c|}{Left-Right} \\
 \hline
\hline
$(c_P)_{ij}$ & \multicolumn{2}{c|}{$c^{CC}_{SRR,21}$} & \multicolumn{1}{c|}{$c^{CC1}_{VLR,21}$} \\ 
$(c_D)_{i^\prime j^\prime}$ & \multicolumn{2}{c|}{$c^{CC}_{SRR,12}$} & \multicolumn{1}{c|}{$c^{CC1}_{VLR,12}$} \\ 
\hline\hline
\LLNF{} Production Modes via $\mathbf{M}$ & & & \\
\hline
$\mathbf{M}\rightarrow \LLNF{}+e^\pm(+X)$& \multicolumn{2}{l|}{$K^\pm,\,K_{\subls},\,D^\pm,\,D^0,\,D^\pm_s$} & $K^\pm,\,K_{\subls},\,D^\pm,\,D^0,\,D^\pm_s$\\
\hline
$\mathbf{M}\rightarrow \LLNF{}+\nu_e(+X)$ & $K^\pm,\,K_{\subls}$ & $\big[V_{ud},\,V_{cs}\big]$ & \\
\hspace{10mm} via CKM $\big[V_{il}\big]$ & $B^\pm,\,B^0,\,B_s^0$ & $\big[V_{ub},\,V_{cb}\big]$ &\\
\hline\hline
\LLNF{} Decay Modes into $\mathbf{M}^\prime$ & & & \\
\hline
$\LLNF{}\rightarrow \mathbf{M}^\prime+e^\pm$ & \multicolumn{2}{l|}{$K^\mp,\,K^{*\mp},\,D^\mp,\,D^{*\mp}$} & $K^\mp,\,K^{*\mp},\,D^\mp,\,D^{*\mp}$\\
\hline
& $\pi^0,\,\rho^0,\,\omega$ & $\big[V_{cd}\big]$ & \\
& $\eta,\,\eta^\prime$ & $\big[V_{us},\,V_{cd}\big]$ & \\
$\LLNF{}\rightarrow \mathbf{M}^\prime+\nu_e$ & $\phi$ & $\big[V_{us}\big]$ &\\
\hspace{10mm}via CKM  $\big[V_{il}\big]$ & $K_{\subls},\,K^{*0}$ & $\big[V_{cs}\big]$ &\\
&$B^0,\,B^{*0}$ & $\big[V_{cb}\big]$ &\\
&$B_s^0,\,B_s^{*0}$ & $\big[V_{ub}\big]$ &\\
\hline
\end{tabular}
	\caption{Summary of flavor benchmark 3 for theoretical scenarios 2, 3, and 4 as described in Sec.~\ref{sec:LLNF_production} and Sec.~\ref{sec:LLNF_decay}.
	}\label{tab:flavor_benchmark_summary_3}
\end{table}

\begin{figure}[t]
\centering
	\includegraphics[width=0.49\textwidth]{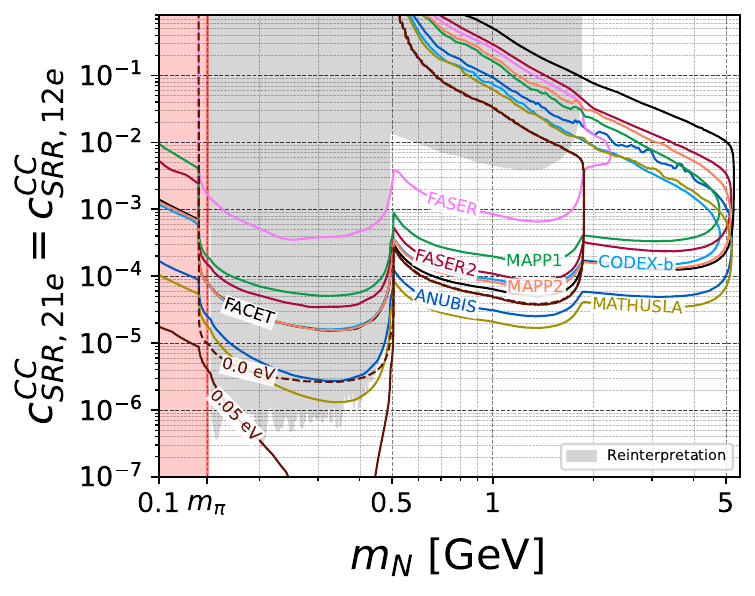}
	\includegraphics[width=0.49\textwidth]{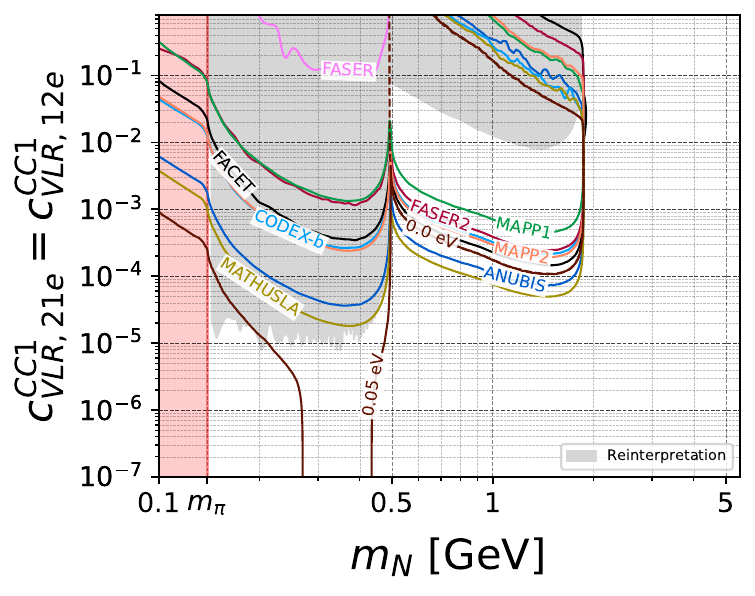}
	\caption{\Dune{} sensitivity isocurves for flavor benchmark 3.
        We do not present the isocurves for $(c_D)_{12}$ vs.~$(c_P)_{21}$ as the sensitivity curves at $m_N=\SI{1}{\giga\eV}$ are very similar to those shown in Fig.~\ref{fig:flavor_benchmark_2}.
        On the left-side (right-side) we display the isocurves in the  $(c_P)_{21}=(c_D)_{12}$ vs.~$m_N$ plane for the leptoquark/RPV-SUSY (left-right symmetric) model.
        The light gray area represents limits on the EFT operators reinterpreted from existing upper bounds on $U_{e4}$.
 }\label{fig:flavor_benchmark_3}
\end{figure}

In the final benchmark scenario, we consider production of the LLNFs through $(c_P)_{21}$ and decays through $(c_D)_{12}$, leading to production and decay modes summarized in Table~\ref{tab:flavor_benchmark_summary_3}.
The sensitivity isocurves are shown in Fig.~\ref{fig:flavor_benchmark_3}.
In the $(c_D)_{12}$ vs.~$(c_P)_{21}$ plane, the structure of the sensitivity curves is very similar to that in Fig.~\ref{fig:flavor_benchmark_2}, so in the interest of brevity, we do not show these plots here.

For the leptoquark and RPV scenarios, the results are similar compared to the previous benchmark except that the \DUNE{} $m_\nu=\SI{0.0}{\eV}$ curve now is somewhat above the  \MATHUSLA{} curve for light LLNFs (this was the other way around in the previous benchmark). For very small EFT couplings, the decay length is large and the probability, see Eq.~\eqref{eq:individual_decay_probability}, for an LLNF to decay inside the detector is approximately 
\begin{eqnarray}
    P_{\LLNF}&=& \exp\big[-L_{T}/\lambda_{N}\big] \cdot \big(1-\exp\big[-L_{I}/\lambda_{N}\big]\big) \simeq L_{I}/\lambda_{N}  \,\, \propto \,\, L_I\cdot\Gamma_{\text{tot}}^N\,,
\end{eqnarray}
and thus proportional to the total decay rate. For light LLNFs and small EFT couplings, in this benchmark the minimal mixing actually dominates $\Gamma_{\text{tot}}^N$. This explains why the \DUNE{} $m_\nu=\SI{0.0}{\eV}$ curve falls below the $\MATHUSLA$ curve where minimal mixing is included. This is more pronounced in the left-right scenario without minimal mixing where for $m_N<m_K$, the LLNF is stable and thus the $m_\nu=\SI{0.0}{\eV}$ curve does not reach this region at all.

For larger masses, however, the limits between flavor benchmarks 2 and 3 are very similar and reach $10^{-4}$ in the couplings, about an order of magnitude weaker than in benchmark 1, because of the lower production rates of the charmed mesons at \DUNE{}. The recasted bounds from existing HNL searches outperform most of the considered experiments for $m_N$ below the kaon threshold, while for $m_K\leq m_N \leq m_{D_s}$, the proposed LHC detectors and \DUNE{} will do a lot better than existing searches.

%!TEX root= dune_eft.tex
\section{Conclusions}\label{sec:conclusions}

In recent years, perhaps driven by the non-observation of heavy beyond-the-Standard-Model (BSM) states at the LHC, more attention has shifted towards hypothetical light and 
feebly-interacting new particles.
These elusive particles can play a role in solving outstanding problems of the Standard Model (SM) such as neutrino masses and/or dark matter.
In this work, we have focused on long-lived neutral fermions (LLNFs), which are predicted in various BSM models.
We show that LLNFs which are SM-singlets, can be described, in a unified way, in terms of neutrino-extended Standard Model Effective Field Theory ($\nu$SMEFT) and its low-energy counterpart $\nu$LEFT.
For illustration purpose, here, we have focused on two sample cases: heavy neutral leptons (HNLs) and the lightest neutralinos in the R-parity-violating supersymmetry (RPV-SUSY).

In minimal models, the HNLs interact solely through mixing with active neutrinos, but in other scenarios they can have very different interactions with SM fields, for example, through the exchange of BSM particles.
Assuming the latter to be heavy, at low energies the HNL phenomenology can be described by the $\nu$SMEFT Lagrangian.
In supersymmetric scenarios, the lightest neutralinos are necessarily bino-like and hence SM-singlets.
They have to decay to avoid overclosing the Universe, which can be realized, \textit{e.g.}~in RPV-SUSY. 
Integrating out heavy supersymmetric fields, the interactions of the lightest neutralino can then also be described by the same $\nu$SMEFT framework.
In this light, the name, neutrino-extended SMEFT, is a bit of a misnomer since it can be efficiently used to describe LLNFs that are not related to neutrino physics at all.
We have derived matching relations between the UV-complete theory (RPV-SUSY) and the $\nu$SMEFT/$\nu$LEFT to illustrate this point.

The main purpose of this work has been to investigate the potential of the neutrino experiment \DUNE{} to detect LLNF's, and compare the \DUNE{} reach to other (proposed) experiments.
At the \DUNE{} experiment, large numbers of strange and  charm mesons are produced that can potentially decay into LLNFs such as the HNLs or the lightest neutralinos. We have investigated four scenarios in detail: the first one is the usual minimal scenario of the HNLs, where we assume that the HNLs mix with the electron neutrinos only.
This scenario has been studied multiple times in the literature before and we mainly use this as a benchmark to verify our calculations and simulation tools: indeed our estimated sensitivities are close to those reported in \textit{e.g.}~Refs~\cite{Krasnov:2019kdc, Coloma:2020lgy}.
In the second and third scenarios we use the $\nu$SMEFT and $\nu$LEFT framework to investigate more general interactions.
In particular we focus on scalar- and tensor-like charged current operators, that are, for example, induced in leptoquark models, and a left-right vector-like operator that is induced in left-right symmetric models.
The fourth scenario is the above-mentioned light bino in RPV-SUSY.
In each scenario we have investigated the role of different flavor assignments which is rather important for the comparison between \DUNE{} and LHC experiments.

We have performed detailed Monte-Carlo simulations and computations to obtain the predicted signal-event rates of the rare decays of the LLNFs in different scenarios inside the \DUNE{} near detector (\texttt{DUNE-ND}).
Given the expected vanishing background, we have presented the sensitivity reach at 95\% confidence level corresponding to three signal events.
For LLNFs produced from kaons that are long-lived themselves, we have developed an algorithm that takes into account the decay position of the kaons instead of assuming that the LLNFs are produced at the interaction point.
With this method, we have estimated the sensitivities of both the \texttt{DUNE-ND} and the proposed LHC far detectors to the HNLs for both minimal scenarios and more complicated extensions in a consistent way.

Our main conclusion is that \DUNE{} will be extremely sensitive to LLNF's with 
masses below roughly $2$ GeV (the $D_s$ threshold).
In particular, for masses below the kaon mass, \DUNE{}'s sensitivity reach even exceeds that of the largest proposed LHC experiment, \MATHUSLA{}.
This holds for the minimal scenario and the RPV-SUSY model, and typically also for general $\nu$SMEFT scenarios.
However, in certain flavor benchmarks where the LLNF production through kaon decays is relatively suppressed, the LHC experiments \MATHUSLA{} and \ANUBIS{} are projected to do better because of the difference in relative production rate of heavier mesons compared to kaons at the LHC.
This also explains why these experiments typically perform somewhat better than \DUNE{} in the $0.5-2$ GeV LLNF mass range. 
However, the difference is not very large and the \DUNE{} sensitivity reach is still excellent.
Our study exemplified the usefulness of studying LLNFs which are SM-singlets in the theoretical framework of effective field theories, and motivates the search for long-lived fermions at the \texttt{DUNE-ND}.

%\bigskip
%%%%%%%%%%%%%%%%%%%%%%%%%%%%%%%%%%%%%%%%%%%%%%%%%%%%%%%%%%%%%%%%%%%%%%

\section*{Acknowledgments}
\bigskip
We thank Martin Hirsch and Oliver Freyermuth for useful discussions.
Financial support for H.K.D.~by the DFG (CRC 110, ``Symmetries and the Emergence of Structure in QCD'') is gratefully acknowledged. JdV acknowledges support from the Dutch Research Council (NWO) in the form of a VIDI grant.
\bigskip

\appendix
\section{Meson decay rate calculation}\label{sec:appendix_decayratecalc}

The dimension-6 operators of the effective Lagrangian (Eq.~\eqref{d6CC} and 
Eq.~\eqref{d6NC}) at tree-level allow for meson decays with two-body or 
three-body final states involving heavy neutral leptons. 
Through the same operators, the heavy neutral leptons can decay into 
final states consisting of lighter mesons and leptons. In this section, 
we explain how we obtain the relevant decay rates.

\subsection{Charged currents - two-body final states}

When calculating decay rates involving an external meson, hadronic matrix 
elements between the meson and the quark currents must be evaluated. 
When the decay only involves a single external meson, such matrix elements 
are parameterized with decay constants. 
For a charged pseudoscalar meson $M_{ij}$ with mass $m_M$ consisting of the 
anti-quark $\bar{q}_i$ and quark $q_j$ exist two non-zero quark-connections 
defined as
\bea
\Braket{0|\qb_i\gamma^\mu\gamma_5q_j|M_{ij}(p)}&=& ip^\mu f_M\label{eq:ps_vectorcurrent_decay_const}\,,\\
\Braket{0|\qb_i\gamma_5q_j|M_{ij}(p)}&=& i f^{PS}_M\label{eq:ps_pseudoscalarcurrent_decay_const}\,,
\eea
where $p^\mu$ is the momentum of the meson $M_{ij}$. Both decay constants 
$f_{M}$ and $f^{PS}_M$ are related via quark current algebra
\bea
-i\frac{m_M^2}{m_{\bar q_i}+m_{q_j}} f_M\equiv if^{PS}_M\,.
\eea
Similarly, we define two decay constants for a charged vector meson $M^*_{ij}$ 
with mass $m_{M^*}$, momentum $p^\mu$ and polarization $\epsilon^\nu$ via
\bea
\Braket{0|\bar q_i\gamma^\mu q_j|M^*_{ij}(p,\epsilon)}&=& m_{M^*}\epsilon^\mu f_{M^*}\label{eq:v_vectorcurrent_decay_const}\,,\\\
\Braket{0|\bar q_i\sigma^{\mu\nu} q_j|M^*_{ij}(p,\epsilon)}&=& i\big(p^\mu\epsilon^\nu-p^\nu\epsilon^\mu\big) f^T_{M^*}\,,\label{eq:v_tensorcurrent_decay_const}
\eea
where $\sigma^{\mu\nu}=\frac{i}{2}\big[\gamma^\mu,\gamma^\nu\big]$. In the 
heavy quark limit, we can relate both vector meson decay constants via 
$f^T_{M^*} \approx f_{M^*}$.

Employing these matrix elements we can evaluate 
the spin-averaged squared matrix elements $|\mathcal{M}|^2$ and obtain the 
decay rates for a process $P1\rightarrow P2+P3$. With the spin-averaging 
factor $n$ for particle $P1$ and the respective masses $m_1$, $m_2$ and $m_3$ 
the decay rates are given by
\bea
\Gamma=\frac{\sqrt{\lambda(m_1^2,m_2^2,m_3^2)}}{16\pi m_1^3}\frac{1}{n}\sum_{\text{spins}}|\mathcal{M}|^2\,,
\eea
where $\lambda(a,b,c)=a^2+b^2+c^2-2ab-2ac-2bc$ is the K\"all\'en function.

The spin-averaged squared matrix elements for a charged pseudoscalar meson 
$M_{ij}^\pm \rightarrow l_k^\pm + \LLNF{}$ 
in terms of the most general couplings in Eq.~\eqref{d6CC} is given by\footnote{These 
decay rates have been calculated in Appendix A of Ref.~\cite{DeVries:2020jbs}, 
but we observe some differences. On 
one hand, we do not explicitly rotate to the neutrino mass basis, but 
instead assume, that mixing is minimal and hence $\nu_4\approx \nu_R$. 
This also means, that we do not express the coupling for the interaction  
$(\bar{u}_L\gamma^\mu d_L)(\bar{e}_L\gamma_\mu \nu_L)$ with a Wilson 
coefficient $c^{CC}_{VLL}$, but instead 
only state the SM contribution $-2V_{ij}$, which changes the signs of 
some mass function $\lambda$ and $\lambda^\prime$ in the interference terms. 
Additionally, because we use a different definition for the decay 
constants for vector mesons $M^*$, the sign of some interference terms 
for the decay mode $\LLNF{}\rightarrow l_k^\pm + M_{ji}^{*\mp}$ changes 
relative to Ref.~\cite{DeVries:2020jbs}. However, the Wilson coefficients 
could in principle be chosen to be negative, so that the sign of interference 
is a matter of convention.}
\bea\label{eq:twobody_charged_psmeson_decay}
\sum_{\text{spins}}|\mathcal{M}(M_{ij}^\pm \rightarrow l_k^\pm + \LLNF{})|^2=\frac{G_F^2}{2}&\bigg\{&f^2_{M}\bigg[|2V_{ij}U_{k4}|^2+|c^{CC1}_{\textrm{VLR}, ijk}-c^{CC}_{\textrm{VRR}, ijk}|^2\bigg]\lambda_{V1}\nn\\
&+&f^2_{M}\text{Re}\big[2V_{ij}U_{k4}(c^{CC1}_{\textrm{VLR}, ijk}-c^{CC}_{\textrm{VRR}, ijk})^*\big]\lambda_{V2}\nn\\
&+&\big(f^{PS}_{M}\big)^2|c^{CC}_{\textrm{SLR}, ijk}-c^{CC}_{\textrm{SRR}, ijk}|^2\lambda_{S}\\
&+&f_{M}f_{M}^{PS}\text{Re}\big[2V_{ij}U_{k4}(c^{CC}_{\textrm{SLR}, ijk}-c^{CC}_{\textrm{SRR}, ijk})^* \big]\lambda_{SV1}\nn\\
&+&f_{M}f_{M}^{PS}\text{Re}\big[(c^{CC1}_{\textrm{VLR}, ijk}-c^{CC}_{\textrm{VRR}, ijk})(c^{CC}_{\textrm{SLR}, ijk}-c^{CC}_{\textrm{SRR}, ijk})^*\big]\lambda_{SV2}\bigg\}\nn
\eea
where we define the mass terms 
\bea\label{eq:twobody_charged_psmeson_decay_mass_terms}
\lambda_{V1}&=&m_{M}^2\big(m_{\LLNF{}}^2+m_{l}^2\big) - \big(m_{\LLNF{}}^2-m_l^2\big)^2\,,\nn\\
\lambda_{V2}&=& 4m_{M}^2m_{\LLNF{}}m_{l}\,,\nn\\
\lambda_{S}&=& m_{M}^2-m_{\LLNF{}}^2-m_{l}^2\,,\\
\lambda_{SV1}&=& 2m_{\LLNF{}}\big( m_{M}^2-m_{\LLNF{}}^2+m_{l}^2\big)\,,\nn\\
\lambda_{SV2}&=& 2m_{l}\big(m_{M}^2+m_{\LLNF{}}^2-m_{l}^2\big)\,,\nn
\eea
where $m_l$ and $m_{\LLNF{}}$ are the masses of the lepton and \LLNF{}, 
respectively. If the initial meson is a vector meson, the most general 
spin-averaged squared matrix element is 
\bea\label{eq:twobody_charged_vmeson_decay}
\frac{1}{3}\sum_{\text{spins}}|\mathcal{M}(M_{ij}^{*\pm} \rightarrow l_k^\pm + \LLNF{})|^2=\frac{G_F^2}{6}&\bigg\{&f^2_{M^*}\bigg[|2V_{ij}U_{k4}|^2+|c^{CC1}_{\textrm{VLR}, ijk}+c^{CC}_{\textrm{VRR}, ijk}|^2\bigg]\lambda^\prime_{V1}\nn\\
&+&f^2_{M^*}\text{Re}\big[2V_{ij}U_{k4}(c^{CC1}_{\textrm{VLR}, ijk}+c^{CC}_{\textrm{VRR}, ijk})^*\big]\lambda^\prime_{V2}\nn\\
&+&\big(f^{T}_{M^*}\big)^2|c^{CC}_{\textrm{T}, ijk}|^2\lambda^\prime_{T}\\
&+&f_{M^*}f_{M^*}^{T}\text{Re}\big[2V_{ij}U_{k4}c^{CC,*}_{\textrm{TRR}, ijk} \big]\lambda^\prime_{VT1}\nn\\
&+&f_{M^*}f_{M^*}^{T}\text{Re}\big[(c^{CC1}_{\textrm{VLR}, ijk}+c^{CC}_{\textrm{VRR}, ijk})c^{CC,*}_{\textrm{TRR}, ijk}\big]\lambda^\prime_{VT2}\bigg\},\nn
\eea
where we define the mass terms 
\bea\label{eq:twobody_charged_vmeson_decay_mass_terms}
\lambda^\prime_{V1}&=&m_{M^*}^2\big(2m_{M^*}^2-m_{\LLNF{}}^2-m_{l}^2\big) - \big(m_{\LLNF{}}^2-m_l^2\big)^2\,,\nn\\
\lambda^\prime_{V2}&=&-12m_{M^*}^2m_{\LLNF{}}m_{l}\,,\nn\\
\lambda^\prime_{T}&=& 16\bigg(m_{M^*}^2\big(m_{M^*}^2+m_{\LLNF{}}^2+m_{l}^2\big) - 2\big(m_{\LLNF{}}^2-m_l^2\big)^2\bigg)\,,\\
\lambda^\prime_{VT1}&=& 24m_{M^*}m_{\LLNF{}}\big( m_{M^*}^2-m_{\LLNF{}}^2+m_{l}^2\big)\,,\nn\\
\lambda^\prime_{VT2}&=& -24m_{M^*}m_{l}\big(m_{M^*}^2+m_{\LLNF{}}^2-m_{l}^2\big)\,.\nn
\eea

For an LLNF decaying into a meson and a lepton, we can relate the spin-averaged 
squared matrix element to the squared matrix element of an identical 
meson decaying into an \LLNF{} and identical lepton. Including the 
charge-conjugate decay modes and appropriate spin-averaging factors, 
we find
\bea\label{eq:twobody_LLNF_charged_decay}
2\times\frac{1}{2}\sum_{\text{spins}}|\mathcal{M}(\LLNF{}\rightarrow l_k^\pm + M^{(*)\mp}_{ij})|^2 &=& -\sum_{\text{spins}}|\mathcal{M}(M^{(*)\pm}_{ij} \rightarrow l_k^\pm + \LLNF{})|^2.
\eea

\subsection{Neutral currents - two-body final states}

If the meson involved in the decay modes are charge-neutral, the contributing couplings are given in Eq.~\eqref{d6NC}. The resulting modes are $M_{ij}^0 \rightarrow \nu_k + \LLNF{}$ and $\LLNF{}\rightarrow \nu_k + M_{ji}^0$. However, as neutral mesons do not always correspond to a single $\bar{q}_iq_j$ state, we must treat the hadronic matrix element differently based on the involved meson. We divide the mesons into heavy neutral mesons (involving a charm or bottom quark), kaons, and other light neutral mesons.

\subsubsection{Heavy neutral mesons}

Heavy neutral mesons such as $D^0$, $B^0$, $B_s^0$ and the related vector mesons do correspond to a pure $\bar{q}_iq_j$ state, so that we can define the decay constants similarly to Eqs.~\eqref{eq:ps_vectorcurrent_decay_const}-\eqref{eq:v_tensorcurrent_decay_const}. As the quark currents must be flavor-changing to allow for these decays, only the three neutral EFT-couplings contribute to the decay modes. For a neutral heavy pseudoscalar mesons, we find
\bea\label{eq:twobody_neutral_heavy_psmeson_decay}
\sum_{\text{spins}}|\mathcal{M}(M_{ij}^0 \rightarrow \nu_k + \LLNF{})|^2&=&\frac{G_F^2}{2}\big(f^{PS}_{M}\big)^2|c^{NC}_{\textrm{SLR}, ijk}-c^{NC}_{\textrm{SRR}, ijk}|^2\lambda_{S}\,,\\
2\times\frac{1}{2}\sum_{\text{spins}}|\mathcal{M}(\LLNF{} \rightarrow \nu_k + M_{ij}^0)|^2&=&-\sum_{\text{spins}}|\mathcal{M}(M_{ij}^0 \rightarrow \nu_k + \LLNF{})|^2\nn\,.
\eea
Similarly, for a heavy neutral vector meson the spin-averaged matrix element squared is
\bea\label{eq:twobody_neutral_heavy_vmeson_decay}
\frac{1}{3}\sum_{\text{spins}}|\mathcal{M}(M_{ij}^{0*} \rightarrow \nu_k + \LLNF{})|^2&=&\frac{G_F^2}{6}\big(f^{T}_{M}\big)^2|c^{NC}_{\textrm{T}, ijk}|^2\lambda^\prime_{T}\,,\\
2\times\frac{1}{2}\sum_{\text{spins}}|\mathcal{M}(\LLNF{} \rightarrow \nu_k + M_{ij}^{0*})|^2&=&-\sum_{\text{spins}}|\mathcal{M}(M_{ij}^{0*} \rightarrow \nu_k + \LLNF{})|^2\nn\,.
\eea

\subsubsection{Kaons}

Neutral pseudoscalar kaons $K^0$ ($\bar{s}d$) and $\bar{K}^0$ ($\bar{d}s$) are flavor eigenstates and can be produced in strong interactions, while their decay implies change of flavor and occurs via weak interactions. However, both are not eigenstates of weak interactions. Knowing that
\begin{align}
\hat{CP}\Ket{K^0}=&-\Ket{\bar{K}^0}\,,\\
\hat{CP}\Ket{\bar{K}^0}=&-\Ket{K^0}\,,
\end{align}
we define the following $CP$ eigenstates (which are simultaneous eigenstates for the total Hamiltonian assuming no $CP$ violation)
\begin{align}
\Ket{K_1}=&\frac{1}{\sqrt{2}}\bigg(\Ket{K^0}-\Ket{\bar{K}^0}\bigg)\,,&\text{with }\hat{CP}\Ket{K_1}=&\Ket{K_1}\,,\\
\Ket{K_2}=&\frac{1}{\sqrt{2}}\bigg(\Ket{K^0}+\Ket{\bar{K}^0}\bigg)\,,&\text{with }\hat{CP}\Ket{K_2}=&-\Ket{K_2}\,.
\end{align}
The actual physical kaons $K_{L}$ and $K_{S}$ are, however, not $CP$ eigenstates, but violate $CP$ slightly. They are mixtures of $K_{1/2}$. We can parameterize this mixture with a parameter $\epsilon\sim\orderof(10^{-3})$ and obtain
\begin{align}
\Ket{K_S}=&\frac{1}{\sqrt{(1+\epsilon^2)}}\bigg[\Ket{K_1}+\epsilon\Ket{K_2}\bigg]\label{eq:matrixelements_kaon_long}\,,\\
\Ket{K_L}=&\frac{1}{\sqrt{(1+\epsilon^2)}}\bigg[\Ket{K_2}+\epsilon\Ket{K_1}\bigg]\label{eq:matrixelements_kaon_short}\,.
\end{align}
In practice, we will interpret $\Ket{K_{S/L}}=\Ket{K_{1/2}}$ and ignore the $CP$ violation (assuming $\epsilon\rightarrow 0$). Then, we have the following decay constants for kaons
\begin{align}
\Braket{0|\bar{d}\gamma^\mu\gamma_5 s|K_{L/S}(p)}=& \pm \frac{i}{\sqrt{2}}p^\mu f_{K^0}\,,\nn\\
\Braket{0|\bar{s}\gamma^\mu\gamma_5 d|K_{L/S}(p)}=& \frac{i}{\sqrt{2}}p^\mu f_{K^0}\,.
\end{align}
As such decay modes necessitate flavor changing quark currents, only the EFT couplings of Eq.~\eqref{d6NC} contribute and the squared matrix element is given by Eq.~\eqref{eq:twobody_neutral_heavy_psmeson_decay} adjusted for the additional factor $1/\sqrt{2}$.

Neutral vector kaons are flavor eigenstates as well as physically propagating states, so that we can use the expression Eq.~\eqref{eq:twobody_neutral_heavy_vmeson_decay} for such decay modes.

\subsubsection{Light neutral mesons}

Light neutral mesons are mixtures of quark-antiquark pairs. The decay constants as defined in Eqs.~\eqref{eq:ps_vectorcurrent_decay_const}-\eqref{eq:v_tensorcurrent_decay_const}, however, connect to a quark current consisting of a single quark-antiquark pair. We can relate the relevant hadronic matrix element to the matrix element of the associated charged meson using $G$-parity and isospin relations\footnote{We follow Appendix B of Ref.~\cite{Bondarenko:2018ptm}.}. $G$-Parity is a multiplicative quantum number combining the charge conjugation and an isospin rotation along the second axis and is conserved by strong interactions. 
\bea
\hat{G}=& \hat{C} \hat{R}_2= \hat{C}\exp(i\pi \hat{I}_2)\,.
\eea 
Pions and the $\omega$-meson are $G$-parity odd, while $\rho$-mesons are $G$-parity even. 
The quark currents involving only light mesons, $u$ und $d$ can be constructed with quark currents, which are also eigenstates of $G$-parity. For example, the down-type axial-vector current is a mix of the $G$-parity even $j_\mu^{A,s}$ and the $G$-parity odd $j_\mu^{A,0}$
\begin{align}
j_\mu^{A,s}=& \frac{1}{\sqrt{2}}\big(\bar{u}\gamma_\mu\gamma_5 u + \bar{d}\gamma_\mu\gamma_5 d\big)\,,\nn\\
j_\mu^{A,0}=& \frac{1}{\sqrt{2}}\big(\bar{u}\gamma_\mu\gamma_5u - \bar{d}\gamma_\mu\gamma_5 d\big)\,,\\
\bar{d}\gamma_\mu \gamma_5 d =&\frac{1}{\sqrt{2}}\big(j_\mu^{A,s}-j_\mu^{A,0} \big)\,.\nn
\end{align}
Assuming $G$-parity symmetry, we can relate matrix elements for light neutral mesons and light charged mesons. Here, we assume ideal mixing between $\omega$ and $\phi$. In such a case $\phi$ is a pure $\bar{s}s$-state.
Lastly, matrix elements for quark currents involving external $\eta$ and $\eta^\prime$ must be handled more 
carefully. Both are neutral unflavored mesons with zero isospin involving the axial singlet state $\phi^0$. 
Because $\phi^0$ is involved, the anomaly of the axial-vector singlet current affects $\eta$-$\eta^\prime$ 
mixing significantly. To account for these effects, we adopt an extended two-parameter mixing scheme as 
proposed in Refs.~\cite{Leutwyler:1997yr,Feldmann:1998vh,Dreiner:2009er}. If we denote light up-type and 
down-type quarks $u$ and $d$ as $q_l$, the decay constants are defined via\footnote{The extension of 
the two-angle mixing scheme by splitting the contributions to the isospin doublet ($u$, $d$) 
and strange contributions was incentivized by the observation that chiral symmetry breaking effects differ because of the larger $s$-mass. }
\renewcommand{\arraystretch}{1.1}
\begin{table}[tb]
\centering
\begin{tabular}{|c|c|c|c|}
\hline
$f_8/f_{\Ppi}$ & $f_1/f_{\Ppi}$ & $\theta^8$ & $\theta^1$\\
\hline
1.26 & 1.17 & -0.37 rad & -0.16 rad \\
\hline
\end{tabular}
\caption{Phenomenological values to determine decay constants for $\Peta$ and $\Petaprime$ \cite{Feldmann:1998vh}.}\label{tab:etaetaprimemixing}
\end{table}
\begin{align}
\big\langle \,0\, \big| \bar q_l\gamma^\mu\gamma_5 q_l\big| \eta^{(\prime)}(p)\big\rangle=ip^\mu  f^{\bar q_lq_l}_{\eta^{(\prime)}}\,,\\
\big\langle \,0\, \big| \bar s\gamma^\mu\gamma_5 s\big| \eta^{(\prime)}(p)\big\rangle=ip^\mu  f^{\APstrange\Pstrange}_{\eta^{(\prime)}}\,,
\end{align}
and can be expressed in terms of the mixing parameters $f^0,$ $f^8$, $\theta^0$, and $\theta^8$ as
\footnote{In $f^{\bar q_lq_l}_{\Peta/\Petaprime}$ we apply an additional factor $1/\sqrt{2}$, as the 
neutral axial current  consists only of $\Pup$ or $\Pdown$. }
\begin{align}
f^{\bar q_lq_l}_{\eta}=& \frac{1}{\sqrt{3}}\bigg[-f_1\sin\theta^1+\frac{1}{\sqrt{2}} f_8\cos\theta^8\bigg]\,,&
f^{\APstrange\Pstrange}_{\eta}=& \frac{1}{\sqrt{3}}\bigg[-f_1\sin\theta^1-\sqrt{2} f_8\cos\theta^8\bigg]\,,\\
 f^{\bar q_lq_l}_{\eta^\prime}=& \frac{1}{\sqrt{3}
}\bigg[f_1\cos\theta^1+\frac{1}{\sqrt{2}} f_8\sin\theta^8\bigg]\,,&
f^{\APstrange\Pstrange}_{\eta^\prime}=& \frac{1}{\sqrt{3}}\bigg[f_1\cos\theta^1-\sqrt{2} f_8\sin\theta^8\bigg]\,.
\end{align}
Phenomenological values for all $\Peta$-$\Petaprime$ mixing-parameters are provided in Table~\ref{tab:etaetaprimemixing}.

\subsection{Three body final states}\label{sec:three_body_formfactors}

Other possible decay modes involve two external mesons, a lepton $l$, and the \LLNF{}.
The relevant hadronic matrix element for such decays are parameterized with form factors, which 
are functions of the squared momentum transfer $q^2$. The $q^2$-dependence is typically 
calculated in a suited framework (\textit{e.g.}~lattice QCD or constituent quark models), the results of which are fitted to a parameterization formula. The number of required form factors 
depends on the number of independent Lorentz covariant structures formed by the available 
Lorentz vectors. For the decay of a pseudoscalar meson  $M$ to a final state involving 
a pseudoscalar meson $M^\prime$,the set of form factors required to describe the 
occurring matrix elements are 
\begin{align}
\Braket{M^\prime(p^\prime)|\bar q_1q_2|M(p)}=& f_S(q^2),\\
\Braket{M^\prime(p^\prime)|\bar q_1\gamma^\mu q_2|M(p)}=& f_+(q^2)P^\mu+f_-(q^2)q^\mu\\
=& f_+(q^2)\bigg(P^\mu-\frac{m_M^2-m_{M^\prime}^2}{q^2}q^\mu\bigg)+f_0(q^2)\frac{m_M^2-m_{M^\prime}^2}{q^2}q^\mu,\nn\\
\Braket{M^\prime(p^\prime)|\bar q_1\sigma^{\mu\nu}q_2|M(p)}=& \frac{2i}{m_M+m_{M^\prime}}\big(p^\mu p^{\prime\nu}-p^\nu p^{\prime\mu}\big)f_T(q^2),
\end{align}
where $P^\mu = p^\mu+p^{\prime \mu}$ and $q^\mu = p^\mu-p^{\prime \mu}$ and 
\begin{equation}
f_0(q^2)=f_+(q^2)+\frac{q^2}{m_M^2-m_{M^\prime}^2}f_-(q^2).
\end{equation}
As the SM does not include tensor currents, $f_T$-parameterizations are only present 
for some meson combinations $M$-$M^\prime$, but we can relate $f_S(q^2)$ and $f_T(q^2)$ to the form factors of the vector current via current-algebra as
\begin{align}
f_S(q^2)=& f_0(q^2)\frac{m_M^2-m_{M^\prime}^2}{m_{q_2}-m_{\overline{q}_1}},\\
f_T(q^2)\simeq& \frac{m_M\big(m_M+m_{M^\prime}\big)}{q^2}\big[f_+(q^2)-f_0(q^2)\big].\label{eq:Formfactor_TensorRelation}
\end{align}
We find that Eq.~\eqref{eq:Formfactor_TensorRelation} agrees reasonably well 
for known $f_T$- parameterizations~\cite{Lubicz:2018rfs,FlavourLatticeAveragingGroup:2019iem, Melikhov:2000yu}. 
We assume the approximation to be justified and apply Eq.~\eqref{eq:Formfactor_TensorRelation} 
to describe tensor form factors of meson pairs, where they are unknown.

If $M^\prime$ is a vector meson, there exist four hadronic matrix elements.
\begin{align}
\Braket{M^{\prime *}(p^\prime,\epsilon)|\overline{q}_1\gamma_5q_2|M(p)}=& i f_{PS}(q^2)\big(\epsilon^*\cdot p\big),\label{eq:formfactor_scalar_current_vector_meson}\\
\Braket{M^{\prime *}(p^\prime,\epsilon)|\overline{q}_1\gamma^\mu q_2|M(p)}=& g(q^2)\epsilon^{\mu\nu\alpha\beta}\epsilon^*_\nu p_\alpha p^\prime_\beta,\label{eq:formfactor_vector_current_vector_meson}\\
\Braket{M^{\prime *}(p^\prime,\epsilon)|\overline{q}_1\gamma^\mu \gamma_5q_2|M(p)}=& i\bigg[ a_+(q^2)P^\mu \big(\epsilon^*\cdot p\big)+a_-(q^2)q^\mu\big(\epsilon^*p\big)+f(q^2)\epsilon^{*\mu}  \bigg],\label{eq:formfactor_axialvector_current_vector_meson}\\
\Braket{M^{\prime *}(p^\prime,\epsilon)|\overline{q}_1\sigma^{\mu\nu}q_2|M(p)}=& i\epsilon^{\mu\nu\alpha\beta}\bigg[g_+(q^2)\epsilon^*_\alpha P_\beta+g_-(q^2)\epsilon^*_\alpha q_\beta+g_0(q^2)p_\alpha p^\prime_\beta(\epsilon^*\cdot p\big)\bigg].\label{eq:formfactor_tensor_current_vector_meson}
\end{align}
Again, $f_{PS}$ can be related to the axial-vector current formfactors via current-algebra via
\begin{align}\label{eq:VectorFFPSRelation}
f_{PS}(q^2)=& -\frac{1}{\big(m_{\overline{q}_1}+m_{q_2}\big)}\bigg[ a_+(q^2)\big(m_M^2-m_{M^\prime}^2\big)+a_-(q^2)q^2+f(q^2) \bigg].
\end{align}
The parameterizations for form factors are usually stated as the mode $M\rightarrow M^\prime$, where charges are not assigned. However, as is the case for decay constants of kaons and other light mesons, we can relate the hadronic matrix elements between various possible modes using isospin symetry. As an example for a decay of the type $D\rightarrow \pi$ with the general Dirac matrix element $\Gamma$, we find
\begin{align}
\Braket{\pi^0|\bar{u}\Gamma c|D^0}=\frac{1}{\sqrt{2}}\Braket{\pi^+|\bar{u}\Gamma c|D^+}=\frac{1}{\sqrt{2}}\Braket{\pi^-|\bar{d}\Gamma c|D^0}=-\Braket{\pi^0|\bar{d}\Gamma c|D^0}.
\end{align}
We observe that the relative matrix elements may differ in sign and in a factor $1/\sqrt{2}$ for every external $\pi^0$, $\rho^0$ or $K_{\scaleto{L/S}{5pt}}$.

Equipped with all possible form factors, we can calculate the decay rate of a semi-leptonic meson decay. This involves lengthy three-body phase-space calculations and can not always be done analytically. Instructions on how to perform the computations numerically with the help of \texttt{Mathematica} are provided in Appendix B of Ref.~\cite{DeVries:2020jbs}.

\section{Phenomenological values}\label{sec:appendix_phenomenological_values}

In this section, we list the phenomenological values used in this paper. This includes specific SM  decay constants, and form factors. For the CKM-matrix elements, the renormalized quark-masses, and other particle masses, we follow the latest results of the Particle Data Group (PDG~\cite{Workman:2022ynf}). The currently stated values for the CKM elements are
\begin{align}
|V_{ud}|&= 0.97373,& |V_{us}|&=	0.2243,& |V_{ub}|&= \SI{3.82}\times 10^{-3},\nn\\
|V_{cd}|&= 0.221,& |V_{cs}|&= 0.975,	& |V_{cb}|&= \SI{40.8}\times 10^{-3}.
\end{align}
The quark masses renormalized with the $\overline{\text{MS}}$ renormalization scheme at the scale $\mu=\SI{2}{\giga\eV}$ are 
\begin{align}
m_u&= \SI{2.16}{\mega\eV}, & m_c &= \SI{1.27}{\giga\eV},\nn\\
m_d&= \SI{4.67}{\mega\eV}, & m_s &= \SI{93.4}{\mega\eV}, & m_b &= \SI{4.18}{\giga\eV}.
\end{align}

\subsection{Decay constants}
Here, we state the numerical values used for the decay constants defined in Eqs.~\eqref{eq:ps_vectorcurrent_decay_const}-\eqref{eq:v_tensorcurrent_decay_const}.
For pseudoscalar decay constants we rely on the the most recent \texttt{FLAG} review \cite{Aoki:2021kgd} and the parameterization of Ref.~\cite{Feldmann:1998vh}. For charmed and bottomed vector mesons we employ the heavy-quark limit approximation $f_{M^\prime}=f_M$. For light vector mesons, we can calculate the decay constants as described in Refs.~\cite{Dreiner:2006gu,Neubert:1997uc}. The numerical values of all decay constants are listed in Table~\ref{tab:DecayConstant}.

\renewcommand{\arraystretch}{1.3}
\begin{table}[H]
\centering
\begin{tabular}{|lrr||lrr|}
\hline
$M$ & $f_{M}$ [\SI{}{\mega\eV}] & Ref. &  $M^*$ & $f_{M^*}$[\SI{}{\mega\eV}] & Ref.\\
\hline
%& & & & & \\
$\pi$  & 130.2 & \cite{Aoki:2021kgd} &$\rho$  & 209.4 & \cite{Dreiner:2006gu,Workman:2022ynf,Neubert:1997uc}\\
$K$   & 155.7 & \cite{Aoki:2021kgd} & $K^*$   & 205.7 & \cite{Dreiner:2006gu,Workman:2022ynf,Neubert:1997uc}\\
$\eta$ - $\bar{q}q$   & 76.5 & \cite{Dreiner:2009er,Feldmann:1998vh}  & $\omega$ & 201.1 & \cite{Dreiner:2006gu,Workman:2022ynf,Neubert:1997uc}\\
$\eta$ - $\bar{s}s$   & -110.8 & \cite{Dreiner:2009er,Feldmann:1998vh}  & $\phi $& 228.1 & \cite{Dreiner:2006gu,Workman:2022ynf,Neubert:1997uc}\\
$\eta^\prime$ - $\bar{q}q$   & 62.6 & \cite{Dreiner:2009er,Feldmann:1998vh} & & &\\
$\eta^\prime$ - $\bar{s}s$   & 135.3 & \cite{Dreiner:2009er,Feldmann:1998vh} & & & \\
$D$  & 212.0 & \cite{Aoki:2021kgd}  & $D^*$ & 212.0 & \cite{Aoki:2021kgd}\\
$D_s$   & 249.9 & \cite{Aoki:2021kgd}  & $D^*$ & 249.9 & \cite{Aoki:2021kgd} \\
$B$    & 190.0 & \cite{Aoki:2021kgd}  & $B^*$ & 190.0 & \cite{Aoki:2021kgd}\\
$B_s$    & 230.3 & \cite{Aoki:2021kgd} & $B_s^* $   & 230.3 & \cite{Aoki:2021kgd}\\
$\eta_c$    & 387.0 & \cite{Becirevic:2013bsa} & $J/\psi $   & 418.0 & \cite{Becirevic:2013bsa}\\
%& & & & & \\
\hline
\end{tabular}
\caption{Decay constants for pseudoscalar and vector mesons defined in Eqs.~\eqref{eq:ps_pseudoscalarcurrent_decay_const} and \eqref{eq:ps_vectorcurrent_decay_const}.}\label{tab:DecayConstant}
\end{table}

\subsection{Form factor parameterizations}

Next, we state the form factor parameterizations that we use for the various decay modes involving an incoming pseudoscalar $M$ and an outgoing $M^\prime$ with respective masses $m_M$ and $m_{M^\prime}$. The parameterizations are stated for a charged quark current, which we then relate to all other charge assignments as described in Sec.~\ref{sec:three_body_formfactors}. 
It is common to expand the form factors in terms of a dimensionless variable $z(q^2)$ defined as
\begin{equation}
z\big(q^2\big)\equiv \frac{\sqrt{t_+-q^2}-\sqrt{t_+-t_0}}{\sqrt{t_+-q^2}+\sqrt{t_+-t_0}}\,.
\end{equation}
We follow the FLAG collaboration \cite{Aoki:2021kgd} in writing $t_+$ and $t_0$ as
\begin{align}
t_\pm=&\big(m_M\pm m_{M^\prime}\big)^2,\\
t_0=t_+\bigg[1-\sqrt{1-\frac{t_-}{t_+}}\bigg]=&\big(m_M+m_{M^\prime}\big)\big(\sqrt{m_M}-\sqrt{m_{M^\prime}}\big)^2\,.
\end{align}

\subsubsection{Kaons: \texorpdfstring{$\mathit{K}\to \pi$}{}}

The scalar and vector form factors for $K\to \pi$ are described reasonably well by a linear $q^2$-dependence of the form
\begin{equation}
f(q^2)=f(0)\cdot \bigg(1+\lambda \frac{q^2}{m_{\Ppi^+}^2}\bigg).
\end{equation}
The optimal fit parameters for this parameterization are listed in Table~\ref{tab:formfactor_Kaon}.
For the tensor form factor, we follow Ref.~\cite{Baum:2011rm} and use a pole-parameterization
\begin{equation}
f_T(q^2)=\frac{f_T(0)}{1-s_T\,q^2}\,,
\end{equation} 
where $f_T(0)=0.417$ and $s_T=\SI{1.10}{\per\giga\eV\squared}$.

\renewcommand{\arraystretch}{1.2}
\begin{table}[tb]
\centering
\begin{tabular}{|c|c|c|c|c|}
\hline
Form factor & $M\to M^\prime$ & $f(0)$ & $\lambda$\\
\hline
$f_{+}(q^2)$ & $K^\pm\to\pi^0$ & 0.9698 & 0.0277\\
$f_{0}(q^2)$ & $K^\pm\to\pi^0$ & 0.9698 & 0.0183 \\
%& & & & \\
$f_{+}(q^2)$ & $K^0\to\pi^\pm$ & 0.9698 & 0.0267 \\
$f_{0}(q^2)$ & $K^0\to\pi^\pm$ & 0.9698 & 0.0117 \\
%& & & & \\
\hline
\end{tabular}
\caption{Form factor fit parameters of $K\to \pi$ using the parameterization of Eq.~\eqref{eq:FF_Parameterization_Lubicz}~\cite{Aoki:2021kgd,NA48:2006xvd,Yushchenko:2003xz,Bondarenko:2018ptm}.}\label{tab:formfactor_Kaon}
\end{table}

\subsubsection{Charmed mesons: \texorpdfstring{$\mathit{D}\to \pi/\mathit{K}$}{}}

The ETM Collaboration has recently published results for the form factors $f_{+,0,T}(q^2)$~\cite{Lubicz:2017syv, Lubicz:2018rfs} in the form
\begin{equation}\label{eq:FF_Parameterization_Lubicz}
f\big(q^2\big)=\frac{f(0)+c\big(z(q^2)-z_0\big)\bigg(1+\frac{z(q^2)+z_0}{2}\bigg)}{1-Pq^2},
\end{equation}
where $z_0=z(q^2=0)$. The fitted results for $f(0)$, $c$ and $P$ are listed in  Table~\ref{tab:formfactor_Lubicz}.
\renewcommand{\arraystretch}{1.2}
\begin{table}[tb]
\centering
\begin{tabular}{|c|c|c|c|c|}
\hline
Form factor & $M\to M^\prime$ & $f(0)$ & $c$ & $P$ [\SI{}{\giga\eV}$^{-2}$]\\
\hline
$f_{+}(q^2)$ & $D\to\pi$ & 0.6117 & -1.985  & 0.1314  \\
%& & & & \\
$f_{0}(q^2)$ & $D\to\pi$ & 0.6117 & -1.188  & 0.0342 \\
%& & & & \\
$f_{T}(q^2)$ & $D\to\pi$ & 0.5063 & -1.10 & 0.1461 \\
%& & & & \\
$f_{+}(q^2)$ &  $D\to K$ & 0.7647 & -0.066 & 0.2241  \\
%& & & & \\
$f_{0}(q^2)$ & $D\to K$ & 0.7647 & -2.084 & 0 \\
%& & & & \\
$f_{T}(q^2)$ & $D\to K$ & 0.6871 & -2.86 & 0.0854 \\
\hline
\end{tabular}
\caption{Form factor fit parameters of $D\to \pi, K$ using the parameterization of Eq.~\eqref{eq:FF_Parameterization_Lubicz}~\cite{Lubicz:2017syv, Lubicz:2018rfs}.}\label{tab:formfactor_Lubicz}
\end{table}

\subsubsection{Charmed strange mesons: \texorpdfstring{$\mathit{D}_s\to \eta/\eta^\prime/\mathit{K}$}{}}

For the decay channels $D_s\to \eta,\eta^\prime,K$, we apply a double-pole parameterization used in Ref.~\cite{Soni:2018adu} of the form 
\begin{equation}\label{eq:FF_Parameterization_Soni}
f\big(q^2\big)=\frac{f(0)}{1-\sigma_1 \frac{q^2}{m_{\PDs}^2}+\sigma_2 \frac{q^4}{m_{\PDs}^4}}.
\end{equation}
To take $\Peta$-$\Petaprime$ mixing into account, the authors of Ref.~\cite{Soni:2018adu} follow the same scheme as used in Ref.~\cite{Feldmann:1998vh}, which  we also employ for the decay constants for $\Peta$-$\Petaprime$. The fit parameters are summarized in  Table~\ref{tab:formfactor_Soni}.

\renewcommand{\arraystretch}{1.2}
\begin{table}[tb]
\centering
\begin{tabular}{|c|c|c|c|c|}
\hline
Form factor & $M\to M^\prime$ & $f(0)$ & $\sigma_1$ & $\sigma_2$ \\
\hline
$f_{+}(q^2)$ & $D_s\to\eta$ &  0.78 & 0.69 & 0.002 \\
%& & & & & \\
$f_{-}(q^2)$ & $D_s\to\eta$ & -0.42 & 0.74 & 0.008 \\
%& & & & & \\
$f_{+}(q^2)$ & $D_s\to\eta^\prime$ & 0.73 & 0.88 & 0.018 \\
%& & & & & \\
$f_{-}(q^2)$ & $D_s\to\eta^\prime$ & -0.28 & 0.92 & 0.009 \\
%& & & & & \\
$f_{+}(q^2)$ & $D_s\to K$ & 0.60 & 1.05  & 0.18 \\
%& & & & & \\
$f_{-}(q^2)$ & $D_s\to K$ & -0.38 & 1.14 & 0.24 \\
\hline
\end{tabular}
\caption{Form factor fit parameters of  $D_s\to\eta,\eta^\prime, K$ using the parameterization of Eq.~\eqref{eq:FF_Parameterization_Soni}~\cite{Soni:2018adu}}\label{tab:formfactor_Soni}
\end{table}

\subsubsection{Bottom Mesons: \texorpdfstring{$\mathit{B}\to \Ppi/\mathit{K}/\mathit{D}$}{} and \texorpdfstring{$\mathit{B}_s\to \mathit{K}/\mathit{D}_s$}{}}
The FLAG collaboration \cite{Aoki:2021kgd} presents form factors using the z-expansion proposal by Bourrely, Caprini, and Lellouch (BCL)
\begin{align}
f_{+,T}\big(q^2\big)=&\frac{1}{1-\frac{q^2}{M^2_{+,T}}}\sum_{n=0}^{N-1}a^{+,T}_n\bigg[z^n-(-1)^{n-N}\frac{n}{N}z^N\bigg],\label{eq:FP_Parameterization_Aoki}\\
f_0\big(q^2\big)=&\frac{1}{1-\frac{q^2}{M^2_{0}}}\sum_{n=0}^{N-1}a^0_n z^n.\label{eq:F0_Parameterization_Aoki}
\end{align}
For $f_0$ one coefficient $a_n$ is restricted by imposing $f_+(0)=f_0(0)$, so that
\begin{equation}
a_{N-1}^0 = \frac{1}{z^{N-1}(0)}\bigg[f_+(0) - \sum_{n=0}^{N-2}a^0_n z^n(0)\bigg].
\end{equation}
Optimal fit values are stated in Table~\ref{tab:formfactor_Aoki}. For the decay modes $B_s\to K$, $t_+$ and $t_-$ are not defined for the same meson masses. Instead, both functions are defined as $t_+=\big(m_B+m_\pi\big)^2$ and $t_-=\big(m_{B_s}+m_K\big)^2$, respectively. 
For $B_s\to D_s$, the dimensionless variable $z(q^2)$ is defined for $t_0=0$.

\renewcommand{\arraystretch}{1.2}
\begin{table}[tb]
\centering
\begin{tabular}{|c|c||c|cccc||c|ccc|}
\hline
$M\to M^\prime$ & $N$ & $M_+$ [\SI{}{\giga\eV}] & $a^+_0$ & $a^+_1$ & $a^+_2$ & $a^+_3$ & $M_0$ [\SI{}{\giga\eV}] &  $a^0_0$ & $a^0_1$ & $a^0_2$
\\
\hline
$B\to \pi$ & 3  & 5.32465 & 0.404 & -0.68 & -0.86 & /  & $\infty$  & 0.49 & -1.61 & /   \\
$B\to K$  & 3  & 5.4154 & 0.471 & -0.74 & 0.32 & /  & 5.711 & 0.301 & 0.40 & /   \\
$B\to D$  & 3  & $\infty$ & 0.896 & -7.94 & 51.4 & / & $\infty$ & 0.7821 & -3.28 & / \\
$B_s\to K$ & 4 & 5.32465 & 0.374 & -0.672 & 0.07 & 1.34 & 5.68 & 0.2203 & 0.089 & 0.24 \\
$B_s\to D_s$ & 3 &  6.329 & -0.075 & -3.24 & 0.7 & / & 6.704 & 0.666 & -0.26 & -0.1 \\
\hline
\end{tabular}
\caption{Vector and scalar form factor fit parameters for the transitions $B\to \pi$, $B_{(s)}\to K$, and $B_{(s)}\to D_{(s)}$  using the parameterization of Eq.~\eqref{eq:FP_Parameterization_Aoki} and Eq.~\eqref{eq:F0_Parameterization_Aoki} \cite{Aoki:2021kgd,Du:2015tda,McLean:2019qcx}.}\label{tab:formfactor_Aoki}
\end{table}

\renewcommand{\arraystretch}{1.2}
\begin{table}[tb]
\centering
\begin{tabular}{|c|c||c|cccc|}
\hline
$M\to M^\prime$ & $N$ & $M_T$ [\SI{}{\giga\eV}] & $a^T_0$ & $a^T_1$ & $a^T_2$ & $a^T_3$  
\\
\hline
$B\to \pi$ & 4 & 5.32465  & 0.393 & -0.65 & -0.6 & 0.1 \\
$B\to K$ & 3 & 5.4154 & 0.455 & -1.00 & -0.9 & / \\
\hline
\end{tabular}
\caption{Tensor form factor fit parameters for the transitions $B\to \pi$ and $B\to K$  using the parameterization of Eq.~\eqref{eq:FP_Parameterization_Aoki}~\cite{Aoki:2021kgd,FermilabLattice:2015cdh,Du:2015tda}.}\label{tab:formfactor_Aoki_tensor}
\end{table}

\subsubsection{Bottom strange meson decays  \texorpdfstring{$\mathit{B}_s\to \eta^{(\prime)}$}{}:}

Such a decay mode is mediated by the $b\to s$ quark current. However, the $\eta^{(\prime)}$ final state is a mixture of light and strange quarks. As was the case for decay constants, we must account for strange component in the final state meson. We follow Ref.~\cite{Melikhov:2000yu} and calculate the decay rates as
\begin{align}
\Gamma\big(D_s\to \eta +l+\nu)  =& \sin^2(\theta) \Gamma\big(D_s\to \eta_s(M_\eta) +l+\nu)\,,\\
\Gamma\big(D_s\to \eta^\prime +l+\nu)  =& \cos^2(\theta) \Gamma\big(D_s\to \eta_s(M_{\eta^\prime}) +l+\nu)\,,
\end{align}
where $\theta \approx 40^\circ$. The parameterizations of the form factors of a hadronic matrix element of the form $\langle   \eta_s(M_{\eta^{(\prime)}})|b\Gamma s| B_s  \rangle $ is then given by
\begin{align}
f_{+,T}(q^2)=&\frac{f(q^2)}{\left(1-\frac{q^2}{M^2}\right)\left[1-\sigma_1 \frac{q^2}{M^2}+\sigma_2 \frac{q^4}{M^4}\right]}\,,\\
f_{0}(q^2)=&\frac{f(q^2)}{\left[1-\sigma_1 \frac{q^2}{M^2}+\sigma_2 \frac{q^4}{M^4}\right]}\,.
\end{align}
The parameters for both decays are stated in Table~\ref{tab:Appendix_Pheno_Semileptonic_Melikhov_eta}.

{\renewcommand{\arraystretch}{1.2}
\begin{table}[tb]
\centering
\begin{tabular}{|l||c|c|c||l||c|c|c||}
\hline
Fit Parameters & $f_+$ & $f_0$ & $f_T$ & Fit Parameters & $f_+$ & $f_0$ & $f_T$ \\
\hline\hline
$\big[B_s\to \eta_s(M_{\eta})\big]$ & & & & $\big[B_s\to \eta_s(M_{\eta^\prime})\big]$ & & &\\
 $f(0)$  & 0.36 & 0.36 & 0.36 &  $f(0)$  & 0.36 & 0.36 & 0.39\\
 $\sigma_1$ & 0.60 & 0.80 & 0.58 & $\sigma_1$ & 0.60 & 0.80 & 0.58\\
 $\sigma_2$ & 0.20 & 0.40 & 0.18 & $\sigma_2$ & 0.20 & 0.45 & 0.18\\
 $M$ [\SI{}{\giga\eV}] & 5.42 & 5.42 & 5.42 & $M$ [\SI{}{\giga\eV}] & 5.42 & 5.42 & 5.42\\
\hline
\end{tabular}
\caption{Vector and tensor form factor fit parameters of $\mathit{B}_s\to \eta_s(M_{\eta^{(\prime)}})$}\label{tab:Appendix_Pheno_Semileptonic_Melikhov_eta}
\end{table}}

\subsubsection{Final state vector mesons}

If $M^\prime$ is a vector meson, we also refer to the form factor parameterizations in Ref.~\cite{Melikhov:2000yu}. There, parameterizations are provided for linear combinations of the form factors defined in Eqs.~\eqref{eq:formfactor_scalar_current_vector_meson}-\eqref{eq:formfactor_tensor_current_vector_meson}
%, $f_{PS}(q^2)$, $a_{\pm}$, $f$ and $g_{\pm,0}$. 
These linear combinations are defined as
\begin{align}\label{eq:Appendix_LinearCombinationFormFactors}
V=&\big(m_{M}+m_{M^\prime}\big)g\,,\nn\\
A_0=& \frac{1}{2m_{M^\prime}{}}\big(f+q^2\cdot a_- + (m_{M}^2-m_{M^\prime}^2)a_+\big)\,,&T_1=&-g_{+}\,,\nn\\
A_1=&\frac{1}{m_{M}+m_{M^\prime}}f\,,&T_2=&-g_{+}-\frac{q^2}{m_{M}^2-m_{M^\prime}^2}g_{-}\,,\nn\\
A_2=&-\big(m_{M}+m_{M^\prime}\big)a_+\,,&T_3=&g_{-}-\frac{m_{M}^2-m_{M^\prime}^2}{2}g_{0}\,.
\end{align}
From Eq.~\eqref{eq:VectorFFPSRelation}, we also see that 
\begin{equation}
A_0= -\frac{m_{\bar q_1}+m_{q_2}}{2M_2}f_{PS}\,.
\end{equation}
The parameterization used for these new variables is similar to Eq.~\eqref{eq:FF_Parameterization_Soni}, where $M_{\PDs}$ is modified for every 
decay mode. For $F_+$, $F_T$, $V$, $T_1$, and $A_0$, resonances contribute strongly 
to the behavior of the form factor.
To accurately include the influence of such resonances, the fit-function is adjusted to  
\begin{equation}
f(q^2)=\frac{f(q^2)}{\left(1-\frac{q^2}{M^2}\right)\left[1-\sigma_1 \frac{q^2}{M^2}+\sigma_2 \frac{q^4}{M^4}\right]}.
\end{equation}
Detailed in Table~\ref{tab:Appendix_Pheno_Semileptonic_Melikhov_B} and Table~\ref{tab:Appendix_Pheno_Semileptonic_Melikhov_D} are the relevant fit parameters for a final state vector meson.

\renewcommand{\arraystretch}{1.2}
\begin{table}[tb]
\centering
\begin{tabular}{|l||c|c|c|c|c|c|c||}
\hline
Fit Parameters & $V$ & $A_0$ & $A_1$ & $A_2$ & $T_1$ & $T_2$ & $T_3$ \\
\hline\hline
$\big[ B\to \rho\big]$ & & & & & & &\\
 $f(0)$  & 0.31 & 0.30 & 0.26 & 0.24 & 0.27 & 0.27 & 0.19 \\
 $\sigma_1$  & 0.59 & 0.54 & 0.73 & 1.40 & 0.60 & 0.74 & 1.42 \\
 $\sigma_2$  & 0.00 & 0.00 & 0.10 & 0.50 & 0.00 & 0.19 & 0.51 \\
 $M$ [\SI{}{\giga\eV}]  & 5.32 & 5.27 & 5.32 & 5.32 & 5.32 & 5.32 & 5.32 \\
 & & & & & & &\\
\hline
$\big[ B\to K^*\big]$ & & & & & & &\\
 $f(0)$  & 0.44 & 0.45 & 0.36 & 0.32 & 0.39 & 0.39 & 0.27\\
 $\sigma_1$  & 0.45 & 0.46 & 0.64 & 1.23 & 0.45 & 0.72 & 1.31\\
 $\sigma_2$  & 0.0 & 0.0 & 0.36 & 0.38 & 0.0 & 0.62 & 0.41\\
 $M$ [\SI{}{\giga\eV}] & 5.42 & 5.37 & 5.42 & 5.42 & 5.42 & 5.42 & 5.42\\
 & & & & & & &\\
\hline
$\big[ B\to D^*\big]$ & & & & & & &\\
 $f(0)$  & 0.76 & 0.69 & 0.66 & 0.62 & 0.68 & 0.68 & 0.33\\
 $\sigma_1$  & 0.57 & 0.58 & 0.78 & 1.40 & 0.57 & 0.64 & 1.46\\
 $\sigma_2$  & 0.0 & 0.0 & 0.0 & 0.41 & 0.0 & 0.0 & 0.50\\
 $M$ [\SI{}{\giga\eV}] & 6.4 & 6.4 & 6.4 & 6.4 & 6.4 & 6.4 & 6.4\\
 & & & & & & &\\
\hline
$\big[B_s\to K^*\big]$ & & & & & & &\\
 $f(0)$  & 0.38 & 0.37 & 0.29 & 0.26 & 0.32 & 0.32 & 0.23\\
 $\sigma_1$  & 0.66 & 0.60 & 0.86 & 1.32 & 0.66 & 0.98 & 1.42\\
 $\sigma_2$  & 0.30 & 0.16 & 0.60 & 0.54 & 0.31 & 0.90 & 0.62\\
 $M$ [\SI{}{\giga\eV}]  & 5.32 & 5.27 & 5.32 & 5.32 & 5.32 & 5.32 & 5.32\\
 & & & & & & &\\
\hline 
$\big[ B_s\to \phi\big]$ & & & & & & &\\
 $f(0)$  & 0.44 & 0.42 & 0.34 & 0.31 & 0.38 & 0.38 & 0.26\\
 $\sigma_1$  & 0.62 & 0.55 & 0.73 & 1.30 & 0.62 & 0.83 & 1.41\\
 $\sigma_2$  & 0.20 & 0.12 & 0.42 & 0.52 & 0.20 & 0.71 & 0.57\\
 $M$ [\SI{}{\giga\eV}] & 5.42 & 5.37 & 5.42 & 5.42 & 5.42 & 5.42 & 5.42\\
 & & & & & & &\\
\hline
\end{tabular}
\caption{Form factor fit parameters of $B_{(s)}\to M^{*}$ of the linear combinations of \ref{eq:Appendix_LinearCombinationFormFactors}. }\label{tab:Appendix_Pheno_Semileptonic_Melikhov_B}
\end{table}

\renewcommand{\arraystretch}{1.2}
\begin{table}[tb]
\centering
\begin{tabular}{|l||c|c|c|c|c|c|c||}
\hline
Fit Parameters & $V$ & $A_0$ & $A_1$ & $A_2$ & $T_1$ & $T_2$ & $T_3$ \\
\hline\hline
$\big[D\to \rho\big]$ & & & & & & &\\
 $f(0)$  & 0.90 & 0.66 & 0.59 & 0.49 & 0.66 & 0.66 & 0.31 \\
 $\sigma_1$ & 0.46 & 0.36  & 0.50 & 0.89 & 0.44 & 0.38 & 1.10 \\
 $\sigma_2$ & 0.00 & 0.00  & 0.00 & 0.00 & 0.00 & 0.50 & 0.17 \\
 $M$ [\SI{}{\giga\eV}]  & 2.01 & 1.87 & 2.01 & 2.01 & 2.01 & 2.01 & 2.01 \\
 & & & & & & &\\
\hline
$\big[D\to K^*\big]$ & & & & & & &\\
 $f(0)$  & 1.03 & 0.76 & 0.66 & 0.49 & 0.78 & 0.78 & 0.45\\
 $\sigma_1$  & 0.27 & 0.17 & 0.30 & 0.67 & 0.25 & 0.02 & 1.23\\
 $\sigma_2$  & 0.00  & 0.00 & 0.20 & 0.16 & 0.00 & 1.80 & 0.34 \\
 $M$ [\SI{}{\giga\eV}]  & 2.11 & 1.97 & 2.11 & 2.11 & 2.11 & 2.11 & 2.11 \\
 & & & & & & &\\
\hline
$\big[D_s\to K^*\big]$ & & & & & & &\\
 $f(0)$  & 1.04 & 0.67 & 0.57 & 0.42 & 0.71 & 0.71 & 0.45 \\
 $\sigma_1$  & 0.24 & 0.20 & 0.29 & 0.58 & 0.22 & -0.06 & 1.08 \\
 $\sigma_2$  & 0.00 & 0.00 & 0.42 & 0.00 & 0.00 & 0.44 & 0.68 \\
 $M$ [\SI{}{\giga\eV}]  & 2.01 & 1.87 & 2.01 & 2.01 & 2.01 & 2.01 & 2.01 \\
 & & & & & & &\\
\hline
$\big[D_s\to \phi \big]$ & & & & & & &\\
 $f(0)$  & 1.10 & 0.73 & 0.64 & 0.47 & 0.77 & 0.77 & 0.46\\
 $\sigma_1$  & 0.26 & 0.10 & 0.29 & 0.63 & 0.25 & 0.02 & 1.34\\
 $\sigma_2$  & 0.00 & 0.00 & 0.00 & 0.00 & 0.00 & 2.01 & 0.45\\
 $M$ [\SI{}{\giga\eV}] & 2.11 & 1.97 & 2.11 & 2.11 & 2.11 & 2.11 & 2.11\\
 & & & & & & &\\
\hline
\end{tabular}
\caption{Form factor fit parameters of $D_{(s)}\to M^{*}$ of the linear combinations of \ref{eq:Appendix_LinearCombinationFormFactors}. }\label{tab:Appendix_Pheno_Semileptonic_Melikhov_D}
\end{table}

\section{RPV derivation}\label{sec:appendix_rpv_derivation}

In this section, we perform the derivation of the 4-fermion vertex interactions in the 
context of R-parity violating supersymmetry.\footnote{We follow the procedure described 
in Ref.~\cite{deVries:2015mfw}. However, instead of using four-component Dirac spinors, 
we employ the two-component Weyl spinor notation used in Ref.~\cite{Martin:1997ns}. 
Further, we correct an erroneous factor in the Fierz-identity in Ref.~\cite{deVries:2015mfw}.}
The resulting dimension-6 operators can be matched to the effective Lagrangian 
Eq.~\eqref{d6CC} and Eq.~\eqref{d6NC}. We start with the chiral interactions in a 
supersymmetric extension of the Standard Model, which can be stated in terms of the 
superpotential $W$ (For an in-depth discussion of supersymmetry, see \textit{e.g.~}Refs.~\cite{Dreiner:2008tw,Martin:1997ns,Drees:2004jm}).
The most general form of the superpotential can be given as
\bea
W&=& W_{\text{MSSM}}+W_{\text{RPV}}\,,\\
W_{\text{MSSM}}&=& (y_u)_{ij} Q_iH_u \bar{u}_j -(y_d)_{ij} Q_iH_d \bar{d}_j-(y_l)_{ij} L_iH_d \bar{e}_j+\mu H_u H_d\,,\\
W_{\text{RPV}}&=& \kappa_i L_i H_u + \lambda_{ijk} L_i L_j \bar{e}_k + \lambda^\prime_{ijk} L_i Q_j \bar{d}_k + \lambda^{\prime\prime}_{ijk} \bar{u}_i\bar{d}_j\bar{d}_k\,.
\eea
A minimal supersymmetric extension of the Standard Model (in terms of interactions) 
imposes a discrete $\mathbb{Z}_2$ symmetry, called R-parity.
For a particle with baryon number $B$, lepton number $L$, and spin $S$, the R-parity is defined as $R_p=(-1)^{3(B-L)+2S}$. 
If R-parity is conserved, only the interactions of $W_{\text{MSSM}}$ are allowed.
If R-parity is violated, the interactions of $W_{\text{RPV}}$, which include three sets of lepton-number-violating and one set of baryon-number-violating terms, can also be non-zero.
To still ensure the stability of the proton, we can, for example, impose a discrete $\mathbb{Z}_3$ symmetry, the baryon triality~\cite{Dreiner:2012ae}, so that the $\lambda^{\prime\prime}\bar{u}\bar{d}\bar{d}$ terms are not allowed. Among the remaining three sets of lepton-number-violating operators, we are interested in the operators $\lambda^\prime LQ\bar{d}$, which contribute to the decay of mesons.
The interaction terms in component form are
\begin{align}
\mathcal{L}_{LQ\bar{d}}=\lambda^\prime_{ijk}\bigg[ &\big(e_i u_j\big)\stilde d^*_{Rk} + \big(e_i \bar{d}_k\big)\stilde u_{Lj} + \big(u_j \bar{d}_k\big)\stilde e_{Li}\nn\\
- &\big(\nu_i d_j\big)\stilde d^*_{Rk} - \big(\nu_i \bar{d}_k\big)\stilde d_{Lj} - \big(d_j \bar{d}_k\big)\stilde \nu_{Li} \bigg]+ \hc\,.
\end{align}
The relevant supersymmetric gauge-interactions are given in component form by 
\begin{align}
\mathcal{L}_{\text{gauge}}
\supset -\sqrt{2}g^\prime\bigg[&Y_{L}\stilde \nu^*_{Li}\big(\nu_i\stilde\chi^0_1\big)+Y_{L}\stilde e^*_{Li}\big(e_i\stilde\chi^0_1\big)\nn\\
+&Y_{Q}\stilde u^*_{Lj}\big(u_j\stilde\chi^0_1\big)+Y_{Q}\stilde d^*_{Lj}\big(d_j\stilde\chi^0_1\big)+Y_{\overline{d}}\stilde d_{Rk}\big(\overline{d}_k\stilde\chi^0_1\big)\bigg]+\hc\,.
\end{align}
We assume the sfermion fields to be heavy ($\sim m_{\mathrm{SUSY}}$) 
compared to the meson mass range ($m_{QCD}$). Hence below $m_{\text{SUSY}}$, 
we can integrate out the heavy degrees of freedom via the equation of 
motion.
The resulting effective tree-level Lagrangian is given by
\begin{align}
\mathcal{L}\supset \sqrt{2}g^\prime\lambda^\prime_{ijk}\bigg[&\frac{ Y_{\overline{d}}}{m^2_{\stilde d_{Rk}}}\big[\big(\nu_id_j\big)\big(\overline{d}_k\stilde\chi^{0}_1\big)-\big(e_iu_j\big)\big(\overline{d}_k\stilde\chi^{0}_1\big)\big]
+\frac{Y_Q}{m^2_{\stilde d_{Lj}}}\big(\nu_i\overline{d}_k\big)\big(d_j\stilde\chi^{0}_1\big)\nn\\
&-\frac{Y_Q}{m^2_{\stilde u_{Lj}}}\big(e_i\overline{d}_k\big)\big(u_j\stilde\chi^{0}_1\big)%\nn\\
+\frac{Y_L}{m^2_{\stilde \nu_{Li}}}\big(d_j\overline{d}_k\big)\big(\nu_i\stilde\chi^{0}_1\big)-\frac{Y_L}{m^2_{\stilde e_{Li}}}\big(u_j\overline{d}_k\big)\big(e_i\stilde\chi^{0}_1\big)\bigg]+\hc\,.
\end{align}
In order to rearrange the currents into pure-quark and lepton-neutralino currents, 
we  apply Fierz identities, assuming the sfermions are mass degenerate $m_{\stilde f}\approx 
m_{\mathrm{SUSY}}$ and insert the explicit hypercharges 
\begin{align}
\mathcal{L}\supset 
\color{black}
 \frac{3\sqrt{2}g^\prime\lambda^\prime_{ijk}}{4m_{\mathrm{SUSY}}^2}\bigg\{&\color{black}\big(\stilde\chi^{0}_1e_i\big)\big(\overline{d}_ku_j\big)-\big(\stilde\chi^{0}_1\nu_i\big)\big(\overline{d}_kd_j\big)\nn\\
\color{black}-&\color{black}\frac{1}{9}\big(\stilde\chi^{0}_1\sigma^{\mu\nu}_{\subtwocomp}e_i\big)\big(\overline{d}_k\sigma_{\subtwocomp,\mu\nu}u_j\big)+\frac{1}{9}\big(\stilde\chi^{0}_1\sigma^{\mu\nu}_{\subtwocomp}\nu_i\big)\big(\overline{d}_k\sigma_{\subtwocomp,\mu\nu}d_j\big)\bigg\}+\hc\,.
\end{align}
To relate the effective interactions in two-component notation to 
four-component Dirac spinors, we use Appendix G of Ref.~\cite{Dreiner:2008tw}.
The final effective interactions in four-component notation in the mass basis (noting that $d^i_L = V^{ij}d^{j,\text{mass}}_L$) are
\begin{align}
\mathcal{L}\supset& \frac{3\sqrt{2}g^\prime\lambda^\prime_{ijk}}{4m_{\mathrm{SUSY}}^2}\bigg\{\big(\bar{\stilde\chi}^{0}_1e_{L}^i\big)\big(\bar d_{R}^ku_{L}^j\big)-\frac{1}{36}\big(\bar{\stilde\chi}^{0}_1\sigma^{\mu\nu}_{\subfourcomp}e_{L}^i\big)\big(\bar d_{R}^k\sigma_{\subfourcomp,\mu\nu}u_{L}^j\big)\bigg\}\\
+&\frac{3\sqrt{2}g^\prime\lambda^\prime_{ilk}}{4m_{\mathrm{SUSY}}^2}V^{lj}\bigg\{-\big(\bar{\stilde\chi}^{0}_1\nu_{L}^i\big)\big(\bar d_{R}^kd_{L}^j\big)+\frac{1}{36}\big(\bar{\stilde\chi}^{0}_1\sigma^{\mu\nu}_{\subfourcomp}\nu_{L}^i\big)\big(\bar d_{R}^k\sigma_{\subfourcomp,\mu\nu}d_{L}^j\big)\bigg\}+\hc\,.
\end{align}

\bibliographystyle{JHEP}
\bibliography{bibliography}
\end{document}